\documentclass[aps,prl,reprint,superscriptaddress,longbibliography,nofootinbib]{revtex4-2}

\usepackage{graphicx}
\usepackage{dcolumn}
\usepackage{bm}

\usepackage[figurename=Figure]{caption}
\usepackage{ragged2e}
\DeclareCaptionJustification{plain}{\justifying}
\usepackage[percent]{overpic}
\usepackage{float}    
\usepackage{verbatim} 
\usepackage{amsmath}  
\usepackage{bbold}
\usepackage{upgreek} 
\usepackage{booktabs}
\usepackage{amssymb}  
\usepackage{svg}
\usepackage{enumitem} 
\usepackage{latexsym,epsfig}
\usepackage{subcaption}
\usepackage{stackrel}
\usepackage{mathtools}

\usepackage{placeins} 

\usepackage[colorlinks=true,breaklinks=true,allcolors=blue]{hyperref}
\usepackage[capitalize]{cleveref}
\usepackage{lipsum}
\usepackage{dsfont}
\usepackage{coffeestains}

\usepackage[scr=boondox]{mathalfa}

\newcommand{\be}{\begin{equation}}
	\newcommand{\ee}{\end{equation}}
\newcommand{\bea}{\begin{equation}\begin{aligned}}
		\newcommand{\eea}{\end{aligned}\end{equation}}
\newcommand{\ben}{\begin{enumerate}}
	\newcommand{\een}{\end{enumerate}}

\DeclareDocumentCommand{\nint}{ O{} O{} m }{\ensuremath{ \int_{\mbox{\scriptsize $#1$}}^{\mbox{\scriptsize$#2$}}\!\!\! \mbox{\small $\,\mathrm{d}#3$\! }}}

\usepackage{tikz,bm} 
\usepackage{circuitikz}
\usetikzlibrary{decorations.markings}
\usetikzlibrary{decorations.pathreplacing,calligraphy}

\definecolor{mycolor}{rgb}{1,0.2,0.3}
\definecolor{brightgreen}{rgb}{0.4, 1.0, 0.0}
\definecolor{britishracinggreen}{rgb}{0.0, 0.26, 0.15}
\definecolor{cadmiumgreen}{rgb}{0.0, 0.42, 0.24}
\definecolor{ceruleanblue}{rgb}{0.16, 0.32, 0.75}
\definecolor{darkelectricblue}{rgb}{0.33, 0.41, 0.47}
\definecolor{darkpowderblue}{rgb}{0.0, 0.2, 0.6}
\definecolor{darktangerine}{rgb}{1.0, 0.66, 0.07}
\definecolor{emerald}{rgb}{0.31, 0.78, 0.47}
\definecolor{palatinatepurple}{rgb}{0.41, 0.16, 0.38}
\definecolor{pastelviolet}{rgb}{0.8, 0.6, 0.79}

\begin{document}
	
	\preprint{APS/123-QED}
	
	\title{Locked Subharmonic Oscillations in the Entanglement Spectrum \\ of a Periodically Driven Topological Chain}
	\author{Rishabh Jha}
	\email{rishabh.jha@usc.edu}
	\affiliation{%
		Department of Physics and Astronomy, University of Southern California, Los Angeles, CA 90089-0484, USA
	}
	
	%


\begin{abstract}
Periodically driven quantum systems can exhibit subharmonic response, usually characterized through physical observables and often discussed in interacting settings. Here we show that a sharp subharmonic signature already appears in the \emph{entanglement spectrum} of a number-conserving free-fermion system. We study a two-step driven Su--Schrieffer--Heeger chain whose Floquet operator supports symmetry-protected edge modes at quasienergies $0$ and $\pi$. When the initial state is a coherent superposition of these two edge sectors, we show that the subsystem correlation matrix alternates between two stroboscopic structures, and the entanglement spectrum is period-doubled as a set, while an overlap-tracked entanglement level shows a robust period-doubling response with Fourier weight concentrated at half the drive frequency. By contrast, diagonal edge densities remain flat by sublattice symmetry, while an off-diagonal edge-bond observable provides the corresponding linear one-body comparator. The effect disappears both when the initial state is replaced by a
stroboscopically stationary Floquet eigenstate built from the same topological mode content, and when the system is placed in the topologically trivial phase where no edge modes exist.
Altogether, these establish zero--$\pi$ Floquet topology as a necessary condition and coherent nonequilibrium preparation as the additional sufficient ingredient. Our results identify entanglement spectroscopy as a sharp subsystem-resolved probe of Floquet topological coherence.
\end{abstract}

	\maketitle
	


\textit{Introduction.---}
Periodic driving provides a route to engineering quantum matter beyond equilibrium. In a time-periodically driven system, the evolution over one driving cycle defines an effective Floquet description in terms of quasienergy bands and stroboscopic dynamics, opening possibilities unavailable in static band structures~\cite{Shirley1965,Sambe1973,Kitagawa2010,Bukov2015}. This perspective has led to the broader program of Floquet engineering, in which periodic drives are used to induce and control topological phases, edge states, and band inversions dynamically~\cite{Lindner2011,Rudner2013,Asboth2014,Nathan2015,Oka2019,RudnerLindner2020}. In particular, driven systems can support protected edge modes at quasienergies zero and $\pi$, and anomalous Floquet phases show that the bulk--edge correspondence in nonequilibrium settings is richer than any direct static analogy would suggest~\cite{Kitagawa2010,Rudner2013,Asboth2014,Nathan2015}. 

A second major theme in periodically driven matter is the possibility of robust subharmonic response. Following the original proposal of time crystals and the subsequent no-go results for equilibrium realizations~\cite{Wilczek2012,Bruno2013,Watanabe2015}, it was recognized that discrete time-translation symmetry can nevertheless be broken in genuinely nonequilibrium Floquet systems~\cite{Khemani2016,Else2016,vonKeyserlingk2016,Yao2017,ElseMonroeNayak2020}. In such phases, observables oscillate with a period that is an integer multiple of the drive period, with the period-doubled response of discrete time crystals providing the paradigmatic example~\cite{Khemani2016,Else2016,Yao2017}. Experimental observations in trapped ions and nitrogen-vacancy spin systems have since established robust subharmonic signatures in interacting quantum platforms~\cite{Zhang2017,Choi2017}. In essentially all of this literature, however, the subharmonic signal is discussed through physical observables, and its stability is tied to many-body mechanisms such as localization or prethermalization~\cite{Else2016,vonKeyserlingk2016,Yao2017,ElseMonroeNayak2020,SachaZakrzewski2017,Zaletel2023}.

At the same time, the entanglement spectrum (ES) has emerged as a powerful diagnostic of quantum structure beyond what is visible in a single scalar entropy. Since the work of Li and Haldane, which showed that the spectrum of the reduced density matrix can encode topological information inaccessible to the von Neumann entropy alone~\cite{LiHaldane2008}, entanglement spectroscopy has become a standard probe of topological order and edge physics~\cite{Laflorencie2016}. For free fermions, the reduced density matrix and its single-particle ES can be obtained directly from correlation functions, making the method especially sharp and transparent in quadratic systems~\cite{Peschel2003,PeschelEisler2009,Fidkowski2010,Turner2010}.
Floquet settings have also revealed that entanglement spectra can carry information not trivially identical to the quasienergy spectrum, as shown for driven Kitaev chains with zero and $\pi$ Majorana structure~\cite{Yates2017}, and that entanglement measures can undergo sharp dynamical transitions in periodically driven spin chains~\cite{BanerjeeSengupta2024,Zhou2024,Verga2023,Gadge2026Mar}.
Yet most uses of the ES remain essentially static: one asks what topology a state contains, not whether the ES itself can provide a clean dynamical probe of coherent Floquet edge-sector interference.

Here we show that such a response already appears in a minimal, number-conserving free-fermion setting. We study a two-step driven Su--Schrieffer--Heeger (SSH) chain~\cite{Su1979,Su1980,Asboth2014} whose Floquet operator supports symmetry-protected zero and $\pi$ edge modes, and we demonstrate that a coherent superposition of these two edge sectors produces a period-$2T$ response in a tracked single-particle entanglement level. The effect is absent for a stroboscopically stationary Floquet eigenstate built from the same topological mode structure (proving not sufficient), and is also absent in the topologically trivial phase where no edge modes exist (showing necessity); coherent nonequilibrium zero--$\pi$ preparation is the additional sufficient ingredient. Our results therefore identify entanglement spectroscopy as a subsystem-defined dynamical probe of Floquet zero--$\pi$ coherence.

\begin{figure*}[t]
\includegraphics[width=\textwidth]{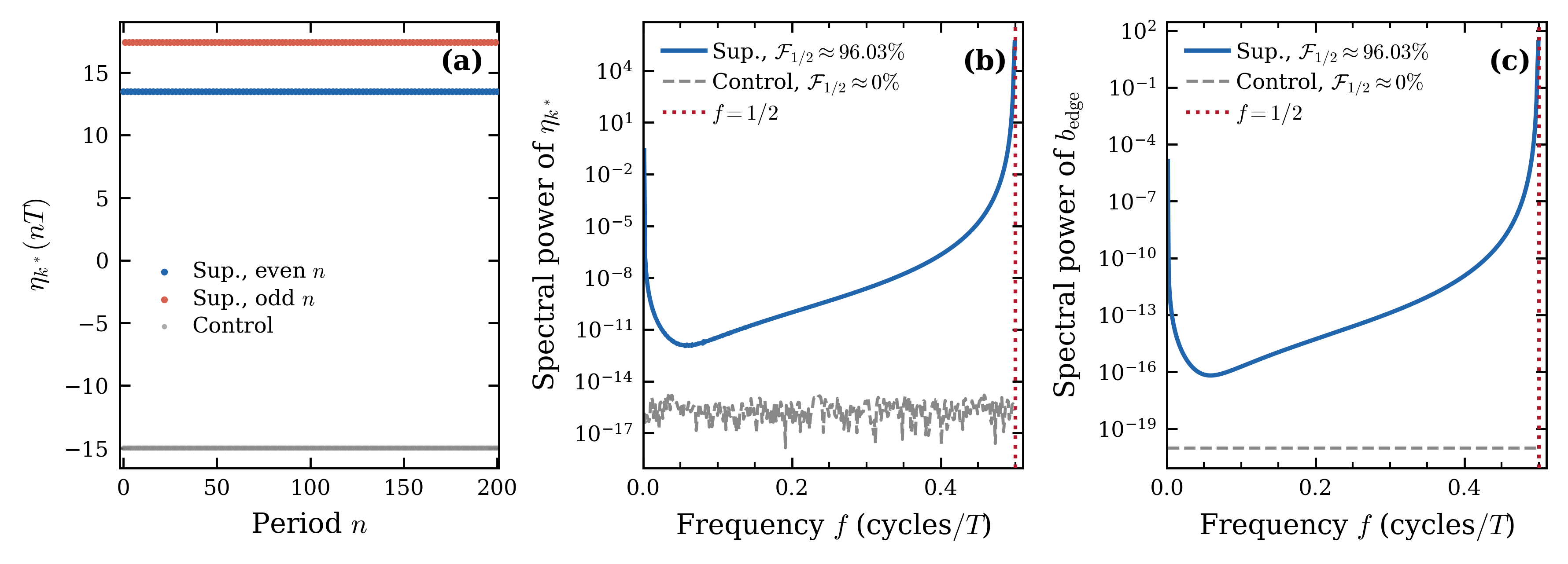}
\caption{
Subharmonic response in the driven SSH chain.
(a)~Tracked entanglement energy $\eta_{k^*}(nT)$ for the coherent zero--$\pi$ superposition state (blue/orange, even/odd periods) and for the $\pi$-mode Floquet-eigenstate control (gray, all periods shown); for the control, even and odd points coincide on a single flat branch, i.e., no even--odd splitting is present. The first 200 of the full $N=800$ stroboscopic periods are shown.
(b)~Hanning-windowed Fourier power of $\eta_{k^*}(nT)$ computed from all $N=800$ periods; the red dotted line marks the subharmonic frequency $f=1/2$, and the normalized fraction $\mathcal{F}_{1/2}=P(1/2)/\sum_{f>0}P(f)$ is stated in the legend.
(c)~Same Fourier diagnostic applied to the left-edge bond observable $b_{\mathrm{edge}}(nT)$, with $w=\lfloor\sqrt{L}\rfloor=22$.
For the coherent superposition state, both $\eta_{k^*}(nT)$ and $b_{\mathrm{edge}}(nT)$ show dominant subharmonic weight, whereas the $\pi$-mode eigenstate control shows no resolved subharmonic response. The diagonal edge density remains flat by a chiral-symmetry selection rule; see Fig.~\ref{fig:supplemental_diagnostics} in the appendix below.
Parameters: $\delta_0=-0.3$, $\delta_{\mathrm{K}}=0.8$, $L=500$, $L_A=\lfloor\sqrt{L}\rfloor=22$, and $N=800$ stroboscopic periods.
}
\label{fig:main}
\end{figure*}

\textit{Model and Protocol.---}
We study an open chain of $L$ spinless fermions at half-filling ($N=L/2$), driven by a two-step protocol. During $0<t\le T/2$ the system evolves under the Su--Schrieffer--Heeger~(SSH) Hamiltonian
\begin{equation}
H_0=\sum_{j=1}^{L-1}t_0\bigl[1+\delta_0(-1)^{j+1}\bigr]
     \bigl(c_j^\dagger c_{j+1}+\mathrm{h.c.}\bigr),
\label{eq:H0}
\end{equation}
and during $T/2<t\le T$ under an SSH Hamiltonian of the same form with $\delta_0\to\delta_{\mathrm K}$. Setting $t_0=t_{\mathrm K}=1$ throughout fixes the energy scale and leaves three free parameters: $\delta_0$, $\delta_{\mathrm K}$, and $T/2$; all quoted half-period values are measured in units of $1/t_0$. The one-period Floquet operator is
\begin{equation}
U(T)=e^{-iH_{\mathrm K}T/2}\,e^{-iH_0T/2}.
\label{eq:UF}
\end{equation}
Both SSH steps respect sublattice (chiral) symmetry, so the Floquet spectrum is symmetric under $\theta\to-\theta$ and the symmetry-fixed quasienergies are $0$ and $\pi$; see SM~\cite{jha_supplemental} for the proof.
For $\delta_0<0$ and suitable $(\delta_{\mathrm K},T/2)$, the bulk Floquet spectrum is gapped at both values and $U(T)$ supports one edge-localized zero mode and one edge-localized $\pi$ mode; their simultaneous presence is the topological necessity for the effect we report.

For the two choices of initial states used in this work, both can be expressed as Slater determinants with $N=L/2$ occupied orbitals, collected as columns of an $L\times N$ matrix $V$, with equal-time correlator $C_{ij}=\langle c_j^\dagger c_i\rangle=(VV^\dagger)_{ij}$; stroboscopic evolution advances $V\mapsto U(T)V$ each period, so $C((n+1)T)=U(T)\,C(nT)\,U(T)^{\dagger}$.
In the \emph{superposition state}, all bulk negative phase Floquet modes are occupied and the zero and $\pi$ edge modes enter as equal-weight combinations; the following equation shows a representative left-localized pair:
\begin{equation}
|\Psi_{\mathrm{sup}}\rangle
  =\tfrac{1}{\sqrt{2}}\bigl(|\Phi_0\rangle+|\Phi_\pi\rangle\bigr).
\label{eq:super}
\end{equation}
Because $U(T)|\Phi_0\rangle=+|\Phi_0\rangle$ and $U(T)|\Phi_\pi\rangle=-|\Phi_\pi\rangle$, the two sectors accumulate a relative sign $(-1)^n$ after $n$ periods; the state is genuinely nonequilibrium. In the \emph{eigenstate control}, $|\Psi_{\mathrm{sup}}\rangle$ is replaced by the $\pi$ mode alone. The occupied subspace is then Floquet-invariant and $C(nT)=C(0)$ exactly at every stroboscopic time.

We restrict $C$ to the leftmost $L_A=\lfloor\sqrt{L}\rfloor$ sites, forming $C_A$. The reduced density matrix can always be written as $\rho_A\propto e^{-H_E}$, thereby defining the many-body entanglement Hamiltonian $H_E$. For a Gaussian state, the associated single-particle entanglement Hamiltonian $h$ is related to $C_A$ by $C_A=(\mathbb{1}+e^{h})^{-1}$~\cite{Peschel2003,PeschelEisler2009}.
If $\{\xi_\ell\}$ are the eigenvalues of $C_A$, the corresponding single-particle entanglement energies, i.e. the eigenvalues of $h$, are
\begin{equation}
\eta_\ell=\ln\!\left[\frac{1-\xi_\ell}{\xi_\ell}\right].
\label{eq:eta}
\end{equation}
The choice $L_A=\lfloor\sqrt{L}\rfloor$ is operational: it resolves the left-edge support while remaining small compared with the full chain; nearby smaller values give the same response, whereas much larger $L_A$ admits sufficient bulk weight to weaken overlap-based tracking.
We therefore track level $\eta_{k^*}$, where $k^*(n)\equiv
\operatorname{arg\,max}_{\ell}\,|\langle\phi_{\ell}^{A}(nT)|\Phi_{\pi}^{A}\rangle|^{2}$, with $|\phi_{\ell}^{A}(nT)\rangle$ the eigenvectors of $C_{A}(nT)$ and $|\Phi_{\pi}^{A}\rangle$ the fixed restriction of $|\Phi_{\pi}\rangle$ to $A$.
Our overlap-based identification of the edge-dominated entanglement level is a numerical tracking heuristic in the spirit of standard maximum-overlap mode-following procedures used to associate eigenstates across avoided crossings and parameter sweeps; analogous overlap-based tracking has been used explicitly in Floquet calculations~\cite{HolderReichl2005}. This operational rule isolates the cleanest edge-dominated oscillator; it is not intended as a unique continuation of an eigenvalue label through level crossings.

The zero--$\pi$ coherence in $|\Psi_{\mathrm{sup}}\rangle$ causes $C_A$ to alternate between two distinct matrices on even and odd periods, guaranteeing period doubling of the ES as a \textit{set} but not of every ordered level individually, since level permutations and mixing with nearby bulk-dominated modes can obscure branch-by-branch tracking. The nontrivial task is therefore not to prove set-wise alternation, which is exact, but to identify a stable edge-dominated branch on which that alternation becomes a clean long-lived single-level signal. We now show that a simple overlap-tracking rule does so in the topological regime and fails in the stroboscopically stationary control state.

\textit{Subharmonic Response.---}
Figure~\ref{fig:main} shows the stroboscopic dynamics for $\delta_0=-0.3$, $\delta_{\mathrm K}=0.8$, $L=500$, $L_A=22$, and $N=800$ drive periods. In addition to the tracked entanglement level $\eta_{k^*}(nT)$, we monitor two left-edge one-body observables over the same window $w=L_A$: the edge density
\(
n_{\mathrm{edge}}(nT)=\frac{1}{w}\sum_{j=1}^{w}\langle c_j^\dagger c_j\rangle(nT)
\)
and the edge bond average
\(
b_{\mathrm{edge}}(nT)=\frac{1}{w-1}\sum_{j=1}^{w-1}
\langle c_j^\dagger c_{j+1}+c_{j+1}^\dagger c_j\rangle(nT).
\)
For any tracked series $x(nT)$ we quantify the subharmonic response by
\(
\mathcal{F}_{1/2}\equiv P(1/2)/\sum_{f>0}P(f),
\)
where $P(f)$ is the Hanning-windowed Fourier power of the mean-subtracted signal; see SM~\cite{jha_supplemental} for further details.

For the superposition state, $\eta_{k^*}(nT)$ separates into two stable branches on alternating periods [Fig.~\ref{fig:main}(a)], and its Fourier weight is concentrated at half the drive frequency, with $\mathcal{F}_{1/2}$ close to unity [Fig.~\ref{fig:main}(b)]. The bond observable $b_{\mathrm{edge}}(nT)$ shows the same subharmonic peak [Fig.~\ref{fig:main}(c)]; the diagonal density $n_{\mathrm{edge}}(nT)$ remains flat by sublattice symmetry; see Fig.~\ref{fig:supplemental_diagnostics} in the appendix below.
For the eigenstate control, $C(nT)=C(0)$ exactly, so $\eta_{k^*}(nT)$, $b_{\mathrm{edge}}(nT)$, and $n_{\mathrm{edge}}(nT)$ are all stroboscopically stationary and the subharmonic weight vanishes completely.

This figure establishes two crucial points. First, the analytic structure fixes the period-$2T$ response of the ES only at the set level, not at the level of an ordered eigenvalue label through crossings. The role of the largest-overlap rule is therefore operational and quantitative: it identifies the edge-dominated branch on which this exact set-wise alternation is realized as the cleanest single-level oscillation. Second, the useful linear comparator is the off-diagonal bond observable, not the diagonal edge density, which remains flat by sublattice symmetry.
The observed response requires coherent nonequilibrium zero--$\pi$ preparation; a complementary trivial-phase check with no edge modes likewise shows no subharmonic signal, confirming necessity of zero--$\pi$ Floquet mode structure (see SM~\cite{jha_supplemental}).

For the $L=500$ data shown here, reducing $L_A$ from $22$ to $12$ leaves the response unchanged, whereas $L_A=100$ suppresses and $L_A=200$ substantially degrades the tracked oscillation; this confirms that the signal is controlled by the edge-localized sector and weakens once the subsystem extends far into the bulk. Moreover, although the top-overlap tracked level is the cleanest one, additional levels can also satisfy the strict even--odd stability diagnostic while others do not (see appendix below), showing that the underlying period-$2T$ structure is not confined to a single cherry-picked branch.

\textit{Mechanism.---}
The origin of the locked $2T$ response is transparent: consider a single occupied zero--$\pi$ superposition orbital on top of the negative-phase bulk modes. At stroboscopic times,
\begin{equation}
|\psi(nT)\rangle=\frac{1}{\sqrt{2}}\Bigl(|\Phi_0\rangle+(-1)^n|\Phi_\pi\rangle\Bigr),
\label{eq:psi_n}
\end{equation}
whereas each occupied bulk Floquet mode contributes only a stationary projector.
The full correlation matrix therefore decomposes as
\begin{equation}
C(nT)=C_{\mathrm{bulk}}+|\psi(nT)\rangle\langle \psi(nT)|.
\label{eq:C_decomp}
\end{equation}
A pure $\pi$-mode eigenstate contributes only the stationary projector $|\Phi_\pi\rangle\langle\Phi_\pi|$ to the full correlation matrix. Restricting to the left subsystem $A$ gives
\begin{equation}
C_A(nT)=C_{A,\mathrm{bulk}}+|\psi_A(nT)\rangle\langle \psi_A(nT)|,
\label{eq:CA_decomp}
\end{equation}
with $|\psi_A(nT)\rangle$ the restriction of $|\psi(nT)\rangle$ to $A$.
Because $|\psi_A(nT)\rangle$ takes only two values on even and odd periods, $C_A(nT)$ alternates between two matrices, so the ES is period-$2T$ as a \textit{set}. This exact statement does not by itself specify a unique ordered level through crossings, nor does it determine how sharply an edge-controlled branch remains isolated once bulk-dominated levels are present. Individual levels therefore need not all display equally clean period doubling, and indeed some do while others do not (see appendix below). The role of the largest-overlap rule is precisely to extract the edge-dominated branch for which the exact set-wise alternation becomes a sharp, long-lived single-level signal, as demonstrated in Fig.~\ref{fig:main}. For several occupied zero--$\pi$ pairs, the same conclusion holds after summing the corresponding projectors.

This same decomposition explains why diagonal density probes remain silent. For any diagonal one-body operator
\(
O_f=\sum_j f_j c_j^\dagger c_j,
\)
the oscillatory contribution is proportional to the interference matrix element $\langle \Phi_0|O_f|\Phi_\pi\rangle$. In the present chiral SSH setting, the zero and $\pi$ edge modes can be chosen on opposite sublattices, so this matrix element vanishes identically. As a result, $n_{\mathrm{edge}}(nT)$ is stroboscopically flat even for the coherent superposition state. By contrast, the bond operator entering $b_{\mathrm{edge}}(nT)$ connects opposite sublattices and is therefore not subject to this selection rule, providing the natural linear one-body comparator to the ES signal.

\begin{figure*}[t]
  \begin{subfigure}{0.49\linewidth}
    \includegraphics[width=\linewidth]{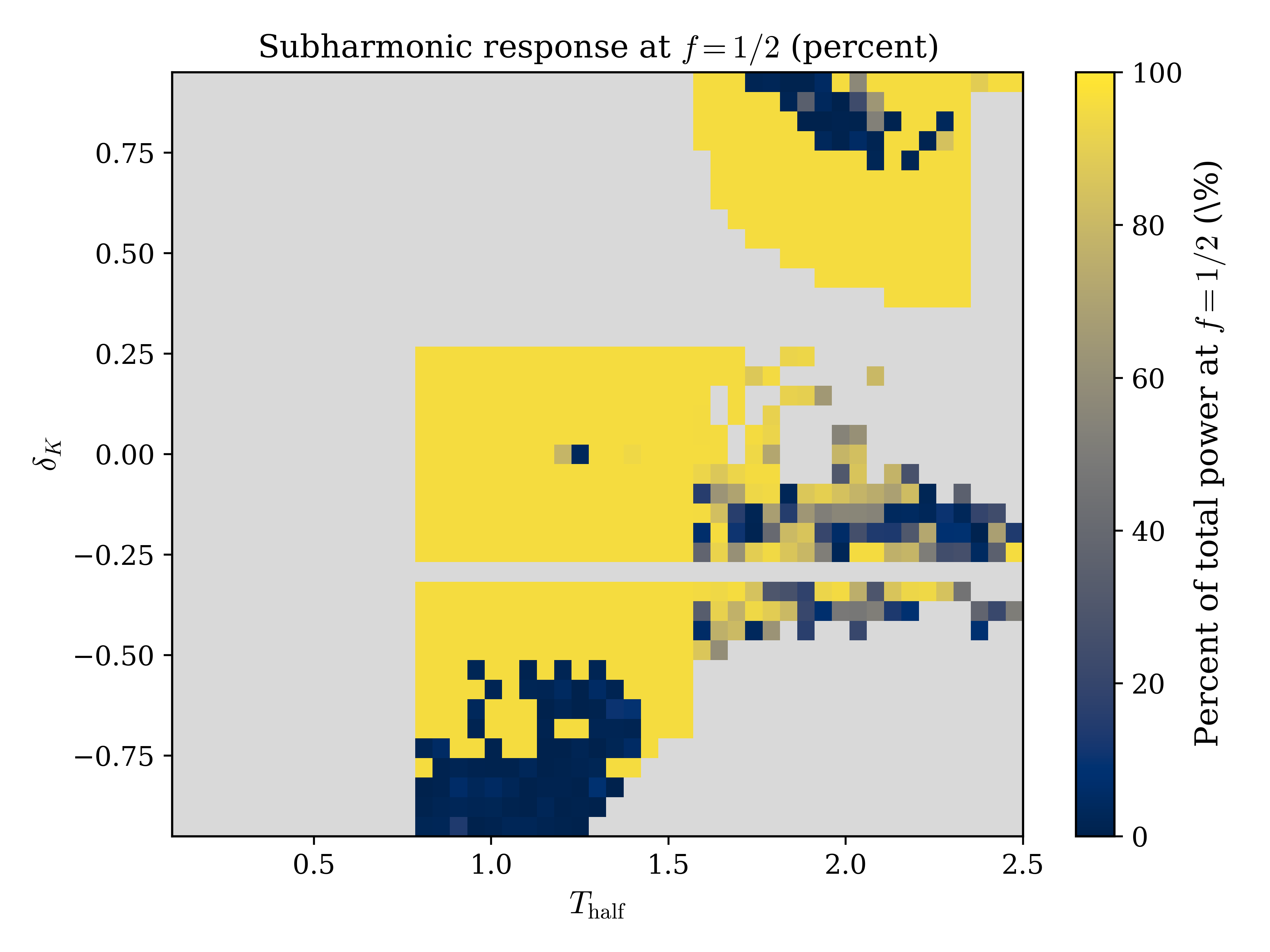}
    \caption{}
    \label{fig:phase_dtc}
  \end{subfigure}
  \hfill
  \begin{subfigure}{0.49\linewidth}
    \includegraphics[width=\linewidth]{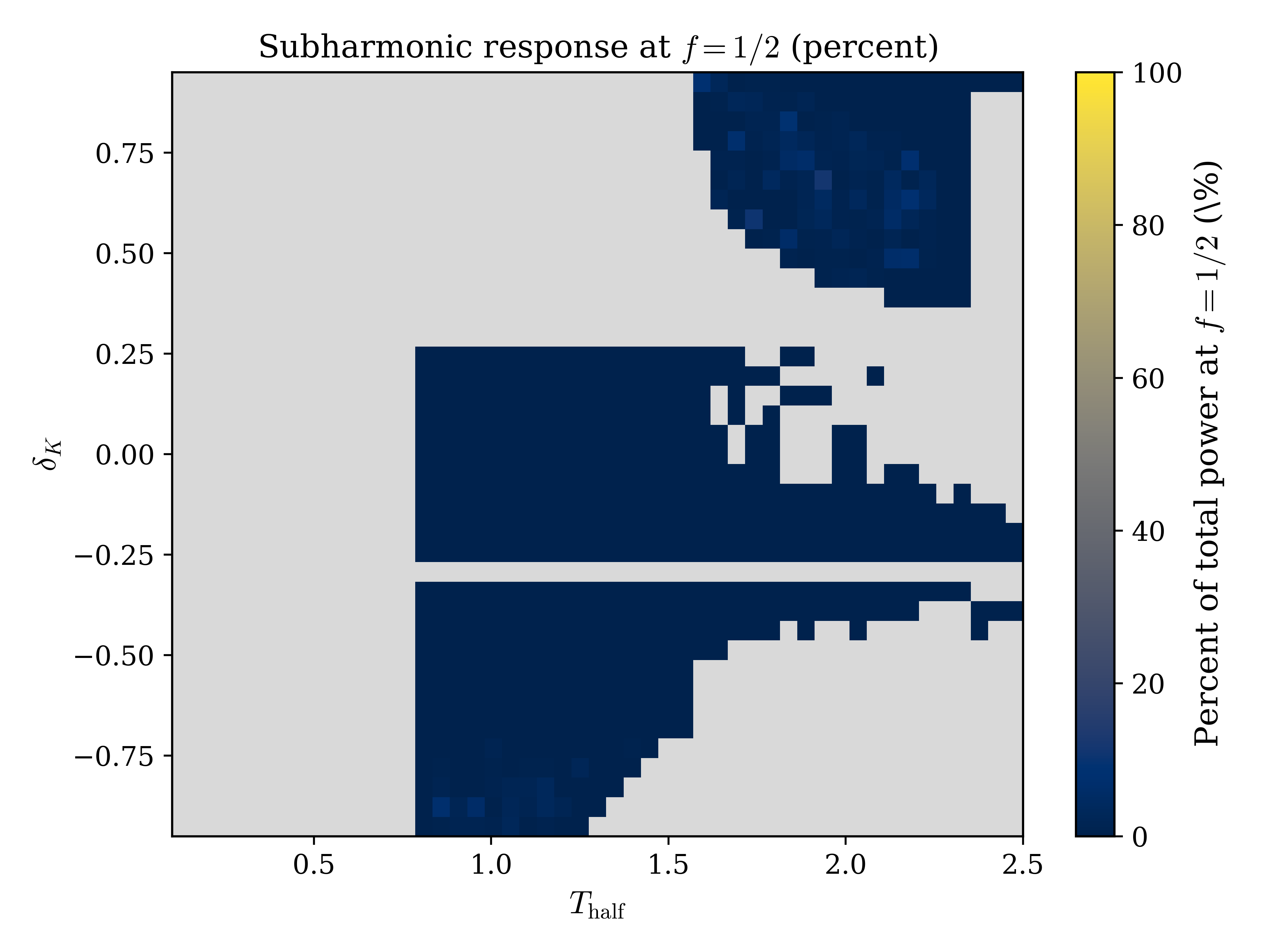}
    \caption{}
    \label{fig:phase_eig}
  \end{subfigure}
  \caption{%
    Phase diagram of the entanglement-spectrum subharmonic response in the $(\delta_{\mathrm K},T/2)$ plane at $\delta_0=-0.3$, $L=500$, $L_A=\lfloor\sqrt{L}\rfloor=22$, and $N=100$ periods.
    Gray cells fail the topological bulk--boundary proxy (see text) and are excluded from both panels; both panels share the same $0$--$100\%$ color scale. The proxy‑active region includes points where the $\pi$ mode is localized only on the right edge; such points show no subharmonic response in our left‑subsystem measurement, so the bright (high‑$\mathcal{F}_{1/2}$) region is a proper subset of the proxy window.
    (a)~Coherent zero--$\pi$ superposition state:
    The subharmonic weight $\mathcal{F}_{1/2}$ is large throughout an extended region of the proxy-active window, reaching almost unity at the optimal working point; the even--odd amplitude $|\Delta_{eo}|\approx3.94$ with $\mathrm{SNR}\sim10^8$ at that point further confirms a clean period-$2T$ even--odd split on the highest-overlap edge-dominated branch, and multiple other entanglement levels independently satisfy the same flatness criterion (see appendix below for the complete overlap-ranked diagnostic table).
    (b)~$\pi$-mode Floquet eigenstate (control):
    The map is uniformly dark on the same scale across the entire proxy-active window: the eigenstate is stroboscopically stationary by construction and carries no period-$2T$ content, even where the proxy is fully satisfied.
    Together with the trivial-phase control in the SM~\cite{jha_supplemental}, the contrast with (a) shows that zero--$\pi$ topological mode structure is necessary but not sufficient for locked ES period doubling; coherent nonequilibrium zero--$\pi$ preparation is additionally required.
  }
  \label{fig:phase}
\end{figure*}

\textit{Phase Diagram.---}%
To establish that the subharmonic response of the ES is not an artifact of a fine-tuned working point, we scan $\mathcal{F}_{1/2}$ across the full $(\delta_{\mathrm K},T/2)$ parameter plane at fixed $\delta_0=-0.3$, $L=500$, and $L_A=\lfloor\sqrt{L}\rfloor=22$ [Fig.~\ref{fig:phase}].
Before any real-time evolution, we evaluate a bulk--boundary proxy from $U(T)$, where $n_0$ and $n_\pi$ count the zero and $\pi$ Floquet modes, $\theta_\alpha$ is the quasienergy eigenphase, and $|\Phi_\alpha\rangle$ is the eigenvector of mode $\alpha$. A point is proxy-active if and only if three conditions hold simultaneously:
(i)~at least one zero--$\pi$ pair is present, $\min(n_0,n_\pi)\ge1$;
(ii)~the bulk Floquet spectrum is gapped at both quasienergy fixed points, $\Delta_0:=\min_{\alpha\notin 0\text{-modes}}|\theta_\alpha|\ge\delta_{\mathrm{tol}}$ and $\Delta_\pi:=\min_{\alpha\notin \pi\text{-modes}}\bigl||\theta_\alpha|-\pi\bigr|\ge\delta_{\mathrm{tol}}$, with $\delta_{\mathrm{tol}}=0.05$; and
(iii)~at least one zero mode and one $\pi$ mode each concentrate more than $30\%$ of their probability within
$w=\lfloor\sqrt{L}\rfloor$ sites of either boundary,
$W_{\mathrm{edge}}(\alpha):=\sum_{j=1}^{w}|\Phi_\alpha(j)|^2+\sum_{j=L-w+1}^{L}|\Phi_\alpha(j)|^2\ge w_{\mathrm{thr}}$
for normalized eigenmodes, with $w_{\mathrm{thr}}=0.30$.
Points failing any condition are masked (gray cells in Fig.~\ref{fig:phase}). The proxy-active region includes points where the $\pi$ mode is localized only on the right edge; such points show no subharmonic response in our left‑subsystem measurement, so the bright (high‑$\mathcal{F}_{1/2}$) region in Fig.~\ref{fig:phase}(a) is a proper subset of the proxy window.

We have verified that the phase-diagram results are robust to physically reasonable variations of the proxy criteria: recomputing the maps with different $\delta_{\mathrm{tol}}$ and $w_{\mathrm{thr}}$ leaves the proxy boundary, the pair-count map, and the $\mathcal{F}_{1/2}$ map unchanged.
For the parameters shown here, the response is stable and numerically converged under changes in system size and subsystem size: reducing $L$ from $500$ to $450$ and $400$ at $L_A=\lfloor\sqrt{L}\rfloor$, or reducing $L_A$ from $22$ to $12$ at fixed $L=500$, produces no detectable change in either $\mathcal{F}_{1/2}$ or the time-domain signal.
The signal weakens only when $L_A$ is taken far into the bulk, where the tracked level is no longer cleanly edge-dominated.
Further robustness checks of the proxy thresholds, finite-size dependence, and subsystem-size dependence are given in SM~\cite{jha_supplemental}.

Within the proxy-active window, the two initial conditions produce strikingly different outcomes [Fig.~\ref{fig:phase}].
For the coherent zero--$\pi$ superposition [Fig.~\ref{fig:phase}(a)], $\mathcal{F}_{1/2}$ is large throughout an extended region, reaching almost unity at the optimal working point.
A complementary even--odd stationarity diagnostic is also useful: writing $\bar\eta_e$ and $\bar\eta_o$ for the means of $\eta_{k^*}(nT)$ over the even- and odd-$n$ stroboscopic subsequences, and $\sigma_e$ and $\sigma_o$ for the corresponding standard deviations, we report $|\Delta_{eo}|:=|\bar\eta_e-\bar\eta_o|$ only when the signal-to-noise ratio $\mathrm{SNR}:=|\Delta_{eo}|/\max(\sigma_e,\sigma_o)\ge 3$, thereby requiring each subsequence to be individually flat rather than merely to have unequal means (see SM~\cite{jha_supplemental}).
By this criterion, the highest-overlap tracked level alternates between two stable values with $|\Delta_{eo}|\approx 3.94$ and $\mathrm{SNR}\sim 10^8$. Several other levels at lower overlap with the edge reference also satisfy the same flatness requirement, while others do not. This is the expected pattern when the exact period-$2T$ structure belongs to the ES as a set, whereas the visibility of any particular ordered level depends on overlap with the edge sector and on local mixing with nearby bulk-dominated modes. The complete overlap-ranked diagnostic table is given in the appendix below. For the $\pi$-mode Floquet eigenstate initialized at the identical topological parameters [Fig.~\ref{fig:phase}(b)], $\mathcal{F}_{1/2}$ is everywhere consistent with zero across the entire proxy-active window, and the even--odd diagnostic finds no clean period-$2T$ oscillator among any tracked level, because the eigenstate is stroboscopically stationary by construction: its correlation matrix is period-$T$ periodic and no observable can acquire a period-$2T$ component.
These results establish that the zero--$\pi$ Floquet mode structure is a \emph{necessary} but not \emph{sufficient} condition: the entanglement-spectrum subharmonic response additionally requires that the initial state be a nonequilibrium coherent superposition of the zero and $\pi$ edge sectors.

\textit{Conclusion and Outlook.---}
We have shown that a periodically driven topological chain can exhibit a clear subharmonic response in its \emph{entanglement spectrum} (ES) even in a number-conserving free-fermion setting. In the two-step SSH drive studied here, the essential ingredients are the coexistence of zero and $\pi$ Floquet edge modes and a genuinely nonequilibrium initial preparation that coherently superposes the two sectors. Under those conditions, the subsystem correlation matrix alternates between two stroboscopic structures, so the ES is period-$2T$ as a set, while an overlap-tracked edge-dominated entanglement level isolates this exact structure as a clean period-$2T$ single-level signal. By contrast, a Floquet eigenstate built from the same topological mode content is exactly stroboscopically stationary, showing that the $\pi$-mode structure alone is not sufficient.
A complementary trivial-phase check confirms necessity: without simultaneous
zero and $\pi$ edge modes the subharmonic signal is absent entirely
(see SM~\cite{jha_supplemental}).
Together, the two controls establish zero--$\pi$ Floquet topology as a
necessary condition and coherent nonequilibrium preparation as the additional
sufficient ingredient.

The central message is therefore conceptual as well as technical: Floquet topological structure can imprint itself not only on observables and quasienergy spectra, but also on the spectrum of the reduced density matrix after subsystem restriction and entanglement spectral reconstruction. Crucially, the exact statement is set-wise: the reduced-density-matrix spectrum alternates with period $2T$ after subsystem restriction, while the numerically nontrivial result is that a simple, fixed-reference tracking rule cleanly extracts an edge-dominated branch carrying that alternation at the single-level level. In that sense, the effect reported here is distinct from ordinary diagonal one-body diagnostics. The edge-bond observable provides a useful linear comparator, while the entanglement-spectrum response is a subsystem-defined nonequilibrium signature of coherent zero--$\pi$ preparation.

A natural next step is to ask whether interactions can promote this free-fermion mechanism into a genuinely many-body phase with the robustness and spontaneous discrete-time-translation-symmetry breaking required of a discrete time crystal~\cite{Else2016,Khemani2016,vonKeyserlingk2016,Yao2017,ElseMonroeNayak2020}. Our results isolate a clean free-fermion baseline whose fate in interacting systems is now sharply posed.
If robust entanglement-sector locking survives beyond Gaussian structure, it could point toward a broader form of Floquet dynamical order not captured by conventional observables alone. It would also be interesting to understand how such edge-coherence-driven entanglement dynamics interface with the temporal entanglement transitions recently identified in periodically driven Ising chains, where the entanglement Hamiltonian itself undergoes symmetry-resolved dynamical critical restructuring~\cite{Gadge2026Mar}.

The experimental outlook is encouraging. Platforms that already realize anomalous Floquet topological edge transport and controlled subharmonic response---including driven photonic lattices and programmable quantum simulators---provide natural arenas in which the present mechanism could be generalized or emulated~\cite{Mukherjee2017,Maczewsky2017,Zhang2017,Choi2017}. Experimental access to entanglement spectra and entanglement Hamiltonians has also advanced rapidly, including direct entanglement-spectrum measurements on quantum hardware, entanglement-Hamiltonian tomography in trapped-ion simulators, and recent realizations of entanglement Hamiltonians in synthetic quantum matter~\cite{Choo2018,Kokail2021,Joshi2023,Redon2024}. Because the free-fermion signal is defined entirely from subsystem correlation data, the present work opens a concrete route toward probing nonequilibrium topology through entanglement spectroscopy rather than through observables alone.

\textit{Acknowledgment.---}
The author acknowledges partial support by the U.S. Department of Energy, Office of Science, Office of Advanced Scientific Computing Research via the Exploratory Research for Extreme Scale Science (EXPRESS) program under Award Number DE-SC0026337. 

\textit{Data Availability.---}
The data that support the findings of this article are openly available~\cite{jha_zenodo}.

\bibliography{refs}

\appendix

\begin{figure*}[h]
    \centering
    \includegraphics[width=\linewidth]{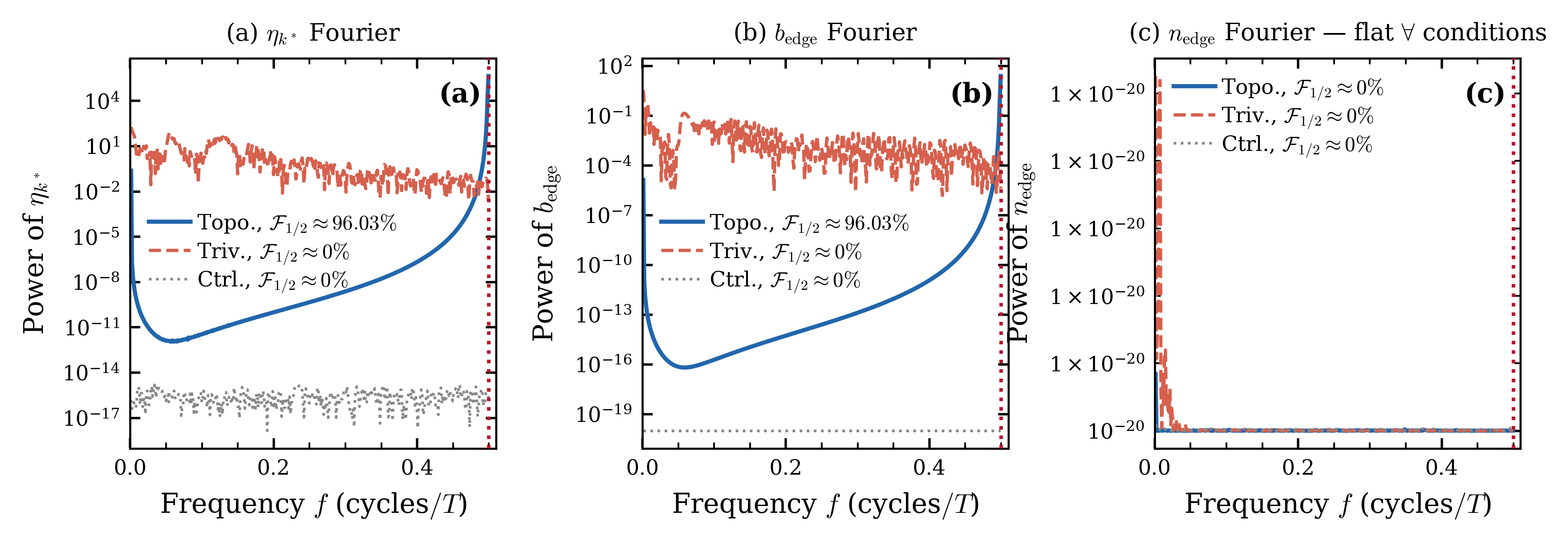}
    \caption{Fourier power spectra of all three left-edge observables for all three initial conditions: topological superposition state ($\delta_0=-0.3<0$, blue), trivial-phase initialization (orange; $\delta_0=+0.3>0$, ground state of $H_0$), and $\pi$-mode Floquet eigenstate control (gray).
    The red dotted line marks the subharmonic frequency $f=1/2$.
    Parameters: $\delta_{\mathrm{K}}=0.8$, $L=500$, $L_A=\lfloor\sqrt{L}\rfloor=22$, $N=800$ periods.
    (a)~$\eta_{k^*}$: sharp $f=1/2$ peak only for the topological superposition.
    (b)~$b_{\mathrm{edge}}$: same conclusion.
    (c)~$n_{\mathrm{edge}}$: flat for all three conditions, confirming that sublattice symmetry forbids a period-$2T$ component in diagonal one-body observables regardless of initial state or phase.}
    \label{fig:supplemental_diagnostics}
\end{figure*}

\section{Flat diagonal edge density and comprehensive observable comparison}
\label{sec:nedge_flat}

The main-text caption of Fig.~\ref{fig:main} notes that the diagonal edge density $n_{\mathrm{edge}}(nT)$ remains flat by sublattice symmetry and is not shown there.
The analytic argument of subsystem chiral symmetry, as provided in the main text as well as in SM~\cite{jha_supplemental}, forces the interference matrix element $\langle \Phi_0|O_f|\Phi_\pi\rangle$ to vanish for any diagonal one-body operator $O_f$, so no period-$2T$ component can appear in $n_{\mathrm{edge}}$ regardless of the initial state.
Figure~\ref{fig:supplemental_diagnostics} provides the numerical confirmation alongside a comprehensive comparison of all three observables and all three initial conditions.
Panel~(c) shows the Fourier power of $n_{\mathrm{edge}}$ for the topological superposition state, the trivial-phase initialization, and the $\pi$-mode Floquet eigenstate control: all three are flat to within numerical noise, with $\mathcal{F}_{1/2}$ consistent with zero.
In the chiral SSH setting the zero and $\pi$ edge modes occupy opposite sublattices, so the interference matrix element $\langle\Phi_0|O_f|\Phi_\pi\rangle$ vanishes identically for any diagonal one-body operator $O_f$, regardless of which initial state is used.

Panels~(a) and~(b) show the Fourier power of $\eta_{k^*}$ and $b_{\mathrm{edge}}$, respectively, for the same three conditions.
A sharp subharmonic peak at $f=1/2$ appears only for the topological superposition state; both the trivial-phase initialization and the eigenstate control show no resolved peak.
Taken together, this figure delivers the full story in one place: the subharmonic response is topological in origin, activated only by coherent nonequilibrium preparation, and absent from diagonal density probes for any initial state.

\section{Overlap-ranked entanglement-level diagnostic}
\label{sec:eo_results}

Table~\ref{tab:overlap_L500} reports the overlap-ranked even--odd
diagnostic at $L=500$, $L_A=22$, $\delta_0=-0.30$, $\delta_K=0.80$,
$N=800$ periods.
Ranks~1--2 are the cleanest period-$2T$ oscillators
($\mathcal{F}_{1/2}\approx0.960$, $|\Delta_{eo}|\approx3.94$,
$\mathrm{SNR}\sim10^8$); ranks~5 and~18 also pass ($\mathrm{SNR}\approx6.5$)
at lower overlap, while remaining ranks fail (\texttt{NaN}).
The $\pi$-mode eigenstate control yields \texttt{NaN} at every rank,
consistent with exact stroboscopic stationarity.
The $L=1000$ counterpart is given in SM~\cite{jha_supplemental}.

\begin{table}[h]
\centering
\caption{Overlap-ranked diagnostic for $L=500$, $L_A=22$,
$\delta_0=-0.30$, $\delta_K=0.80$, $N=800$ periods.
$\bar{o}$: mean overlap with reference vector (\%).
$|\Delta_{eo}|$: even-odd amplitude (reported only when
$\mathrm{SNR}\geq 3.0$; otherwise \texttt{NaN}).
$\mathcal{F}_{1/2}$: subharmonic fraction.
Top~5 and bottom~5 overlap-ranked levels are shown.}
\label{tab:overlap_L500}
\begin{tabular}{ccccccccc}
\toprule
& \multicolumn{4}{c}{Superposition state}
& \multicolumn{4}{c}{$\pi$-mode eigenstate} \\
\cmidrule(lr){2-5}\cmidrule(lr){6-9}
Rank & $\bar{o}$ & $|\Delta_{eo}|$ & SNR & $\mathcal{F}_{1/2}$
     & $\bar{o}$ & $|\Delta_{eo}|$ & SNR & $\mathcal{F}_{1/2}$ \\
\midrule
1  & 24.05 & 3.944  & $9.6\times10^8$ & 0.960 & 59.69 & NaN & 0.15 & 0.000 \\
2  & 24.05 & 3.944  & $3.8\times10^8$ & 0.960 & 21.34 & NaN & 0.06 & 0.000 \\
3  & 14.10 & NaN    & 0.89            & 0.288 & 11.48 & NaN & 0.03 & 0.000 \\
4  & 14.10 & NaN    & 0.89            & 0.288 &  4.05 & NaN & 0.03 & 0.000 \\
5  &  5.77 & 35.878 & 6.53            & 0.915 &  2.04 & NaN & 0.08 & 0.003 \\
\midrule
18 &  0.08 & 26.689 & 6.59            & 0.910 &  0.00 & NaN & 0.14 & 0.000 \\
19 &  0.05 & NaN    & 1.64            & 0.524 &  0.00 & NaN & 0.04 & 0.000 \\
20 &  0.05 & NaN    & 1.64            & 0.524 &  0.00 & NaN & 0.05 & 0.000 \\
21 &  0.00 & NaN    & 0.90            & 0.257 &  0.00 & NaN & 0.11 & 0.001 \\
22 &  0.00 & NaN    & 0.90            & 0.257 &  0.00 & NaN & 0.11 & 0.009 \\
\bottomrule
\end{tabular}
\end{table}

\end{document}


\preprint{APS/123-QED}
	
\title{Supplemental Material --- Locked Subharmonic Oscillations in the Entanglement Spectrum of a Periodically Driven Topological Chain}

\author{Rishabh Jha}
\email{rishabh.jha@usc.edu}
\affiliation{%
	Department of Physics and Astronomy, University of Southern California, Los Angeles, CA 90089-0484, USA
}%

	\maketitle
	
	\onecolumngrid

	\tableofcontents


\section{Driven SSH chain: model and Floquet protocol}
\label{sec:ssh_model}

This section defines the periodically driven Su–Schrieffer–Heeger (SSH) chain studied in the main text.
The model is number conserving; a chain of $L$ sites is fully specified by an $L\times L$ single‑particle hopping matrix in the site basis.

\subsection{Fermions, geometry, and parameters}
\label{subsec:ssh_geometry_notation}

We consider an open chain of $L$ sites with spinless fermions $c_j$ ($j=1,\dots,L$) satisfying $\{c_i,c_j^\dagger\}=\delta_{ij}$ and $\{c_i,c_j\}=0$.
We set $\hbar=1$.

The drive is specified by:
(i) a static dimerization parameter $\delta_0$ for the first step, 
(ii) a second‑step dimerization parameter $\delta_{\mathrm K}$ for the second step, and
(iii) the half‑period duration $T_{\mathrm{half}}=T/2$.

\subsection{Step‑1 Hamiltonian $H_0$ (static SSH)}
\label{subsec:ssh_H0}

The Hamiltonian generating the first half of each period is
\begin{equation}
H_0
=
\sum_{j=1}^{L-1} t^{(0)}_{j}\left(c_j^\dagger c_{j+1}+c_{j+1}^\dagger c_j\right),
\label{eq:ssh_H0}
\end{equation}
where all onsite potentials are set to zero, and the nearest-neighbor hoppings alternate as
\begin{equation}
t^{(0)}_{j}=
\begin{cases}
t_0(1+\delta_0), & j\ \text{odd} \quad \text{(bond }(1,2),(3,4),\dots),\\[2pt]
t_0(1-\delta_0), & j\ \text{even}\quad \text{(bond }(2,3),(4,5),\dots).
\end{cases}
\label{eq:ssh_t_static}
\end{equation}
Here $t_0>0$ sets the overall hopping scale and $\delta_0$ controls the dimerization.

With this convention, we define the \emph{intracell} and \emph{intercell} hoppings as
\begin{equation}
t_1^{(0)} := t_0(1+\delta_0),\qquad
t_2^{(0)} := t_0(1-\delta_0).
\label{eq:ssh_t1t2_static}
\end{equation}

\subsection{Step‑2 Hamiltonian $H_{\mathrm K}$ (dimerization step)}
\label{subsec:ssh_HK}

The second half‑step is generated by another SSH‑type hopping Hamiltonian,
\begin{equation}
H_{\mathrm K}
=
\sum_{j=1}^{L-1} t^{(\mathrm K)}_{j}\left(c_j^\dagger c_{j+1}+c_{j+1}^\dagger c_j\right),
\label{eq:ssh_HK}
\end{equation}
with
\begin{equation}
t^{(\mathrm K)}_{j}=
\begin{cases}
t_{\mathrm K}(1+\delta_{\mathrm K}), & j\ \text{odd},\\[2pt]
t_{\mathrm K}(1-\delta_{\mathrm K}), & j\ \text{even}.
\end{cases}
\label{eq:ssh_t_kick}
\end{equation}
Here $t_{\mathrm K}>0$ is the overall hopping scale during the second step, and $\delta_{\mathrm K}$ is its dimerization.
Throughout this work we set $t_0 = t_{\mathrm{K}} = 1$, which fixes the overall energy scale and leaves $\delta_0$, $\delta_{\mathrm{K}}$, and $T_{\mathrm{half}}$ as the only free parameters, where all quoted $T_{\mathrm{half}}$ values are measured in units of $1/t_0$.

\paragraph{Terminology.}
We refer to $H_{\mathrm K}$ as the \emph{second‑step} (or \emph{dimerization‑step}) Hamiltonian.
It is applied for a finite time $T_{\mathrm{half}}$ and is therefore not an instantaneous ``kick.''

\subsection{Two‑step drive and Floquet operator}
\label{subsec:ssh_floquet}

Let $T$ be the drive period and $T_{\mathrm{half}}=T/2$.
The protocol is piecewise constant:
\begin{equation}
H(t)=
\begin{cases}
H_0, & 0<t\le T_{\mathrm{half}},\\[2pt]
H_{\mathrm K}, & T_{\mathrm{half}}<t\le T.
\end{cases}
\label{eq:ssh_protocol}
\end{equation}
Because each step is time independent, the time‑ordered evolution over one full period is exactly
\begin{equation}
U(T)
=
e^{-i H_{\mathrm K} T_{\mathrm{half}}}\;
e^{-i H_0 T_{\mathrm{half}}}.
\label{eq:ssh_UF}
\end{equation}
The stroboscopic dynamics at times $t=nT$ is generated by repeated application:
\begin{equation}
|\psi(nT)\rangle = U(T)^n |\psi(0)\rangle.
\label{eq:ssh_stroboscopic}
\end{equation}
The corresponding stroboscopic evolution of the correlation matrix is derived in Sec.~\ref{subsec:corr_evolution}.

\subsection{Floquet eigenphases and the meaning of $0$ and $\pi$ modes}
\label{subsec:ssh_zero_pi_modes}

Since $U(T)$ is unitary, its eigenvalues lie on the unit circle.
We define Floquet eigenmodes by
\begin{equation}
U(T)\,|\Phi_\alpha\rangle = e^{-i \varepsilon_\alpha T}\,|\Phi_\alpha\rangle,
\label{eq:ssh_quasienergy_def}
\end{equation}
where the quasienergy $\varepsilon_\alpha$ is defined modulo $2\pi/T$.
Equivalently, it is often convenient to work with eigenphases
\begin{equation}
\theta_\alpha := \varepsilon_\alpha T \in (-\pi,\pi],
\qquad
U(T)\,|\Phi_\alpha\rangle = e^{-i\theta_\alpha}\,|\Phi_\alpha\rangle.
\label{eq:ssh_eigenphase}
\end{equation}
Here $|\Phi_\alpha\rangle$ denotes a \emph{single‑particle} eigenstate of the one‑period evolution operator $U(T)$; we call such states Floquet modes.

\paragraph{Why $0$ and $\pi$ are special in a Floquet spectrum.}
The points $\theta=0$ and $\theta=\pi$ correspond to the two real unit‑modulus eigenvalues of $U(T)$:
\begin{equation}
e^{-i\theta}=+1 \quad (\theta=0),\qquad
e^{-i\theta}=-1 \quad (\theta=\pi).
\label{eq:ssh_pm1}
\end{equation}
We will refer to eigenmodes near $\theta\simeq 0$ as ``$0$ modes'' and those near $\theta\simeq \pi$ as ``$\pi$ modes.''

\paragraph{Stroboscopic sign flip and period doubling.}
A $\pi$ mode satisfies $U(T)|\Phi_\pi\rangle = -|\Phi_\pi\rangle$ and therefore picks up a factor $(-1)^n$ after $n$ periods:
\begin{equation}
U(T)^n|\Phi_\pi\rangle = (-1)^n |\Phi_\pi\rangle.
\label{eq:ssh_pi_sign_flip}
\end{equation}
By contrast, a $0$ mode satisfies $U(T)^n|\Phi_0\rangle = |\Phi_0\rangle$.
Therefore, an initial state containing a coherent superposition of the two sectors,
\begin{equation}
|\psi(0)\rangle \propto |\Phi_0\rangle + |\Phi_\pi\rangle,
\label{eq:ssh_superposition}
\end{equation}
is not an eigenstate of $U(T)$ and exhibits an exact period-$2T$ component in suitable observables, because the relative phase between the two contributions alternates sign each period. A pure $\pi$-mode Floquet eigenstate, by contrast, is stroboscopically stationary: its correlation matrix is unchanged after each period, so no one-body observable acquires a period-$2T$ component.

\paragraph{Remark (edge localization and gaps).}
In an open dimerized SSH chain, boundary‑localized modes (when present) appear at quasienergies/energies lying inside a \emph{bulk} spectral gap, i.e., an interval where the bulk supports no extended states.
In the driven setting, the analogous condition is a quasienergy gap of the bulk Floquet spectrum around $\theta=0$ and/or $\theta=\pi$, in which case boundary‑localized Floquet modes can reside inside that gap.

\subsection{Sublattice (chiral) symmetry in the onsite‑free limit}
\label{subsec:ssh_chiral_sym}

This subsection makes explicit a symmetry that is present in our SSH steps because they contain only nearest‑neighbor hoppings between the two sublattices (and no onsite terms).
Its main role for us is to explain why the Floquet spectrum is symmetric under $\theta\to-\theta$ and why
$\theta=0$ and $\theta=\pi$ are the symmetry‑fixed points.

\paragraph{Sublattice operator.}
We partition the sites into sublattices $A$ (odd $j$) and $B$ (even $j$), and define the single‑particle operator
\begin{equation}
\Gamma := \sum_{j=1}^{L} s_j\,|j\rangle\langle j|,
\qquad
s_j=
\begin{cases}
+1, & j\ \text{odd}\ (j\in A),\\
-1, & j\ \text{even}\ (j\in B).
\end{cases}
\label{eq:ssh_Gamma_sp}
\end{equation}
By construction, $\Gamma^\dagger=\Gamma$ and $\Gamma^2=\mathds{1}$ (its eigenvalues are $\pm1$).

\paragraph{Chiral (sublattice) symmetry of the step Hamiltonians.}
Because $H_0$ and $H_{\mathrm K}$ only connect opposite sublattices (odd $\leftrightarrow$ even) and have no onsite terms, they anticommute with $\Gamma$:
\begin{equation}
\{\Gamma,H_0\}=0,\qquad
\{\Gamma,H_{\mathrm K}\}=0.
\label{eq:ssh_chiral}
\end{equation}
A short proof is obtained by evaluating matrix elements in the site basis: for a nonzero hopping element $\langle m|H|n\rangle$, the sites $m$ and $n$ have opposite parity, hence $s_ms_n=-1$, which implies $\langle m|\Gamma H\Gamma|n\rangle = s_ms_n\langle m|H|n\rangle = -\langle m|H|n\rangle$ and therefore $\Gamma H\Gamma=-H$ (here $H$ represents either of two Hamiltonians in the two-step drive).

\paragraph{Floquet consequence and the fixed points $\theta=0,\pi$.}
From $\{\Gamma,H\}=0$ and $\Gamma^2=\mathds{1}$, one obtains for each step
\begin{equation}
\Gamma\,e^{-iH T_{\mathrm{half}}}\,\Gamma
=
e^{+iH T_{\mathrm{half}}}
=
\left(e^{-iH T_{\mathrm{half}}}\right)^\dagger.
\label{eq:ssh_chiral_step}
\end{equation}
Because the hopping amplitudes in $H_0$ and $H_{\mathrm K}$ are real, we have $\big(e^{-iH T_{\mathrm{half}}}\big)^* = e^{+iH T_{\mathrm{half}}}$.
Applying these facts to the Floquet operator
$U(T)=e^{-iH_{\mathrm K}T_{\mathrm{half}}}e^{-iH_0T_{\mathrm{half}}}$ gives
\begin{align}
\Gamma\,U(T)\,\Gamma
&= \big(\Gamma e^{-iH_{\mathrm K}T_{\mathrm{half}}}\Gamma\big)
   \big(\Gamma e^{-iH_0T_{\mathrm{half}}}\Gamma\big) \nonumber\\
&= e^{+iH_{\mathrm K}T_{\mathrm{half}}} e^{+iH_0T_{\mathrm{half}}}
 = U(T)^* .
\label{eq:ssh_chiral_floquet_real}
\end{align}
Thus $\Gamma$ conjugates $U(T)$ to its complex conjugate.  
If $U(T)|\Phi_\alpha\rangle = e^{-i\theta_\alpha}|\Phi_\alpha\rangle$, then $U(T)^*(\Gamma|\Phi_\alpha\rangle) = e^{-i\theta_\alpha}(\Gamma|\Phi_\alpha\rangle)$.
Complex conjugation of this eigenvalue equation yields $U(T)(\Gamma|\Phi_\alpha\rangle)^* = e^{+i\theta_\alpha}(\Gamma|\Phi_\alpha\rangle)^*$,
so $e^{+i\theta_\alpha}$ is also an eigenvalue. 
The spectrum is therefore symmetric under $\theta\mapsto -\theta$. The only eigenphases unchanged by this map are $\theta=0$ and $\theta=\pi$; these are the symmetry‑fixed points.


\subsection{Static topological convention (open chain)}
\label{subsec:ssh_static_topology}

For the static SSH Hamiltonian $H_0$ defined above, the two dimerized regimes are distinguished by whether the intercell hopping exceeds the intracell hopping.
With the convention in Eq.~\eqref{eq:ssh_t1t2_static}, the topological regime corresponds to
\begin{equation}
t_2^{(0)} > t_1^{(0)}
\qquad \Longleftrightarrow \qquad
\delta_0<0,
\label{eq:ssh_topo_condition}
\end{equation}
while $\delta_0>0$ is trivial.
In a finite open chain, the topological edge‑mode splitting is exponentially small in system size.

\subsection{Time evolution of the correlation matrix}
\label{subsec:corr_evolution}

We now describe how the single-particle correlation matrix $C(t)$ evolves under the two-step drive and how the stroboscopic data $C(nT)$ used throughout this work are generated.

\paragraph{Equation of motion.}
For any quadratic, number-conserving Hamiltonian \(H(t)=\sum_{i,j}h_{ij}(t)\,c_i^\dagger c_j\), the Heisenberg equations of motion for the fermion operators close at the two-point level and yield
\begin{equation}
  i\,\frac{d}{dt}C(t) = \bigl[h(t),\,C(t)\bigr],
  \label{eq:corr_eom}
\end{equation}
where $h(t)$ is the $L\times L$ single-particle Hamiltonian matrix with entries $h_{ij}(t)$, and $C_{ij}(t)=\langle c_j^\dagger c_i\rangle(t)$ is defined in Eq.~\eqref{eq:Cij_def}.
This is the single-particle analogue of the Liouville--von~Neumann equation and is exact for the quadratic Hamiltonians studied here.
Its formal solution reads
\begin{equation}
  C(t) = \mathcal{U}(t,0)\,C(0)\,\mathcal{U}(t,0)^\dagger,
  \label{eq:corr_formal}
\end{equation}
where $\mathcal{U}(t,0)=\mathcal{T}\exp\!\bigl(-i\int_0^t h(t')\,dt'\bigr)$ is the single-particle time-evolution operator.

\paragraph{Exact stroboscopic update.}
Because $H(t)$ is piecewise constant [Eq.~\eqref{eq:ssh_protocol}], the time-ordered exponential in $\mathcal{U}(T,0)$ reduces to ordinary matrix exponentials, and $\mathcal{U}(T,0)$ is \emph{exactly} the Floquet operator $U(T)$ defined in Eq.~\eqref{eq:ssh_UF}. Eq.~\eqref{eq:corr_formal} therefore yields the stroboscopic recurrence
\begin{equation}
  C\bigl((n+1)T\bigr) = U(T)\,C(nT)\,U(T)^\dagger,
  \qquad n=0,1,2,\dotsc
  \label{eq:corr_stroboscopic}
\end{equation}
No approximation is involved beyond the piecewise-constant protocol and the quadratic structure of the Hamiltonian.

\paragraph{Orbital representation and numerical procedure.}
For a Slater determinant, $C = VV^\dagger$ [Eq.~\eqref{eq:C_from_V}], where $V$ is the $L\times N$ matrix of occupied single-particle orbitals.
Substituting into Eq.~\eqref{eq:corr_stroboscopic} shows that the orbital matrix evolves as
\begin{equation}
  V\bigl((n+1)T\bigr) = U(T)\,V(nT),
  \label{eq:V_update}
\end{equation}
after which $C((n+1)T)=V((n+1)T)\,V((n+1)T)^\dagger$.
To prevent the accumulation of floating-point round-off that would gradually destroy the orthonormality of the columns of $V$ over hundreds of periods, the columns are re-orthonormalized after each step via a $QR$ decomposition,
\begin{equation}
  U(T)\,V(nT) \;=:\; Q(nT)\,R(nT),
  \qquad
  V\bigl((n+1)T\bigr) := Q(nT),
  \label{eq:V_qr}
\end{equation}
where $Q(nT)$ has orthonormal columns spanning the same occupied subspace.
Because $C=VV^\dagger$ depends only on the column space of $V$ and not on the particular orthonormal basis chosen for it, the re-orthogonalization does not alter $C((n+1)T)$; it only stabilizes the numerics. The subsystem correlator $C_A(nT)$ is then extracted as the $L_A\times L_A$ principal submatrix of $C(nT)$ [Eq.~\eqref{eq:CA_def}], and the entanglement energies $\{\eta_\ell(nT)\}$ are computed via the Peschel--Eisler formula of Sec.~\ref{sec:peschel_eisler}.




\section{Initial state preparation}
\label{sec:initial_states}

In all our numerical simulations the many-body state at stroboscopic times is a Gaussian (Slater-determinant) state, fully specified by an orthonormal set of occupied single-particle orbitals.
Collecting these orbitals as the columns of an \(L \times N\) matrix \(V\) (with \(N\) the number of occupied orbitals), the equal-time correlation matrix in our convention is
\begin{equation}
C = V V^{\dagger},
\end{equation}
as discussed in Sec.~\ref{sec:peschel_eisler}.
This choice corresponds to \(C_{ij} = \langle c_j^{\dagger} c_i \rangle\); the alternative convention \(C_{ij} = \langle c_i^{\dagger} c_j \rangle\) would simply give the transpose of the matrix used here and leads to the same entanglement spectrum.

The initial occupied orbitals are constructed from the Floquet modes of the one-period evolution operator.
For the two-step protocol defined in Sec.~\ref{subsec:ssh_floquet}, the one-period Floquet operator reads
\begin{equation}
U(T) = e^{-i H_{\mathrm{K}} T_{\mathrm{half}}} \, e^{-i H_0 T_{\mathrm{half}}},
\qquad T = 2 T_{\mathrm{half}},
\end{equation}
in direct correspondence with Eq.~\eqref{eq:ssh_UF}.
Diagonalizing \(U(T)\) yields complex eigenvalues \(e^{-i \theta_{\alpha}}\) with quasienergies (eigenphases) \(\theta_{\alpha} \in (-\pi,\pi]\) and associated Floquet modes \(\lvert \Phi_{\alpha} \rangle\).
In the numerics we classify the modes using a fixed phase tolerance \(\delta_{\mathrm{tol}} = 0.05\):
\begin{itemize}
\item a \emph{zero mode} if \(|\theta_{\alpha}| < \delta_{\mathrm{tol}} \),
\item a \(\pi\) \emph{mode} if \(\bigl||\theta_{\alpha}| - \pi\bigr| < \delta_{\mathrm{tol}}\),
\item a \emph{negative-phase bulk mode} if \(\theta_{\alpha} < 0\) and it is
      neither a zero nor a \(\pi\) mode,
\item a \emph{positive-phase bulk mode} if \(\theta_{\alpha} > 0\) and it is
      neither a zero nor a \(\pi\) mode.
\end{itemize}
Throughout, all negative-phase bulk modes are always occupied and provide a stationary ``Fermi-sea'' background.
Positive-phase bulk modes are only used, when needed, to reach the desired filling, as described below.

For the representative parameter sets used in this work (for example, $\delta_0=-0.3$, $\delta_{\mathrm K}=0.8$, and the corresponding optimized $T_{\mathrm{half}}$), the Floquet spectrum contains a small number of zero and $\pi$ modes, often one or two of each, that are well localized near the edges.
The constructions below are formulated for general numbers $n_0$ and $n_\pi$ of such modes and automatically reduce to the simple case $n_0=n_\pi=1$.

We quantify localization near the left edge by the weight on the first \(w = \lfloor\sqrt{L}\rfloor\) sites,
\begin{equation}
w_{\alpha} = \sum_{j=1}^{w} \bigl|\Phi_{\alpha}(j)\bigr|^2,
\qquad w = \bigl\lfloor\sqrt{L}\bigr\rfloor.
\label{eq:weight_edge}
\end{equation}
This single weight serves a dual role.
First, it ranks zero and \(\pi\) modes for the construction of the superposition initial state (Eq.~\eqref{eq:init_superposition_orb}), ensuring that the selected orbital has appreciable amplitude in the first \(w\) sites where the edge density is measured (Sec.~\ref{subsec:edge_density}).
Second, because the subsystem \(A\) is chosen to have size \(L_A = w = \lfloor\sqrt{L}\rfloor\) (see Sec.~\ref{subsec:restrict_subsystem}), the same weight also governs mode selection for the reference vector used in entanglement tracking (Sec.~\ref{subsec:reference_vector}).
The two roles are therefore automatically consistent, and no separate, larger localization weight over subsystem \(A\) is needed.

\paragraph{Choice of the edge window $w$ and subsystem size $L_A$.}
The edge density is evaluated in the first $w$ sites, and we take $w=\lfloor\sqrt{L}\rfloor$ as a convenient boundary window that is large compared with the localization length of the edge modes but still small compared with the full chain. The same scale is used for the default subsystem size, $L_A=w$, so that the mode used for entanglement-level tracking is selected and probed within the same left-edge region.

The motivation for this choice is operational rather than exact.
The tracked entanglement level is defined as the eigenvector of $C_A$ with the largest overlap with a fixed reference vector obtained from the most left-localized $\pi$ mode restricted to subsystem $A$.
When $L_A$ is chosen near the edge region, this procedure isolates a clean tracked entanglement level. Other levels can also carry a subharmonic component, but the largest-overlap rule gives the most stable signal.
When $L_A$ is made much larger, additional bulk-dominated degrees of freedom are included in $C_A$, and the overlap-based identification of one distinguished level becomes less sharp.

For $L=500$, where $\lfloor\sqrt{L}\rfloor=22$, reducing $L_A$ from $22$ to $12$ produces no visible change in the tracked signal or in the subharmonic fraction $\mathcal{F}_{1/2}$, whereas increasing $L_A$ to $100$ suppresses the signal and $L_A=200$ degrades the period-$2T$ oscillation substantially; see Fig.~\ref{fig:LA_robustness}.
We therefore use $L_A=\lfloor\sqrt{L}\rfloor$ as a simple and well-converged default choice.
This choice is not unique: nearby values give the same results, but taking $L_A$ much larger than the edge region weakens the tracking.

\subsection{Superposition state (zero--\texorpdfstring{\(\pi\)}{π} combination)}
\label{subsec:init_superposition}

Our main nonequilibrium initial condition is a coherent superposition of zero and \(\pi\) edge modes on top of the negative-phase bulk background.
It is designed to produce a robust period-\(2T\) signal in the entanglement spectrum and in off-diagonal one-body observables that connect opposite sublattices. By contrast, diagonal site-density observables are symmetry-blind to this coherence, as shown explicitly in Sec.~\ref{subsec:edge_density}.

Let \(\{\lvert \Phi_{0,k} \rangle\}\) and \(\{\lvert \Phi_{\pi,k} \rangle\}\) denote the sets of zero and \(\pi\) modes, respectively, with total numbers \(n_0\) and \(n_\pi\).
We first sort each set independently in descending order of their left-edge weight \(w_{\alpha}\) defined in Eq.~\eqref{eq:weight_edge}, and then pair them by rank.
We form \(n_{\mathrm{pairs}} = \min(n_0, n_\pi)\) such pairs: the \(k^{\rm th}\) most left-localized zero mode is paired with the \(k^{\rm th}\) most left-localized \(\pi\) mode. Any excess unpaired zero or \(\pi\) modes are left unoccupied.
For each pair we define the equal-weight superposition
\begin{equation}
\lvert \Psi_{\mathrm{sup},k} \rangle
  = \frac{1}{\sqrt{2}}\bigl(\lvert \Phi_{0,k} \rangle
  + \lvert \Phi_{\pi,k} \rangle\bigr).
\label{eq:init_superposition_orb}
\end{equation}
The occupied single-particle orbitals that define the initial state then consist of: (i) all negative-phase bulk modes, and (ii) all the \(n_{\mathrm{pairs}}\) superposed edge orbitals \(\lvert \Psi_{\mathrm{sup},k} \rangle\) constructed in this way.
The corresponding many-body state can be written as the Slater determinant
\begin{equation}
\lvert \Psi_{\mathrm{sup}}(0) \rangle
 = \Bigl(\prod_{\text{bulk-}} b_{\beta}^{\dagger}\Bigr)
   \Bigl(\prod_{k=1}^{n_{\mathrm{pairs}}} d_{k}^{\dagger}\Bigr) \lvert 0 \rangle,
\label{eq:full_superposition}
\end{equation}
where \(b_{\beta}^{\dagger}\) creates a negative-phase bulk mode and \(d_{k}^{\dagger}\) creates the superposed orbital of the \(k^{\rm th}\) zero--\(\pi\) pair.

By construction, the zero and \(\pi\) modes satisfy
\begin{equation}
U(T)\lvert \Phi_{0,k} \rangle = \lvert \Phi_{0,k} \rangle, \qquad
U(T)\lvert \Phi_{\pi,k} \rangle = - \lvert \Phi_{\pi,k} \rangle,
\end{equation}
so that the superposition \(\lvert \Psi_{\mathrm{sup},k} \rangle\) is \emph{not} an eigenstate of the Floquet operator.
Instead,
\begin{equation}
U(T)\lvert \Psi_{\mathrm{sup},k} \rangle
  = \frac{1}{\sqrt{2}}\bigl(\lvert \Phi_{0,k} \rangle
  - \lvert \Phi_{\pi,k} \rangle\bigr),
\end{equation}
which is orthogonal to the original state in Eq.~\eqref{eq:init_superposition_orb}.
After one period the relative sign between the zero and \(\pi\) components flips, and after two periods it returns to its original value.
This alternating relative phase produces a clean period-\(2T\) structure in the many-body state. In the present chiral SSH setting, diagonal density observables do not resolve this coherence, whereas off-diagonal bond observables and a distinguished level of the entanglement spectrum do.
We refer to this choice as the \emph{superposition initial state}.
If the total number of occupied orbitals obtained in this way differs from half‑filling \(L/2\), we add the required number of positive‑phase bulk modes (or trim the least negative negative‑phase bulk modes) to reach exactly \(L/2\); the procedure is detailed in Sec.~\ref{subsec:orthonormalisation}.
By symmetry, one may equally start from the orthogonal superposition \(\lvert \Psi_{\mathrm{sup},k}^{-} \rangle = (\lvert \Phi_{0,k} \rangle - \lvert \Phi_{\pi,k} \rangle)/\sqrt{2}\), which is precisely the state the system occupies after one period; this variant produces identical period-\(2T\) physics.
Throughout this work the \(+\) superposition, Eq.~\eqref{eq:init_superposition_orb}, is used as the canonical initial state.

\subsection{\texorpdfstring{\(\pi\)}{π}-mode Floquet eigenstate}
\label{subsec:init_pi_eigen}

As a control we also consider an initial state that is itself an eigenstate of the Floquet operator. Because it spans an invariant subspace of $U(T)$, it is stroboscopically stationary and cannot exhibit period doubling in any observable.
Here we again occupy all negative-phase bulk modes, but instead of forming superpositions we add a single \(\pi\) edge mode and, if necessary, a minimal number of positive-phase bulk modes so as to reach half-filling.
Concretely, among all \(\pi\) modes, we select the one with the largest left-edge weight \(w_{\alpha}\) (Eq.~\eqref{eq:weight_edge}); let \(p^{\dagger}\) denote the corresponding creation operator.
We begin with all negative-phase bulk modes and the selected $\pi$ mode.
If the total number of these orbitals is less than $L/2$, we add positive-phase bulk modes in order of increasing positive eigenphase until half-filling is reached.
If the total exceeds $L/2$, we instead discard the least negative negative-phase bulk modes (those with quasienergy closest to zero) until exactly half-filling is achieved.
The construction fails (and the fallback ground state is used) only if there are insufficient positive-phase bulk modes to fill a deficit, or if the surplus exceeds the number of negative-phase bulk modes available.
All these orbitals are Floquet eigenmodes, so they span an invariant subspace of \(U(T)\).
The many-body state can be written as
\begin{equation}
\lvert \Psi_{\pi}(0) \rangle
  = \Bigl(\prod_{\text{bulk-}} b_{\beta}^{\dagger}\Bigr)
    \, p^{\dagger} \,
    \Bigl(\prod_{\text{bulk+}} c_{\gamma}^{\dagger}\Bigr) \lvert 0 \rangle,
\label{eq:full_pi}
\end{equation}
where \(b_{\beta}^{\dagger}\) creates a negative-phase bulk mode, \(p^{\dagger}\) creates the chosen \(\pi\) edge mode, and \(c_{\gamma}^{\dagger}\) creates a positive-phase bulk mode (included only if needed to achieve half-filling).
The construction fails and is replaced by the fallback described below if either there are not enough positive-phase bulk modes to reach \(N = L/2\), or if the number of negative-phase bulk modes together with the single \(\pi\) mode already exceeds \(L/2\) before any positive-phase modes are added.

In practice we first assemble the corresponding occupied-orbital matrix \(V_{\text{raw}}\) and then perform a QR decomposition to obtain a perfectly orthonormal set of occupied orbitals.
Although the columns of the resulting matrix \(V\) are not individually Floquet eigenmodes, they still span the same invariant subspace, so the projector \(C = V V^{\dagger}\) commutes with \(U(T)\).
At stroboscopic times the many-body state is therefore stationary up to a global phase, and no subharmonic oscillation can appear in any observable.
This $\pi$-mode eigenstate confirms that the presence of a $\pi$ mode in the Floquet spectrum is not sufficient to produce period doubling; the effect requires a nonequilibrium, coherent superposition between the zero and $\pi$ sectors.

\subsection{Orthonormalisation and half-filling}
\label{subsec:orthonormalisation}

After selecting the desired orbitals (negative-phase bulk modes together with either the superposed edge orbitals or a single \(\pi\) mode plus, if necessary, some positive-phase bulk modes), we collect them into a matrix \(V_{\text{raw}}\), whose columns represent the occupied single-particle states.
In exact arithmetic the Floquet modes are orthonormal, and any linear combination of mutually orthogonal modes---such as the equal-weight superpositions in Eq.~\eqref{eq:init_superposition_orb}---remains orthonormal.
However, numerical diagonalisation and subsequent floating-point operations introduce small deviations from perfect orthogonality.

To restore orthonormality to machine precision we perform a QR decomposition
\begin{equation}
V_{\text{raw}} = Q R,
\end{equation}
where the columns of \(Q\) form an orthonormal basis for the occupied subspace and \(R\) is an upper-triangular matrix.
We then set
\begin{equation}
V = Q,
\end{equation}
which leaves the occupied subspace (and hence the projector \(C\)) unchanged but guarantees that \(C = V V^{\dagger}\) is an exact projector in numerical precision.
The stroboscopic time evolution is implemented by propagating the occupied-orbital matrix directly,
\begin{equation}
V \;\leftarrow\; U(T)\,V,
\label{eq:evolve_V}
\end{equation}
which maps one orthonormal set of occupied orbitals to the next. The correlation matrix \(C = V V^{\dagger}\) is then computed from this updated \(V\).
A QR factorization is applied both when constructing the initial state and immediately after each update~\eqref{eq:evolve_V} during time evolution: although \(U(T)\) is unitary, floating-point errors accumulate over many matrix multiplications and gradually spoil orthonormality; the QR factorization restores an orthonormal basis for the occupied subspace at each step without changing the projector \(C = VV^{\dagger}\), thereby ensuring that the Slater determinant remains properly normalized at all stroboscopic times.

After forming \(C = V V^{\dagger}\), we enforce exact numerical Hermiticity as a standard step by symmetrization,
\begin{equation}
C \leftarrow \frac{C + C^{\dagger}}{2},
\end{equation}
applied unconditionally every time \(C\) is constructed: in exact arithmetic \(V V^{\dagger}\) is already Hermitian, but floating-point rounding introduces deviations of order \(\varepsilon_{\mathrm{machine}}\) that, if left uncorrected, can accumulate over many time steps and eventually spoil the reality of the entanglement spectrum.
We then form \(C_A\) by restricting to the \(L_A \times L_A\) block corresponding to the left subsystem.
Since \(C_A\) is Hermitian by construction, we diagonalize it with a dedicated Hermitian eigensolver, which returns a real-valued eigenvalue array \(\{\xi_{\ell}\}\) directly. As an explicit guard against numerical breakdown (for instance, if the symmetrization of \(C\) were inadvertently skipped), we additionally verify that \(\max_{\ell} |\operatorname{Im}\xi_{\ell}| < 10^{-12}\) and abort the simulation if this condition is violated.
To avoid spurious divergences when converting \(\xi_{\ell}\) to entanglement energies
\begin{equation}
\eta_{\ell} = \ln\biggl(\frac{1 - \xi_{\ell}}{\xi_{\ell}}\biggr),
\end{equation}
we clip the eigenvalues to the interval \(\xi_{\ell} \in [\varepsilon,1-\varepsilon]\) with \(\varepsilon = 10^{-15}\).

All simulations are performed at half-filling, $N = L/2$.
The particle number is fixed by construction as the number of columns of $V$ and is conserved under the subsequent evolution.
For the superposition initial state, after constructing the zero--$\pi$ superpositions, we have a certain number of occupied orbitals.
If this number is less than $L/2$, we add the required number of positive-phase bulk modes (those with the smallest positive quasienergy) to reach half-filling.
If it exceeds $L/2$, we instead discard the least negative (i.e., those with quasienergy closest to zero) negative-phase bulk modes to reduce the count to exactly $L/2$.
For the $\pi$-mode eigenstate, we begin with all negative-phase bulk modes and the selected $\pi$ mode.
If the total is less than $L/2$, we add positive-phase bulk modes as needed; if it exceeds $L/2$, we trim the excess by removing the least negative negative-phase bulk modes (those with quasienergy closest to zero).
In either case, if the required number of positive-phase bulk modes is insufficient, or if the surplus exceeds the number of negative-phase bulk modes, the construction fails and we fall back to the half-filled ground state of $H_0$.
Any data point where this fallback occurs is excluded from the main analysis; in parameter scans such points are skipped, while in the main simulation they are flagged.

\subsection{Role of the initial state in the observed dynamics}
\label{subsec:init_role}

The superposition initial state is the crucial ingredient that produces a locked subharmonic response in the entanglement spectrum.
Because it combines two Floquet sectors whose quasienergies differ by \(\pi/T\), the relative phase between the zero and \(\pi\) components alternates with period \(2T\), imprinting a period-doubled structure on the many-body wavefunction.
This structure is inherited by the reduced density matrix of subsystem \(A\) and, as shown in the main text, leads to at least one entanglement level that oscillates with a clean period-\(2T\) signature.
By contrast, the \(\pi\)-mode eigenstate is stationary at stroboscopic times and shows no clean period-\(2T\) response in the tracked signal, demonstrating that the effect is not simply a property of the Floquet spectrum itself but requires coherent superposition between the two edge sectors.

Operationally, we determine the optimal half-period $T_{\mathrm{half}}$ by scanning the drive parameter while using the superposition initial state and maximizing the Fourier power at the subharmonic bin $f=1/2$ (in units of cycles per driving period $T$, corresponding to period $2T$).
The scan is performed over a grid of $T_{\mathrm{half}}$ values ranging from $0.1$ to $4.0$ (in units of $1/t_0$) with $80$ equally spaced points.
For each candidate $T_{\mathrm{half}}$, the system is evolved for $200$ periods using the superposition initial state, and the Fourier power at $f=1/2$ is computed from the resulting entanglement-spectrum time series.
The value that yields the highest power is selected as the optimal $T_{\mathrm{half}}$.
Concretely, the time series $\eta_{k^*}(nT)$ is first mean-subtracted and then multiplied element-wise by a Hann window
\begin{equation}
w_n=\frac12\left[1-\cos\left(\frac{2\pi n}{N_T-1}\right)\right],
\qquad n=0,1,\dots,N_T-1,
\end{equation}
before the discrete Fourier transform (DFT) is taken; the power at $f=1/2$ is defined as the squared modulus of the resulting DFT coefficient at the bin closest to frequency $1/2$.
The value of $T_{\mathrm{half}}$ that maximizes this signal is then kept fixed for all subsequent simulations, including those that use the $\pi$-mode eigenstate and the trivial control case.
For trivial dimerization $\delta_0 > 0$ the topologically protected, edge-localized zero and $\pi$ modes are generically absent, so the superposition initial state cannot be constructed; the system is instead initialized in the half-filled ground state of \(H_0\).
This provides a meaningful control: the ground state of \(H_0\) is a generic bulk state with no special coherence between edge modes, and under the Floquet evolution it does not exhibit a subharmonic response.

Importantly, the superposition initial state is a genuinely nonequilibrium configuration: it is not an eigenstate of the Floquet operator, so the system necessarily undergoes non-trivial dynamics, even when observed only at stroboscopic times.
In the present free-fermion setting, the same mechanism operates at the level of Gaussian states: the relative phase accumulated between the zero and \(\pi\) sectors is invisible to diagonal site-density observables, but it is resolved by off-diagonal bond observables and, more strikingly, by a locked period-doubled response in the entanglement spectrum.
The \(\pi\)-mode eigenstate, being stationary, represents an effective equilibrium situation and exhibits no such response.
Thus the observed subharmonic oscillation is a genuinely nonequilibrium effect that relies on the coherence between topologically protected edge modes.

\section{Peschel–Eisler reduction}
\label{sec:peschel_eisler}

This section explains how we compute the \emph{single‑particle entanglement spectrum} (ES) of a spatial subsystem from equal‑time two‑point correlators in a free, number‑conserving fermion system.
The method, due to Peschel and Eisler \cite{Peschel2003, PeschelEisler2009}, applies to any Slater‑determinant many‑body state and expresses the reduced density matrix $\rho_A$ entirely in terms of the correlation matrix restricted to $A$.

\subsection{Slater determinants and the one‑body correlator}
\label{subsec:slater_corr}

We work in the site basis with spinless fermions $c_i$ ($i=1,\dots,L$) satisfying $\{c_i,c_j^\dagger\}=\delta_{ij}$ and $\{c_i,c_j\}=0$.
The one‑body correlation matrix is defined in this work as
\begin{equation}
C_{ij} := \langle c_j^\dagger c_i\rangle,
\qquad i,j=1,\dots,L,
\label{eq:Cij_def}
\end{equation}
where the expectation value is taken in the many‑body state of interest. 

A Slater determinant is specified by an orthonormal set of $N$ occupied single‑particle orbitals $\{|\psi_k\rangle\}_{k=1}^{N}$ ($N\le L$).  We denote their site amplitudes by $\psi_k(i)=\langle i|\psi_k\rangle$, which are complex numbers in general.  We assemble these orbitals into an $L\times N$ matrix $V$ with entries
\begin{equation}
V_{ik} := \psi_k(i),\qquad i=1,\dots,L,\;\; k=1,\dots,N.
\label{eq:V_def}
\end{equation}
The columns of $V$ are the occupied orbitals expressed in the site basis.

To relate $V$ to the correlation matrix, we expand the physical fermion operators in the basis that diagonalizes the occupation.  Extend $\{|\psi_k\rangle\}_{k=1}^{N}$ to a complete orthonormal basis $\{|\psi_k\rangle\}_{k=1}^{L}$ of the single‑particle space, where the additional $L-N$ orbitals are empty in the Slater determinant. The corresponding fermionic operators are
\begin{equation}
a_k^\dagger := \sum_{i=1}^{L} \psi_k(i)\, c_i^\dagger,\qquad
a_k   := \sum_{i=1}^{L} \psi_k(i)^*\, c_i ,
\label{eq:ak_def}
\end{equation}
which satisfy $\{a_k,a_l^\dagger\}=\delta_{kl}$ and $\{a_k,a_l\}=0$.
The many‑body state is $|\Psi\rangle = \prod_{k=1}^{N} a_k^\dagger |0\rangle$.

The inverse transformation gives the expansion of the physical operators in this basis:
\begin{equation}
c_i = \sum_{k=1}^{L} \psi_k(i)\, a_k,\qquad
c_i^\dagger = \sum_{k=1}^{L} \psi_k(i)^*\, a_k^\dagger .
\label{eq:c_expansion}
\end{equation}
These relations follow from the completeness of the orthonormal single‑particle basis, namely $\sum_{k=1}^{L} \psi_k(i) \psi_k(j)^* = \delta_{ij}$.

Now evaluate the correlation matrix element using the definition \eqref{eq:Cij_def}:
\begin{align}
C_{ij}
&= \langle c_j^\dagger c_i \rangle
= \sum_{k,l=1}^{L} \psi_k(j)^* \psi_l(i) \; \langle a_k^\dagger a_l\rangle .
\end{align}
In the Slater determinant, only the occupied orbitals contribute:
\[
\langle a_k^\dagger a_l\rangle = \delta_{kl}\quad (k,l\le N),\qquad
\langle a_k^\dagger a_l\rangle = 0\quad\text{otherwise}.
\]
Hence
\begin{equation}
C_{ij} = \sum_{k=1}^{N} \psi_k(j)^* \psi_k(i).
\label{eq:Cij_sum}
\end{equation}

This is precisely the $(i,j)$ element of the matrix product $V V^\dagger$:
\[
(V V^\dagger)_{ij} = \sum_{k=1}^{N} V_{ik} (V^\dagger)_{kj}
= \sum_{k=1}^{N} \psi_k(i)\,\psi_k(j)^* = C_{ij},
\]
where the last equality follows because $C_{ij}$ in Eq.~\eqref{eq:Cij_sum} has the indices in the order $\psi_k(j)^* \psi_k(i)$, which is the same as $\psi_k(i)\psi_k(j)^*$ (they are just complex numbers and complex multiplication is commutative).
Thus we obtain the compact matrix relation which is also used in our numerical implementation:
\begin{equation}
C = V V^\dagger\, .
\label{eq:C_from_V}
\end{equation}
For a pure Slater determinant, the correlation matrix $C$ is the orthogonal projector onto the occupied single‑particle subspace; consequently, it satisfies $C^2 = C$ and $\operatorname{Tr} C = N$.

\paragraph{Note on the convention.}
If one instead adopts the other convention for the correlation matrix $\tilde C_{ij} = \langle c_i^\dagger c_j\rangle$, then a short calculation using the same expansion shows that $\tilde C = V^* V^T = (V V^\dagger)^T$, i.e., the transpose of our $C$.  Since the eigenvalues of a matrix are invariant under transposition, the entanglement spectrum obtained from $\tilde C_A$ is identical to that obtained from $C_A$ (defined below).
Thus our choice of convention is a matter of convenience and does not affect any physical results.

\paragraph{Note on notation.}
The occupied orbitals $|\psi_k\rangle$ that appear in $V$ are not to be confused with the Floquet modes $|\Phi_\alpha\rangle$ introduced in Sec.~\ref{subsec:ssh_zero_pi_modes}.  Floquet modes are eigenstates of the one‑period evolution operator $U(T)$ and describe stroboscopic dynamics; the occupied orbitals are simply a basis for the subspace filled in the many‑body state at a given time.  In our simulations we often choose the occupied orbitals as superpositions of Floquet modes (e.g., the superposition initial state) or as eigenstates of $H_0$; the relation $C = V V^\dagger$ holds for any orthonormal set spanning the occupied subspace.

Finally, the opposite operator ordering follows from fermionic anticommutation:
\begin{equation}
\langle c_i c_j^\dagger\rangle = \delta_{ij} - \langle c_j^\dagger c_i\rangle
= \delta_{ij} - C_{ij}.
\label{eq:ccdagger_ordering}
\end{equation}
\subsection{Subsystem restriction}
\label{subsec:restrict_subsystem}

We bipartition the chain into a subsystem $A$ and its complement $\bar A$.
In this work, $A$ is the left segment of length $L_A$,
\begin{equation}
A=\{1,2,\dots,L_A\}.
\label{eq:A_segment}
\end{equation}
Throughout this work we set \(L_A = \lfloor\sqrt{L}\rfloor\) (e.g.\ \(L_A = 22\) for \(L = 500\)), the same scale as the edge-density window \(w\) (Sec.~\ref{subsec:edge_density}).
The motivation for this choice and its numerical validation are discussed in Sec.~\ref{sec:initial_states}; in brief, the signal is stable for nearby values of $L_A$ and is suppressed when $L_A$ is taken much larger.
The subsystem correlation matrix is the $L_A\times L_A$ principal submatrix
\begin{equation}
C_A := \left. C \right|_{A}.
\label{eq:CA_def}
\end{equation}
Let $\{\xi_\ell\}_{\ell=1}^{L_A}$ be the eigenvalues of $C_A$,
\begin{equation}
C_A \,\phi_\ell = \xi_\ell\,\phi_\ell,
\qquad 0\le \xi_\ell \le 1,
\label{eq:CA_eigs}
\end{equation}
where $\phi_\ell$ are the corresponding eigenvectors (they live on $A$).

\subsection{Reduced density matrix and entanglement energies}
\label{subsec:entanglement_hamiltonian}

For a Slater determinant, the reduced density matrix on $A$ is Gaussian and can be written as
\begin{equation}
\rho_A = \frac{1}{Z_A}\exp(-H_E),\qquad
H_E = \sum_{m,n\in A} h_{mn}\, f_m^\dagger f_n,
\label{eq:rhoA_HE}
\end{equation}
with $Z_A=\mathrm{Tr}_A\,e^{-H_E}$.  The matrix $h$ (known as the \textit{entanglement Hamiltonian}) is diagonal in the same basis as $C_A$, and its eigenvalues $\eta_\ell$, called the \emph{single‑particle entanglement energies}, are related to the $\xi_\ell$ by the Fermi function
\begin{equation}
\xi_\ell = \frac{1}{e^{\eta_\ell}+1}.
\label{eq:xi_fermi}
\end{equation}
Inverting this gives the central formula used throughout our work:
\begin{equation}
\eta_\ell = \ln\!\left(\frac{1-\xi_\ell}{\xi_\ell}\right)
\;  .
\label{eq:eta_from_xi}
\end{equation}
Equivalently, at the matrix level one may write
\begin{equation}
h = \ln\!\left[(\mathds{1}_{A}-C_A)\,C_A^{-1}\right],
\label{eq:h_from_CA}
\end{equation}
understood in the eigenbasis of $C_A$.

\paragraph{Numerical stability.}
To avoid divergences in Eq.~\eqref{eq:eta_from_xi} when $\xi_\ell$ is extremely close to $0$ or $1$, we clip eigenvalues to $\xi_\ell\in[\varepsilon,1-\varepsilon]$ before evaluating the logarithm.
In the numerics presented here we use $\varepsilon=10^{-15}$.

\paragraph{Notational remark.}
The entanglement energies $\eta_\ell$ are eigenvalues of the entanglement Hamiltonian $H_E$ defined in Eq.~\eqref{eq:rhoA_HE}; they are evaluated from the instantaneous subsystem correlator $C_A(t)$ at any time $t$ (in particular at stroboscopic times $t=nT$ generated by Eq.~\eqref{eq:corr_stroboscopic}).
These should not be confused with the Floquet quasienergies $\varepsilon_\alpha$ (or eigenphases $\theta_\alpha$) introduced in Sec.~\ref{subsec:ssh_zero_pi_modes}, which describe the one‑period evolution operator $U(T)$.
The two sets of energies refer to different operators and are conceptually distinct.

\section{Details of $\pi$-Overlap Tracking}
\label{sec:tracking}

In this section we explain how we extract a single, representative time series from the many‑body state that can reveal subharmonic (period‑$2T$) oscillations.  
Because the system is non‑interacting and the state remains a Slater determinant at all stroboscopic times, all information is contained in the instantaneous correlation matrix $C(t)$ (Sec.~\ref{sec:peschel_eisler}).  
We focus primarily on the entanglement spectrum: a specific entanglement energy \(\eta_{k^*}(nT)\) obtained from the reduced density matrix of the left subsystem \(A\) of size \(L_A = \lfloor\sqrt{L}\rfloor\).
This quantity is the central diagnostic of our work; it exhibits a clear period-doubling signal when the initial state is a coherent superposition of $0$ and $\pi$ Floquet modes, and shows no such signal when the initial state is a pure $\pi$-mode Floquet eigenstate.
For completeness, we also monitor simple one-body observables near the left edge. In the present chiral setting the edge density \(n_{\mathrm{edge}}(nT)\) is symmetry-blind to the zero--\(\pi\) coherence, whereas an off-diagonal nearest-neighbor bond observable provides a meaningful linear comparator to the entanglement-spectrum signal; nevertheless, the entanglement spectrum remains the key focus.

\subsection{Why track a single level?}
\label{subsec:tracking_motivation}

The entanglement spectrum $\{\eta_\ell(nT)\}_{\ell=1}^{L_A}$ contains $L_A$ eigenvalues.  
If the many‑body state contains a superposition of a $0$ mode and a $\pi$ mode that are well localized near the boundary of $A$, the entanglement spectrum will oscillate as a \textit{set} with period $2T$ but this does not guarantee that a clean distinguished entanglement level can be extracted that inherits the same property due to near degeneracies, and mixing with bulk-dominated eigenvectors which can obscure the even--odd structure at the level of individual branches. This is exactly what we show here that indeed it does and there is a clean operational way to extract that level. Other levels may also carry a subharmonic component; at larger system sizes some additional levels pass the same strict criterion, though typically with smaller amplitude or lower signal-to-noise ratio. The largest-overlap level gives the cleanest and most stable signal.
To detect this oscillation cleanly we must follow the same level over time; simply looking at all levels together would obscure the effect.  
This level‑specific oscillation is the hallmark of the subharmonic response we aim to characterize.  
In the present chiral SSH setting, diagonal density probes do not in fact provide such a signal. For a coherent zero--\(\pi\) superposition, the cross-term in any operator of the form \(\sum_j f_j c_j^\dagger c_j\) vanishes exactly because the zero and \(\pi\) modes occupy opposite sublattices. The simplest one-body observable that can resolve the coherence is instead a nearest-neighbor bond operator connecting the two sublattices, for example \(\hat b_{12}=c_1^\dagger c_2+c_2^\dagger c_1\), whose expectation value can oscillate with period \(2T\). In contrast, the appearance of a subharmonic response in the entanglement spectrum is less trivial: the entanglement energies are derived from the eigenvalues of the subsystem correlation matrix, which encode how particles are correlated across the cut. 
A periodic doubling in a specific entanglement level reveals that the superposition of $0$ and $\pi$ modes imprints itself not only on local densities but also on the structure of quantum correlations — a feature that is not simply a linear function of the wavefunction amplitudes.  
We therefore treat the entanglement-spectrum tracking as the principal tool of our analysis. The useful linear comparator is not the edge density but an off-diagonal bond observable, while the main point of the paper is that the entanglement spectrum still displays a sharp locked period-\(2T\) response despite its nonlinear relation to the correlation matrix.

\subsection{Reference vector from Floquet $\pi$ modes}
\label{subsec:reference_vector}

The first step is to identify a $\pi$ mode of the Floquet operator $U(T)$ that is strongly localized near the left edge of the chain.  
From the eigen decomposition
\begin{equation}
U(T)\,|\Phi_\alpha\rangle = e^{-i\theta_\alpha}|\Phi_\alpha\rangle,
\qquad \theta_\alpha\in(-\pi,\pi],
\end{equation}
we extract all modes whose eigenphase satisfies
\(\bigl||\theta_\alpha|-\pi\bigr|<\delta_{\mathrm{tol}}\) (we use $\delta_{\mathrm{tol}}=0.05$ in practice).  
Among these \(\pi\) modes we select the one with the largest edge weight \(w_{\alpha}\) defined in Eq.~\eqref{eq:weight_edge}, i.e.\ the mode most localized in the first \(w = L_A = \lfloor\sqrt{L}\rfloor\) sites.
Denoting its amplitudes by $\Phi_{\pi}(i)=\langle i|\Phi_\pi\rangle$ ($i=1,\dots,L$), we define a reference vector on $A$ by restricting and normalizing:
\begin{equation}
\label{eq:ref_vec}
\mathbf{v}_{\mathrm{ref}} \;=\; \frac{1}{\sqrt{\sum_{i=1}^{L_A} |\Phi_\pi(i)|^{2}}}
      \bigl( \Phi_\pi(1),\,\Phi_\pi(2),\,\dots,\,\Phi_\pi(L_A) \bigr)^{\mathsf T}\;\in\;\mathbb{C}^{L_A}.
\end{equation}
This vector is computed once from the Floquet operator and stored.  
Its role is to mark the entanglement level that is expected to oscillate: because the initial state contains a superposition of $0$ and $\pi$ modes, the eigenvector of $C_A$ that has large overlap with $\mathbf{v}_{\mathrm{ref}}$ will predominantly originate from the $\pi$ sector.  
Tracking the entanglement energy associated with this eigenvector is an operational rule that yields the cleanest and most stable period-$2T$ signal; it does not imply that no other level can exhibit a subharmonic response.

If no $\pi$ mode exists (i.e. the Floquet spectrum has no eigenphase near $\pi$), the construction of a superposition state with a $\pi$ partner is impossible, and we do not expect any robust period doubling.  
In that case we nevertheless need a prescription to select a level to plot, to provide a baseline time series.  
We then simply track the level whose entanglement energy is closest to zero; this choice is arbitrary but gives a well‑defined quantity that, in the topological phase, often corresponds to the mid‑gap edge mode.  
This fallback time series serves as a control, confirming the absence of a subharmonic signal when the required $\pi$ mode is missing.

\subsection{Overlap‑based selection at each stroboscopic time}
\label{subsec:overlap_selection}

At each stroboscopic time $t=nT$ we compute the correlation matrix $C(nT)$ from the occupied orbitals $V(nT)$ (see Sec.~\ref{sec:peschel_eisler}) and extract its restriction to $A$:
\begin{equation}
C_A(nT) = \bigl[ C(nT) \bigr]_{i,j=1}^{L_A}.
\end{equation}
Solving the eigenvalue problem
\begin{equation}
C_A(nT)\,\phi_\ell(nT) = \xi_\ell(nT)\,\phi_\ell(nT),\qquad
\ell=1,\dots,L_A,
\end{equation}
gives the eigenvectors $\phi_\ell$ (living on $A$) and the corresponding entanglement energies
\begin{equation}
\eta_\ell(nT) = \ln\frac{1-\xi_\ell(nT)}{\xi_\ell(nT)}.
\end{equation}
Among all $\ell$ we select the index $k^*(n)$ that maximizes the squared overlap with the reference vector:
\begin{equation}
k^*(n) = \operatorname{argmax}_{\ell}\,
\bigl| \mathbf{v}_{\mathrm{ref}}^\dagger \phi_\ell(nT) \bigr|^{2}.
\end{equation}
The tracked entanglement energy is then
\begin{equation}
\eta_{k^*}(nT) \equiv \eta_{k^*(n)}(nT).
\end{equation}
This procedure guarantees that we follow the level most closely related to the originally localized $\pi$ mode, even if the eigenvectors of $C_A$ change slowly with time.  
The resulting time series $\{\eta_{k^*}(nT)\}$ is the principal object of our analysis.

If no $\pi$ mode existed in the Floquet spectrum, the reference vector is undefined; we instead fall back to tracking the level with entanglement energy closest to zero:
\begin{equation}
k^*(n) = \operatorname{argmin}_{\ell} |\eta_\ell(nT)|.
\end{equation}
This fallback serves two purposes. First, it gives a continuous time series for plotting, avoiding undefined values. Second, it provides a control case: in the absence of a $\pi$ mode, we do not expect any robust $2T$ oscillation. The Fourier transform of this fallback time series should show no peak at $f=1/2$, confirming that the subharmonic response is indeed tied to the presence of the $\pi$ mode. This expectation is indeed confirmed by our numerical checks.


\subsection{Edge density, analytical structure, and the control experiment}
\label{subsec:edge_density}

To connect our entanglement-based diagnostic with more conventional probes, we also monitor the average occupancy in a window of width $w$ adjacent to the left boundary,
\begin{equation}
n_{\mathrm{edge}}(t) = \frac{1}{w}\sum_{i=1}^{w} \langle c_i^\dagger c_i\rangle(t),
\end{equation}
which for a Slater determinant reduces to
\begin{equation}
n_{\mathrm{edge}}(t) = \frac{1}{w}\sum_{i=1}^{w} C_{ii}(t),
\end{equation}
since $\langle c_i^\dagger c_i \rangle = C_{ii}$ by the definition $C_{ij} = \langle c_j^\dagger c_i \rangle$.
We take $w = \lfloor\sqrt{L}\rfloor$ (e.g.\ $w=22$ for $L=500$), large enough to suppress short-range fluctuations but small compared to the system size.

\paragraph{Superposition initial state: diagonal density probes are symmetry-blind.}

For clarity, we first discuss the common case of a single occupied zero--\(\pi\) superposition orbital on top of the stationary negative-phase bulk background. The superposition orbital
\(
\ket{\Psi_{\mathrm{sup}}}
=
(\ket{\Phi_0}+\ket{\Phi_\pi})/\sqrt{2}
\)
evolves at stroboscopic time \(nT\) as
\(
\ket{\Psi_n}
=
(\ket{\Phi_0}+(-1)^n\ket{\Phi_\pi})/\sqrt{2}
\),
since
\(U(T)^n\ket{\Phi_0}=\ket{\Phi_0}\)
and
\(U(T)^n\ket{\Phi_\pi}=(-1)^n\ket{\Phi_\pi}\).

For any Hermitian one-body operator \(\hat O\), the full many-body expectation value can therefore be written as
\begin{equation}
\langle \hat O \rangle(nT)
=
\langle \hat O \rangle_{\mathrm{bulk}}
+
\frac{1}{2}\mel{\Phi_0}{\hat O}{\Phi_0}
+
\frac{1}{2}\mel{\Phi_\pi}{\hat O}{\Phi_\pi}
+
(-1)^n \Re \mel{\Phi_0}{\hat O}{\Phi_\pi},
\label{eq:Oexp_superposition}
\end{equation}
where \(\langle \hat O \rangle_{\mathrm{bulk}}\) is the time-independent contribution from the occupied bulk Floquet modes. Thus the entire period-\(2T\) content is carried by the interference matrix element \(\mel{\Phi_0}{\hat O}{\Phi_\pi}\).

For isolated symmetry-fixed edge modes at quasienergies \(0\) and \(\pi/T\), in the chiral SSH setting, $|\Phi_0\rangle$ and $|\Phi_\pi\rangle$ necessarily have support on opposite sublattices (the specific assignment of which mode occupies which sublattice is determined by the model parameters, but the relative placement on opposite sublattices is a consequence of the chiral symmetry established in Sec.~\ref{subsec:ssh_chiral_sym}). Hence
\begin{equation}
\Phi_0(j)^* \Phi_\pi(j)=0
\qquad \text{for every site } j ,
\label{eq:sitewise_zero_overlap}
\end{equation}
and therefore any diagonal one-body operator of the form
\begin{equation}
\hat O_f= \sum_{j=1}^{L} f_j\, c_j^\dagger c_j
\end{equation}
has vanishing zero--\(\pi\) matrix element,
\begin{equation}
\mel{\Phi_0}{\hat O_f}{\Phi_\pi}
=
\sum_{j=1}^{L} f_j\,\Phi_0(j)^*\Phi_\pi(j)
=0.
\label{eq:diag_selection_rule}
\end{equation}
Thus all diagonal site-density probes are blind to the coherent zero--\(\pi\) mixing. The extension to several occupied zero--\(\pi\) superposition orbitals is immediate: each pair contributes an analogous interference term, and each such term vanishes identically for diagonal density operators.

In particular, for the edge-density operator
\begin{equation}
\hat n_{\mathrm{edge}}
=
\frac{1}{w}\sum_{j=1}^{w} c_j^\dagger c_j ,
\end{equation}
the oscillatory cross-term in Eq.~\eqref{eq:Oexp_superposition} vanishes exactly as is numerically verified as well,
\begin{equation}
\mel{\Phi_0}{\hat n_{\mathrm{edge}}}{\Phi_\pi}=0,
\end{equation}
so \(n_{\mathrm{edge}}(nT)\) is stroboscopically time independent even for the coherent superposition initial state.

\paragraph{Off-diagonal bond observables can resolve the coherence.}

The correct linear comparator is an operator that connects opposite sublattices. To match the numerics, we define the edge-bond average over the same left-edge window used for the density probe and for the default subsystem size,
\begin{equation}
\hat b_{\mathrm{edge}}
=
\frac{1}{w-1}\sum_{i=1}^{w-1}
\left(c_i^\dagger c_{i+1}+c_{i+1}^\dagger c_i\right),
\label{eq:bedge_op}
\end{equation}
with expectation value
\begin{equation}
b_{\mathrm{edge}}(t)
=
\langle \hat b_{\mathrm{edge}} \rangle(t)
=
\frac{1}{w-1}\sum_{i=1}^{w-1}
\langle c_i^\dagger c_{i+1}+c_{i+1}^\dagger c_i\rangle(t).
\label{eq:bedge_def}
\end{equation}
To connect Eq.~\eqref{eq:bedge_def} to the correlation matrix, recall that in our convention
\(
C_{ij}(t)=\langle c_j^\dagger c_i\rangle(t)
\).
Hence
\(
\langle c_i^\dagger c_{i+1}\rangle(t)=C_{i+1,i}(t)
\)
and
\(
\langle c_{i+1}^\dagger c_i\rangle(t)=C_{i,i+1}(t)
\).
Since \(C(t)\) is Hermitian for a Gaussian state,
\(
C_{i,i+1}(t)=C_{i+1,i}(t)^*
\),
and therefore
\begin{equation}
\langle c_i^\dagger c_{i+1}+c_{i+1}^\dagger c_i\rangle(t)
=
C_{i+1,i}(t)+C_{i+1,i}(t)^*
=
2\,\Re\,C_{i+1,i}(t),
\end{equation}
which yields Eq.~\eqref{eq:bedge_C} after averaging over \(i=1,\dots,w-1\). Using the convention \(C_{ij}(t)=\langle c_j^\dagger c_i\rangle(t)\), each bond term satisfies
\begin{equation}
\langle c_i^\dagger c_{i+1}+c_{i+1}^\dagger c_i\rangle(t)
=
C_{i+1,i}(t)+C_{i,i+1}(t)
=
2\,\Re\, C_{i+1,i}(t),
\end{equation}
so that
\begin{equation}
b_{\mathrm{edge}}(t)
=
\frac{2}{w-1}\sum_{i=1}^{w-1}\Re\, C_{i+1,i}(t).
\label{eq:bedge_C}
\end{equation}
This observable is off-diagonal in the site basis and is not subject to the diagonal selection rule in Eq.~\eqref{eq:diag_selection_rule}. Accordingly, its zero--\(\pi\) interference term is generically nonzero, so \(b_{\mathrm{edge}}(nT)\) can exhibit a clear period-\(2T\) response even though \(n_{\mathrm{edge}}(nT)\) remains flat.

\paragraph{Superposition initial state: analytic structure of $C_A$.}

The same coherence drives a period-\(2T\) structure in the correlation matrix, which we now make precise. For clarity, consider first the case of a single occupied zero--\(\pi\) superposition orbital on top of the occupied negative-phase bulk modes. Since each bulk mode is a Floquet eigenmode, the phase \(e^{-i\theta_\beta n}\) it acquires under \(U(T)^n\) cancels exactly in the outer product \(\ket{\chi_\beta(nT)}\bra{\chi_\beta(nT)}\) for the bulk states, so the bulk contribution
\begin{equation}
C_{\mathrm{bulk}}=\sum_\beta \ket{\chi_\beta}\bra{\chi_\beta}
\end{equation}
is strictly time independent. The superposition orbital is not a Floquet eigenmode, so its phases do not cancel, and the full correlation matrix decomposes as
\begin{equation}
C(nT)=C_{\mathrm{bulk}}+\ket{\Psi_n}\bra{\Psi_n}.
\label{eq:C_rank1}
\end{equation}

Restricting to the \(L_A\times L_A\) subsystem block gives
\begin{equation}
C_A(nT)=C_{A,\mathrm{bulk}}+\ket{\psi_n^A}\bra{\psi_n^A},
\label{eq:CA_rank1}
\end{equation}
where \(\ket{\psi_n^A}\) is the restriction of \(\ket{\Psi_n}\) to subsystem \(A\). Because \(\ket{\psi_n^A}\) takes only two values, one for even \(n\) and one for odd \(n\), the matrix \(C_A(nT)\) alternates between exactly two matrices. Hence the entanglement spectrum of \(C_A\), regarded as a set of eigenvalues, is exactly period-\(2T\). This is true as a \textit{set}. However, due to the possible presence of degeneracies or level permutations, each ordered eigenvalue is not guaranteed to individually have period-\(2T\) as well. This is the result we establish numerically it does and clean entanglement levels can be extracted that inherit the period-doubling. 

If several zero--\(\pi\) superposition orbitals are occupied, the generalization is immediate:
\begin{equation}
C(nT)=C_{\mathrm{bulk}}+\sum_{k=1}^{n_{\mathrm{pairs}}}\ket{\Psi_{n,k}}\bra{\Psi_{n,k}},
\end{equation}
with the same even/odd alternation inherited from the common factor \((-1)^n\). The subsystem matrix \(C_A(nT)\) therefore still alternates between two matrices at stroboscopic times, so its spectrum as a set remains exactly period-\(2T\).

This is, however, precisely where the analytic argument ends. The algebra above guarantees exact period doubling of the correlation matrix and of the entanglement spectrum as a set, but it does not determine which level carries the dominant visible oscillation, how large that oscillation is, or how sharply it is isolated from the rest of the spectrum. Those features depend on the localization of the zero and \(\pi\) modes inside \(A\), on their separation from bulk-dominated eigenvectors of \(C_{A,\mathrm{bulk}}\), and on the finite-size realization of the Floquet gaps at \(\theta=0\) and \(\theta=\pi\). The persistence of a sharp, single-level subharmonic signal over long times, together with its disappearance in the trivial phase outside the topological phase boundary, is therefore a genuinely nontrivial numerical result rather than a consequence of Eq.~\eqref{eq:CA_rank1}.

\paragraph{Eigenstate initial state: all one-body observables are stationary.}

When the superposition is replaced by a pure \(\pi\)-mode Floquet eigenstate, the occupied single-particle subspace contains only genuine Floquet eigenmodes. Each occupied orbital then acquires only a scalar phase \(e^{-i\theta_\alpha n}\) under \(U(T)^n\), and these phases cancel between bra and ket in the expectation value of any one-body operator. Consequently, all one-body observables are exactly stationary at stroboscopic times, including diagonal densities such as \(n_{\mathrm{edge}}\) and off-diagonal bond observables such as \(b_{\mathrm{edge}}\).

At the matrix level, if the columns of \(V(0)\) span the occupied Floquet eigenspace, then
\begin{equation}
V(nT)=V(0)\,\Lambda_n,
\qquad
\Lambda_n=\mathrm{diag}\!\left(e^{-i\theta_1 n},e^{-i\theta_2 n},\dots\right),
\end{equation}
with \(\Lambda_n\) diagonal and unitary. Therefore,
\begin{equation}
C(nT)=V(nT)V(nT)^\dagger
=
V(0)\Lambda_n\Lambda_n^\dagger V(0)^\dagger
=
V(0)V(0)^\dagger
=
C(0),
\end{equation}
so \(C_A(nT)=C_A(0)\) exactly and all entanglement energies are frozen. The control state is thus stroboscopically stationary, which makes it the appropriate equilibrium-like baseline against which the nonequilibrium period doubling of the superposition state should be judged.

\paragraph{The control experiment: a clean binary switch.}

The control experiment consists of replacing the superposition initial state with a pure \(\pi\)-mode Floquet eigenstate and running the same analysis for the tracked entanglement level together with the one-body probes introduced above. In the superposition state, the off-diagonal edge-bond observable exhibits a clear period-\(2T\) oscillation, whereas the diagonal edge density remains flat by symmetry. In the \(\pi\)-mode eigenstate control, however, both types of one-body probes are stroboscopically stationary: the density is stationary, and the off-diagonal bond oscillation disappears as well. Under the same tracking rule, the entanglement signal shows no clean period-$2T$ response. This clean before/after contrast shows that the observed subharmonic response is genuinely a nonequilibrium coherence effect tied to the zero--\(\pi\) superposition. The control state is stroboscopically stationary, so the period-doubled response cannot be attributed to the mere presence of Floquet edge modes or to the static structure of the Floquet spectrum alone.

\subsection{Numerical stability and fallbacks}
\label{subsec:tracking_numerics}

Several practical points ensure the robustness of the tracking:

\begin{itemize}
\item The reference vector $\mathbf{v}_{\mathrm{ref}}$ is renormalized after restriction, so its norm is unity.  
  Overlaps $|\mathbf{v}_{\mathrm{ref}}^\dagger \phi_\ell|^2$ are therefore bounded between $0$ and $1$, and the maximisation is unambiguous.

\item When multiple $\pi$ modes exist, we choose the one with the largest left‑edge weight; this typically picks the mode most relevant for a left‑subsystem observable.  
  The same reference vector is used for all times.

\item If at some time the correlation matrix becomes nearly singular (eigenvalues $\xi_\ell$ extremely close to $0$ or $1$), the clipping $\xi_\ell\to\max(\varepsilon,\min(1-\varepsilon,\xi_\ell))$ with $\varepsilon=10^{-15}$ avoids divergences in $\eta_\ell$.  
  This affects only the energy values, not the eigenvectors, so the overlap selection remains well‑defined.

\item For the edge density, no special treatment is needed; $C_{ii}$ are always between $0$ and $1$ and are computed directly from $V V^\dagger$.

\end{itemize}

The tracked time series \(\{\eta_{k^*}(nT)\}_{n=0}^{N_T}\), \(\{n_{\mathrm{edge}}(nT)\}_{n=0}^{N_T}\), and \(\{b_{\mathrm{edge}}(nT)\}_{n=0}^{N_T}\) are then analysed by Fourier transform (Sec.~\ref{sec:observable}) to quantify the strength of the subharmonic component at frequency \(f=1/2\) (i.e. period \(2T\)).
The entanglement-spectrum series is the central result. The edge density is included to demonstrate the symmetry-based absence of any diagonal one-body response, while the edge-bond series provides the appropriate linear comparator on the same left-edge region. Taken together, these probes show that the locked period doubling is a genuinely nonequilibrium coherence effect and appears in the entanglement spectrum, which is itself a nonlinear function of the correlation matrix, and is therefore not an artifact of any linear one-body observable.

\section{Quantifying Subharmonic Response}
\label{sec:observable}

After obtaining the tracked time series \(\{\eta_{k^*}(nT)\}_{n=0}^{N_T}\), \(\{n_{\mathrm{edge}}(nT)\}_{n=0}^{N_T}\), and \(\{b_{\mathrm{edge}}(nT)\}_{n=0}^{N_T}\), we need a quantitative measure of the period-\(2T\) oscillation.
We employ Fourier analysis, which extracts the frequency content of the signal.  
Because the system is stroboscopically sampled at integer multiples of the driving period $T$, the natural frequency unit is cycles per period, i.e., a frequency $f$ corresponds to an oscillation with period $1/f$ in units of $T$.  
A perfect period‑$2T$ signal therefore appears as a peak at $f = 1/2$.

\subsection{Fourier analysis of the tracked time series}
\label{subsec:fourier_method}

Let \(x_n\) denote any of the tracked stroboscopic series, in particular \(\eta_{k^*}(nT)\), \(n_{\mathrm{edge}}(nT)\), or \(b_{\mathrm{edge}}(nT)\), for \(n = 0,1,\dots,N_T\), where \(N_T\) is the total number of stroboscopic periods recorded.
We first subtract the mean value to remove the zero‑frequency (DC) component:
\begin{equation}
\tilde{x}_n = x_n - \frac{1}{N_T+1}\sum_{m=0}^{N_T} x_m .
\end{equation}
To reduce spectral leakage caused by the finite length of the time series, we multiply $\tilde{x}_n$ by a Hanning window:
\begin{equation}
w_n = \frac{1}{2}\left[1 - \cos\left(\frac{2\pi n}{N_T}\right)\right],\qquad
y_n = w_n \, \tilde{x}_n .
\end{equation}
The discrete Fourier transform is then computed using the real fast Fourier transform (rFFT):
\begin{equation}
Y_k = \sum_{n=0}^{N_T} y_n \, e^{-2\pi i k n / (N_T+1)},\qquad
k = 0,1,\dots,\lfloor N_T/2\rfloor
\end{equation}
The power at frequency $f_k = k/(N_T+1)$ (in units of $1/T$) is $P_k = |Y_k|^2$.

\subsection{Subharmonic strength metrics}
\label{subsec:metrics}

We use two closely related quantities to characterise the strength of the $f=1/2$ component.

\paragraph{Raw power at $f=1/2$.}
The simplest measure is the value of the power spectrum at the frequency bin closest to $f=1/2$:
\begin{equation}
P_{1/2} = P_{k^*},\qquad k^* = \operatorname{argmin}_k |f_k - 0.5|.
\end{equation}
While useful for tasks such as optimising $T_{\mathrm{half}}$ (where only relative values matter), the raw power depends on the overall amplitude of the signal and is not easily comparable between different simulations.

\paragraph{Fraction of total oscillatory power.}
A more robust metric normalises $P_{1/2}$ by the total power contained in all non‑zero frequencies, i.e. the sum over $k\ge 1$:
\begin{equation}
\mathcal{F}_{1/2} = \frac{P_{1/2}}{\displaystyle\sum_{k\ge 1} P_k}.
\label{eq:F12_def}
\end{equation}
This fraction lies between $0$ and $1$ (or $0$ and $100\%$ when multiplied by $100$) and quantifies how much of the time‑dependent signal is concentrated at the subharmonic frequency.  
A value close to $1$ (or $100\%$) indicates a clean period‑$2T$ oscillation; values near $0$ signify no detectable subharmonic response.

\subsection{Floating-point regularisation of $\mathcal{F}_{1/2}$}
\label{subsec:guard}

The estimator $\mathcal{F}_{1/2}$ defined in Eq.~\eqref{eq:F12_def} is a ratio of powers computed from the mean-subtracted series $\tilde{x}_n$. If the physical series $x_n$ has no time dependence at all---so that $\tilde{x}_n = 0$ for every $n$ in exact arithmetic---then both the numerator and the denominator of Eq.~\eqref{eq:F12_def} are identically zero, and the ratio is undefined. In a finite-precision computation this manifests as follows: $\tilde{x}_n$ is not exactly zero but instead contains rounding noise at the level of machine precision $\varepsilon_{\rm mach}\approx 10^{-14}$, accumulated over the QR orthogonalisation steps of the time evolution. The ratio $P_{1/2}/\sum_{k\ge 1}P_k$ then amounts to dividing noise by noise and can take any value in $[0,1]$ without physical meaning.

To detect this situation we compare the standard deviation of the mean-subtracted series,
\begin{equation}
  \sigma_x = \left[\frac{1}{N_T+1}\sum_{n=0}^{N_T}\tilde{x}_n^{\,2}\right]^{1/2},
\end{equation}
to the mean $\bar{x} = (N_T+1)^{-1}\sum_{n=0}^{N_T} x_n$. The ratio $\sigma_x/|\bar{x}|$ measures how large the fluctuations are relative to the mean value. We set
\begin{equation}
  \mathcal{F}_{1/2} = 0 \quad \text{whenever} \quad \frac{\sigma_x}{|\bar{x}|} < 10^{-8},
\label{eq:F12_guard}
\end{equation}
and apply this guard to every evaluated time series. For the superposition state the physical oscillation gives $\sigma_x/|\bar{x}|\simeq 1.5\times 10^{-2}$, so the guard never triggers. For the $\pi$-eigenstate control, whose observables are stroboscopically constant by construction, the residual fluctuations satisfy $\sigma_x/|\bar{x}|\lesssim 10^{-11}$, and the guard correctly returns $\mathcal{F}_{1/2}=0$. The threshold $10^{-8}$ therefore ensures no false negatives.

\subsection{Masking phase diagrams with bulk–boundary proxy criteria}
\label{subsec:masking}

In constructing phase diagrams that map the subharmonic response across parameter space, we are interested only in regions where the system satisfies the necessary conditions for hosting robust $0$ and $\pi$ modes.  
These conditions are diagnosed using stroboscopic time evolution via the Floquet operator $U(T)$.  
We introduce a \textit{bulk–boundary proxy flag} that is true if and only if the following three criteria are met:

\begin{enumerate}
\item \textbf{Pair count:} at least one pair of zero and $\pi$ modes exists, i.e. $\min(n_0,n_\pi) \ge 1$ (with a tolerance $\delta_{\mathrm{tol}}=0.05$ on the eigenphases).
\item \textbf{Bulk gaps:} the bulk quasienergy spectrum is gapped around both $\theta=0$ and $\theta=\pi$.  
  Denoting the set of eigenphases not classified as zero or $\pi$ modes, we define
  \begin{align}
  \Delta_0 &= \min_{\theta \notin \text{zero}} |\theta|,\\
  \Delta_\pi &= \min_{\theta \notin \text{$\pi$}} \bigl|\,|\theta|-\pi\,\bigr|.
  \end{align}
  The gaps are considered open if $\Delta_0 \ge \delta_{\mathrm{tol}}$ and $\Delta_\pi \ge \delta_{\mathrm{tol}}$ (with the same tolerance $\delta_{\mathrm{tol}}=0.05$).
\item \textbf{Edge localization:} among the zero modes, at least one has an edge weight exceeding a threshold $w_{\mathrm{thr}}=0.30$, and similarly for the $\pi$ modes.  
  The edge weight for a mode $\Phi$ is defined as the fraction of its probability density in the first and last $w$ sites (with $w = \lfloor\sqrt{L}\rfloor$), i.e.
  \begin{equation}
W_{\mathrm{edge}}(\Phi)
  = \frac{\displaystyle\sum_{i=1}^{w} |\Phi(i)|^2
          + \sum_{i=L-w+1}^{L} |\Phi(i)|^2}
         {\displaystyle\sum_{i=1}^{L} |\Phi(i)|^2}.
  \end{equation}
  A mode is considered edge‑localized if $W_{\mathrm{edge}} \ge w_{\mathrm{thr}}$.
\end{enumerate}


Only when all three conditions are satisfied do we regard the system as a candidate for exhibiting a robust subharmonic response.
However, the subharmonic signal in the left subsystem also requires the \(\pi\) mode to be left‑localized; therefore the region where a strong response appears can be a proper subset of the proxy region, as indeed seen in the phase diagram heat maps. 
This shows a stronger constraint that despite a point in the parameter space being bulk-boundary active proxy, there is an additional requirement needed for the locked subharmonic response in the entanglement level. This is indeed what we find: for any point where a left‑localized \(\pi\) mode exists together with a zero mode and bulk gaps, the subharmonic response is clean and persistent within the bulk-boundary proxy active area.
For each point in the $(\delta_{\mathrm{K}},T_{\mathrm{half}})$ plane, we run the full stroboscopic simulation and compute $\mathcal{F}_{1/2}$; the result is then included in the final map only if the proxy flag is true, and is otherwise masked out.
This masking ensures that the displayed subharmonic strength is only shown where the underlying Floquet spectrum supports the required edge modes and gaps, providing a clear visual separation of the “active” region.

The resulting maps --- the pair‑count map, the proxy flag map, and the masked subharmonic fraction map --- together establish the correspondence between the topological Floquet properties and the observed period‑doubling in the entanglement spectrum.

\subsection{Numerical protocol and parameter summary}
\label{sec:numerical_protocol}

All simulations use the fixed conventions and parameters as shown in Table \ref{tab:params}, unless stated otherwise in a figure caption.

\begin{table}[h]
\centering
\caption{Default parameters used throughout the simulations.}
\begin{tabular}{lcl}
\toprule
Parameter & Value & Description \\
\midrule
\multicolumn{3}{l}{\textit{Model and geometry}} \\
\midrule
System size \(L\) & 500 & Number of sites (open chain, OBC) \\
Subsystem \(A\) size \(L_A\) = edge window \(w\) & \(\lfloor\sqrt{L}\rfloor\)
  & \makecell[l]{Left segment; both quantities use the same \\
    scale (e.g.\ \(L_A = w = 22\) for \(L = 500\))} \\
Filling & \(N = L/2\) & Half-filling; fixed by construction \\
Onsite potential & \(0\) & No onsite terms in \(H_0\) or \(H_K\) \\
Hopping scales & \(t_0 = t_K = 1\) & \makecell[l]{Sets the energy and time unit; all \(T_{\mathrm{half}}\) \\
    values are in units of \(1/t_0\)} \\
Static dimerization \(\delta_0\) & \(-0.30\) (topo) & \makecell[l]{Dimerization of \(H_0\); \(\delta_0 < 0\) is topological, \\
    \(\delta_0 = +0.30\) used for trivial control} \\
Kick dimerization \(\delta_K\) & \(0.80\) & Dimerization of the kick step \(H_K\) \\
Floquet operator &
  \(U(T) = e^{-iH_K T_{\mathrm{half}}}\) 
    \(\times\, e^{-iH_0 T_{\mathrm{half}}}\) &
  \makecell[l]{\(H_0\) acts in the first half-period, \\
    \(H_K\) acts in the second} \\
\midrule
\multicolumn{3}{l}{\textit{Floquet mode classification and proxy}} \\
\midrule
Phase tolerance \(\delta_{\mathrm{tol}}\) & \(0.05\) & Classification of zero/\(\pi\) modes \\
Edge weight threshold \(w_{\mathrm{thr}}\) & \(0.30\) & \makecell[l]{Minimum edge weight for a mode to be \\
    considered localized} \\
Pair threshold & \(\min(n_0, n_\pi) \geq 1\) & \makecell[l]{Minimum number of zero--\(\pi\) pairs required \\
    for a point to be classified as topological} \\
\midrule
\multicolumn{3}{l}{\textit{Time evolution and numerics}} \\
\midrule
Clipping constant \(\varepsilon\) & \(10^{-15}\) &
  Prevents divergences in \(\eta = \ln[(1-\xi)/\xi]\) \\
QR re-orthonormalization & every time step &
  \makecell[l]{Applied after each application of \(U(T)\) to \\
    maintain exact projector properties} \\
\hline
\multicolumn{3}{l}{\textit{\(T_{\mathrm{half}}\) scan and main runs (observable tests)}} \\
\hline
\(T_{\mathrm{half}}\) scan range & \([0.1, 4.0]\) & 80 equally spaced points \\
Number of periods for scan & 200 & Used to find optimal \(T_{\mathrm{half}}\) \\
Number of periods for main runs & 800 & Final data for figures \\
\hline
\multicolumn{3}{l}{\textit{Phase diagram scans}} \\
\hline
\(\delta_K\) grid & \([-0.95,\, 0.95]\), step \(0.05\) & 39 points \\
\(T_{\mathrm{half}}\) grid & \([0.10,\, 2.50]\), step \(0.05\) & 49 points \\
Number of periods (phase diagrams) & 100 & \makecell[l]{Distinct from the 200/800 used in \\
    observable tests} \\
Initial states (phase diagrams) & superposition state, $\pi$-mode eigenstate &
  Two separate subharmonic maps produced \\
\hline
\multicolumn{3}{l}{\textit{Fourier analysis}} \\
\hline
Pre-processing & Mean subtraction &
  \(\tilde{x}_n = x_n - \bar{x}\) before windowing \\
Window function & Hanning &
  \makecell[l]{\(w_n = \tfrac{1}{2}\!\left(1 - \cos\tfrac{2\pi n}{N_T}\right)\), \\
    applied element-wise to reduce spectral leakage} \\
Subharmonic metric \(F_{1/2}\) &
  \(P_{1/2}\big/\!\displaystyle\sum_{k \geq 1} P_k\) &
  \makecell[l]{Fraction of total non-DC Fourier power at \\
    \(f = 1/2\); lies in \([0, 1]\)} \\
\hline
\multicolumn{3}{l}{\textit{Fallback and tie-breaking}} \\
\hline
Fallback handling & Exclude (scan) / flag (main) &
  \makecell[l]{Points where the requested initial state \\
    cannot be constructed are omitted from \\
    analysis} \\
Tie-breaking for pairing &
  \makecell[l]{Sort by descending left-edge weight \\ (first \(w\) sites)} &
  \makecell[l]{\(k^{\rm th}\) most localized zero mode paired with \\
    \(k^{\rm th}\) most localized \(\pi\) mode} \\
Reference vector for tracking &
  \makecell[l]{Most left-localized \(\pi\) mode \\
    (largest \(w_{\alpha}\), Eq.~\eqref{eq:weight_edge})} &
  \makecell[l]{Normalized restriction to first \(L_A = \lfloor\sqrt{L}\rfloor\) \\
    sites} \\
\bottomrule
\end{tabular}
\label{tab:params}
\end{table}

The Floquet operator is always constructed as  
\(U(T) = e^{-i H_{\mathrm{K}} T_{\mathrm{half}}}\, e^{-i H_0 T_{\mathrm{half}}}\)  
with \(H_0\) and \(H_{\mathrm{K}}\) defined in Eqs.~\eqref{eq:ssh_H0} and \eqref{eq:ssh_HK}. For the trivial dimerization \(\delta_0>0\) the superposition initial state cannot be built; the system is then initialized in the half‑filled ground state of \(H_0\), and these data are clearly flagged as a control.

\section{Trivial--phase control}
\label{sec:trivial_control}

In the main text the $\pi$-mode Floquet eigenstate serves as the primary
control, showing that Floquet topology is not \emph{sufficient} for the
observed subharmonic response.
A complementary check addresses necessity: if the system is placed in the
topologically trivial phase ($\delta_0=+0.30>0$), the Floquet operator
supports no edge-localized zero or $\pi$ modes, and the superposition initial
state of Sec.~\ref{subsec:init_superposition} cannot be formed.
In that regime the system is initialized in the half-filled ground state of
$H_0$ instead, which is a generic bulk state with no edge-sector coherence.

\begin{figure}[htbp]
    \centering
    \includegraphics[width=\linewidth]{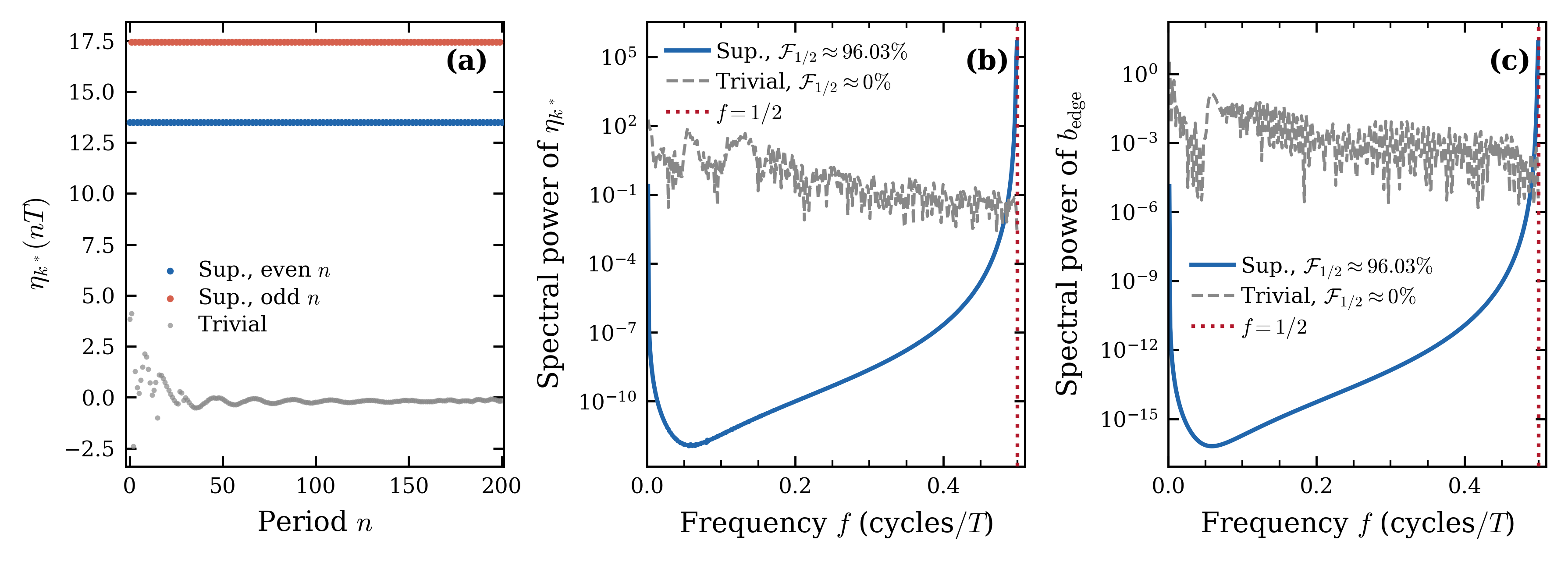}
    \caption{Three-panel subharmonic diagnostic comparing the topological
    superposition state (blue/orange, even/odd stroboscopic periods) with the
    trivial-phase initialization (gray; $\delta_0=+0.30$, no edge modes,
    half-filled ground state of $H_0$).
    The topological step uses $\delta_0=-0.30$; both cases share
    $\delta_{\mathrm{K}}=0.8$, $L=500$,
    $L_A=\lfloor\sqrt{L}\rfloor=22$, $N=800$ periods.
    (a)~Stroboscopic time trace of $\eta_{k^*}(nT)$: the topological state
    separates into two stable even/odd branches while the trivial
    initialization shows no even--odd splitting.
    (b)~Hanning-windowed Fourier power of $\eta_{k^*}$: a sharp peak at
    $f=1/2$ (red dotted) appears only for the topological superposition.
    (c)~Fourier power of $b_{\mathrm{edge}}$: same conclusion.
    The absence of any subharmonic signal in the trivial phase confirms that
    zero--$\pi$ Floquet topology is a necessary condition for the effect.}
    \label{fig:trivial_control}
\end{figure}

Figure~\ref{fig:trivial_control} shows the three-panel observable layout for
the topological superposition state alongside the trivial-phase initialization,
with the second-step dimerization $\delta_{\mathrm{K}}=0.8$ and all other
parameters identical ($L=500$, $L_A=22$, $N=800$ periods).
For the trivial initialization, $\mathcal{F}_{1/2}$ is consistent with zero
in both the entanglement-level and bond-observable diagnostics; no resolved
subharmonic peak appears at $f=1/2$.
This confirms that the simultaneous presence of zero and $\pi$ Floquet edge
modes is a \emph{necessary} ingredient: together with the eigenstate control
of the main text, the two checks establish that zero--$\pi$ Floquet topology
is necessary and, when combined with coherent nonequilibrium preparation,
sufficient for the locked subharmonic response.

\subsection{Flat diagonal edge density and comprehensive observable comparison}
\label{subsec:nedge_flat}

\begin{figure}[htbp]
    \centering
    \includegraphics[width=\linewidth]{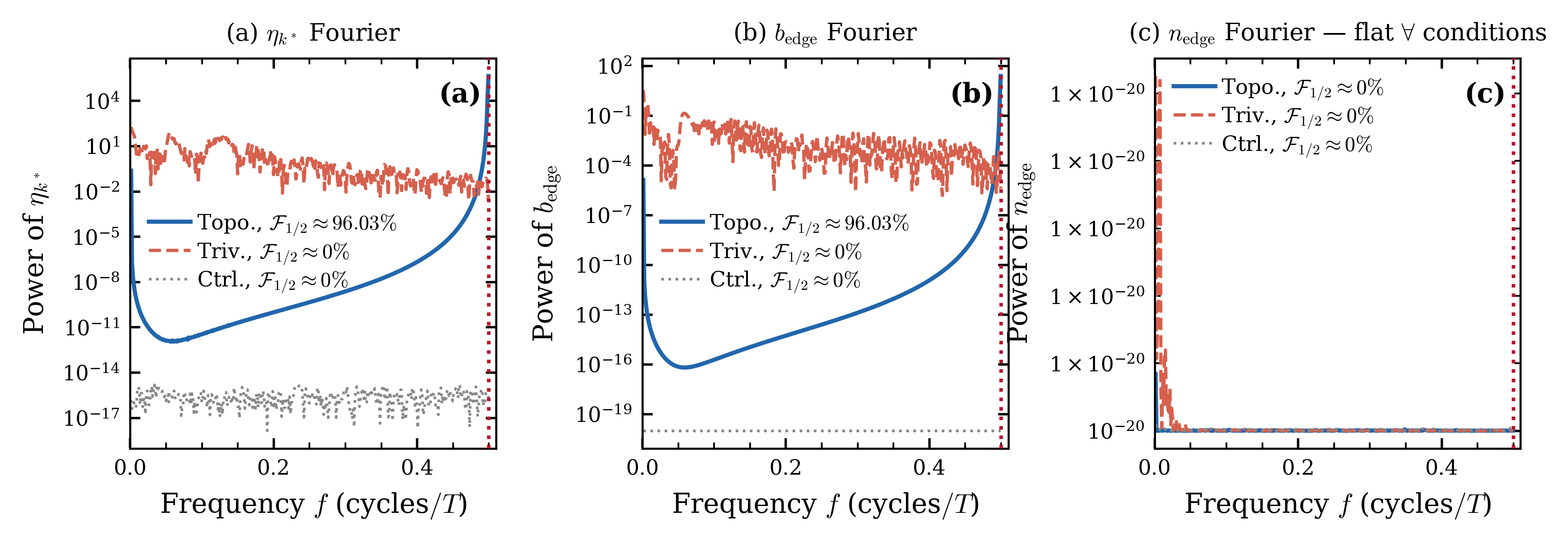}
    \caption{Fourier power spectra of all three left-edge observables for all
    three initial conditions: topological superposition state ($\delta_0=-0.3<0$, blue), trivial-phase
    initialization (orange; $\delta_0=+0.3>0$, ground state of $H_0$), and
    $\pi$-mode Floquet eigenstate control (gray).
    The red dotted line marks the subharmonic frequency $f=1/2$.
    Parameters: $\delta_{\mathrm{K}}=0.8$, $L=500$,
    $L_A=\lfloor\sqrt{L}\rfloor=22$, $N=800$ periods.
    (a)~$\eta_{k^*}$: sharp $f=1/2$ peak only for the topological superposition.
    (b)~$b_{\mathrm{edge}}$: same conclusion.
    (c)~$n_{\mathrm{edge}}$: flat for all three conditions, confirming that
    sublattice symmetry forbids a period-$2T$ component in diagonal one-body
    observables regardless of initial state or phase.}
    \label{fig:supplemental_diagnostics}
\end{figure}

The main-text caption of Fig.~1 notes that the diagonal edge density $n_{\mathrm{edge}}(nT)$ remains flat by sublattice symmetry and is not shown there.
The analytic argument of subsystem chiral symmetry as provided in the main text as well as above forces the interference matrix element $\langle\Phi_0|O_f|\Phi_\pi\rangle$ to vanish for any diagonal one-body operator $O_f$, so no period-$2T$ component can appear in $n_{\mathrm{edge}}$ regardless of the initial state.
Figure~\ref{fig:supplemental_diagnostics} provides the numerical confirmation
alongside a comprehensive comparison of all three observables and all three
initial conditions.
Panel~(c) shows the Fourier power of $n_{\mathrm{edge}}$ for the topological
superposition state, the trivial-phase initialization, and the $\pi$-mode
Floquet eigenstate control: all three are flat to within numerical noise, with
$\mathcal{F}_{1/2}$ consistent with zero.
The mechanism is provided above as well as in the main text: in the chiral SSH setting the zero
and $\pi$ edge modes occupy opposite sublattices, so the interference matrix
element $\langle\Phi_0|O_f|\Phi_\pi\rangle$ vanishes identically for any
diagonal one-body operator $O_f$, regardless of which initial state is used.

Panels~(a) and~(b) show the Fourier power of $\eta_{k^*}$ and
$b_{\mathrm{edge}}$, respectively, for the same three conditions.
A sharp subharmonic peak at $f=1/2$ appears only for the topological
superposition state; both the trivial-phase initialization and the eigenstate
control show no resolved peak.
This single figure therefore consolidates three results: the subharmonic
signal requires topology (Fig.~\ref{fig:trivial_control}), requires coherent
nonequilibrium preparation (main text), and is invisible to diagonal density
probes regardless of the initial state (panel~(c)).

\section{Robustness of the Bulk--Boundary Proxy Criteria}
\label{sec:robustness}

The bulk--boundary proxy used to identify the topological window in the phase diagrams depends on three numerical analysis choices: the quasienergy tolerance $\delta_{\mathrm{tol}}$ used to classify a Floquet mode as a zero or $\pi$ mode, the edge-weight threshold $w_{\mathrm{thr}}$ used to classify such a mode as boundary-localized, and the subsystem size $L_A$/system size $L$.
For the parameter regime studied here, the resulting proxy maps and masked subharmonic-response maps are robust under controlled variations of all three choices.
The reason is that, inside the topological regime, the Floquet spectrum exhibits clear bulk gaps around $\theta=0$ and $\theta=\pi$ while the corresponding edge modes remain strongly localized, so modest changes in the classification rules do not alter the identified window at the numerical resolution used here.
We verify this explicitly below.

\begin{figure}[htbp]
    \centering
    \begin{subfigure}{0.49\linewidth}
        \includegraphics[width=\linewidth]{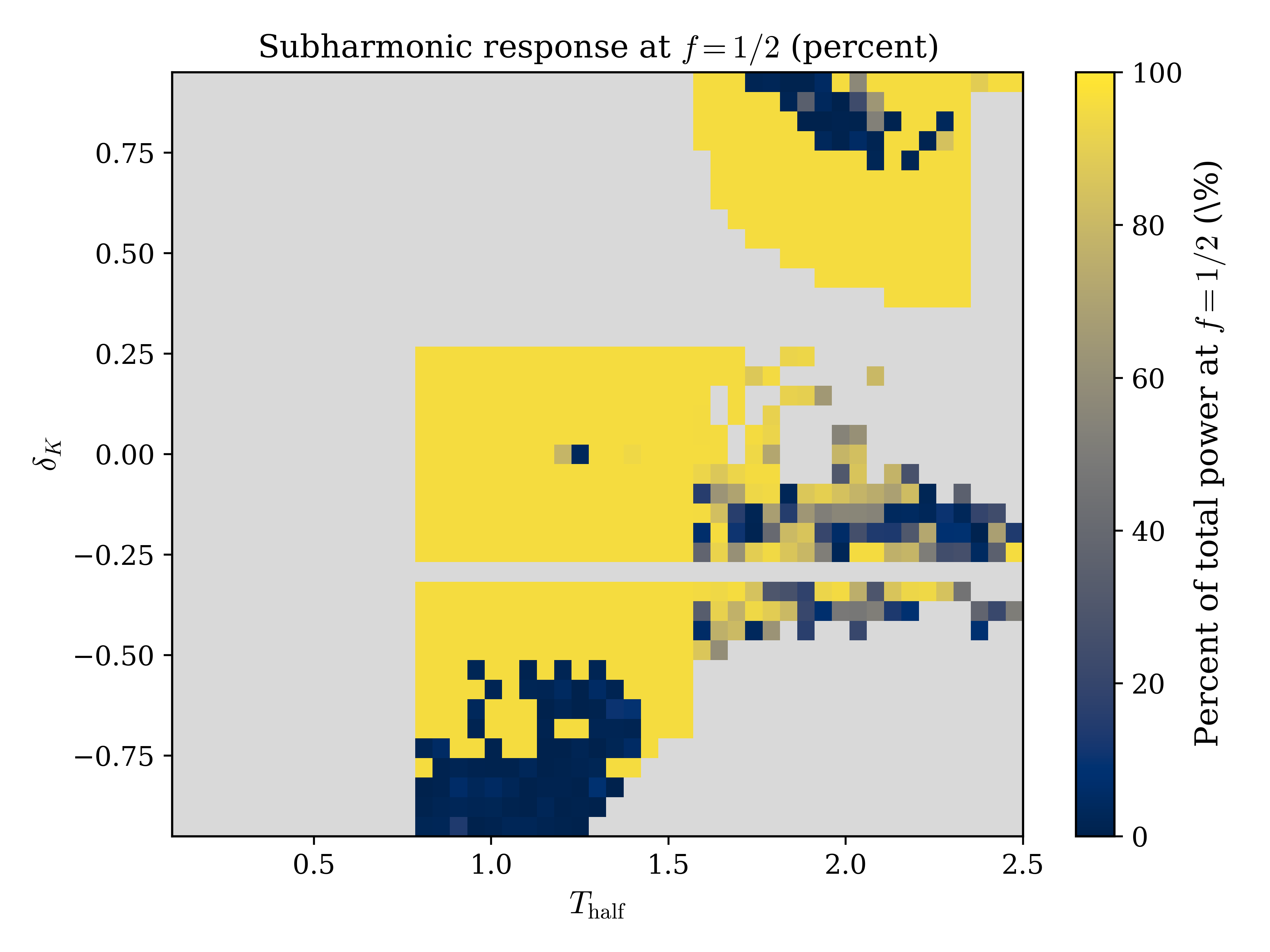}
        \caption{Coherent zero--$\pi$ superposition state.}
        \label{fig:tol025_dtc}
    \end{subfigure}
    \hfill
    \begin{subfigure}{0.49\linewidth}
        \includegraphics[width=\linewidth]{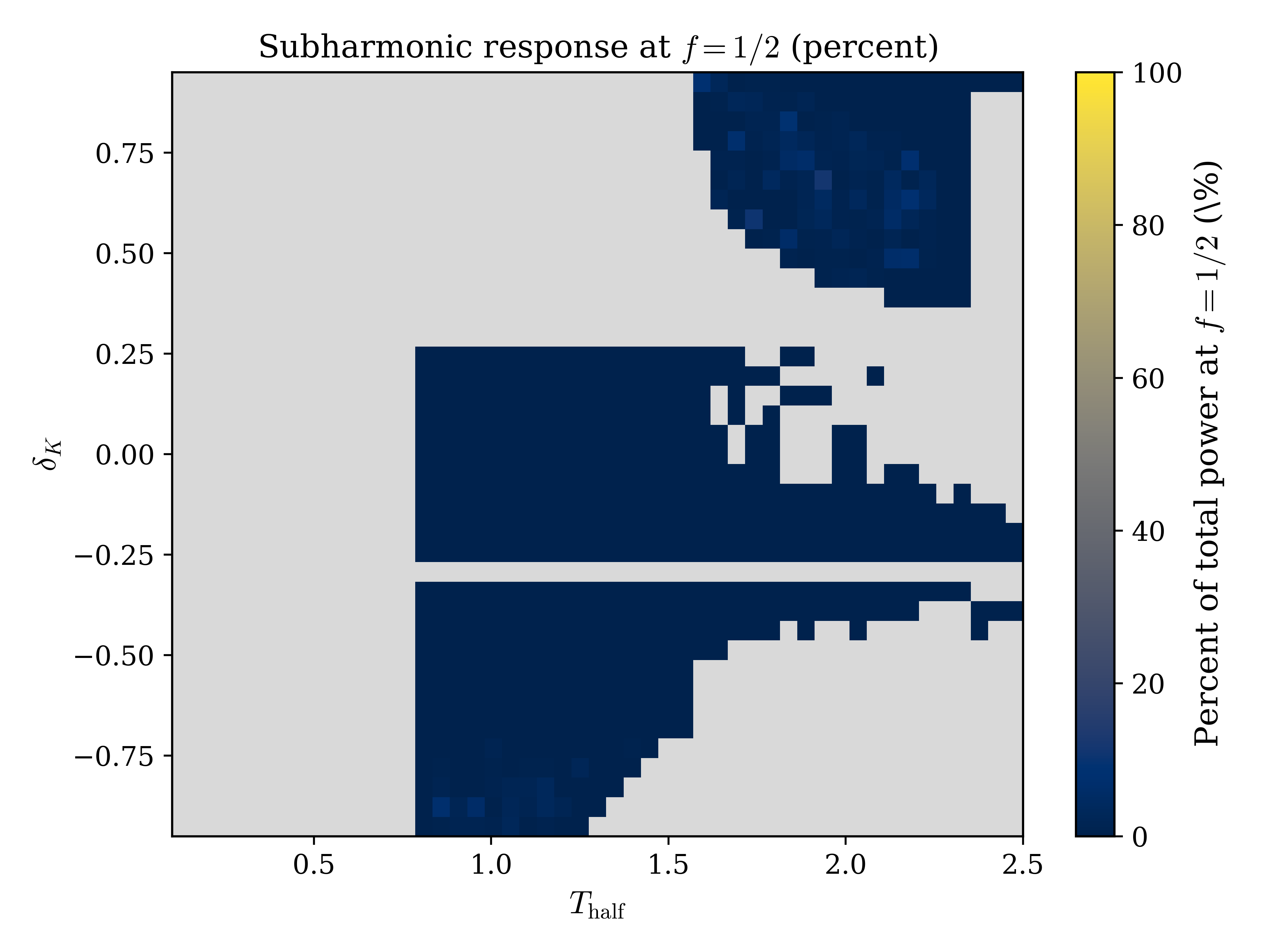}
        \caption{$\pi$-mode Floquet eigenstate (control).}
        \label{fig:tol025_eig}
    \end{subfigure}
    \caption{Phase diagrams recomputed with $\delta_{\mathrm{tol}}=0.025$,
    in the same format as Fig.~2 of the main text (which used the default
    $\delta_{\mathrm{tol}}=0.05$).
    (a)~Coherent zero--$\pi$ superposition: the masked $\mathcal{F}_{1/2}$ map
    is indistinguishable from the main-text result, confirming that the
    identified topological window and the subharmonic response are unaffected
    by halving the tolerance.
    (b)~$\pi$-mode Floquet eigenstate control: the map remains uniformly dark,
    as expected for a stroboscopically stationary state.
    The stark superposition--eigenstate contrast is preserved, demonstrating
    that $\delta_{\mathrm{tol}}=0.05$ is not fine-tuned.}
    \label{fig:tol_robustness}
\end{figure}

\paragraph{Quasienergy tolerance $\delta_{\mathrm{tol}}$.}
A Floquet mode is classified as a zero (or $\pi$) mode if its eigenphase satisfies $|\theta_\alpha| < \delta_{\mathrm{tol}}$ (or $\bigl||\theta_\alpha| - \pi\bigr| < \delta_{\mathrm{tol}}$), with default $\delta_{\mathrm{tol}} = 0.05$.
This parameter matters only if bulk modes approach the symmetry-fixed quasienergies closely enough to enter the classification window.
In the topological regime studied here, the bulk Floquet spectrum remains gapped around both $\theta = 0$ and $\theta = \pi$, while the corresponding edge modes lie well inside those gaps.
We verified robustness by recomputing the phase-diagram diagnostics with $\delta_{\mathrm{tol}} \in \{0.025,0.05\}$.
At the resolution of the computed grids, the pair-count map, the binary proxy flag, and the masked subharmonic fraction $\mathcal{F}_{1/2}$ are unchanged; see Fig.~\ref{fig:tol_robustness}.

\begin{figure}[htbp]
    \centering
    \begin{subfigure}{0.49\linewidth}
        \includegraphics[width=\linewidth]{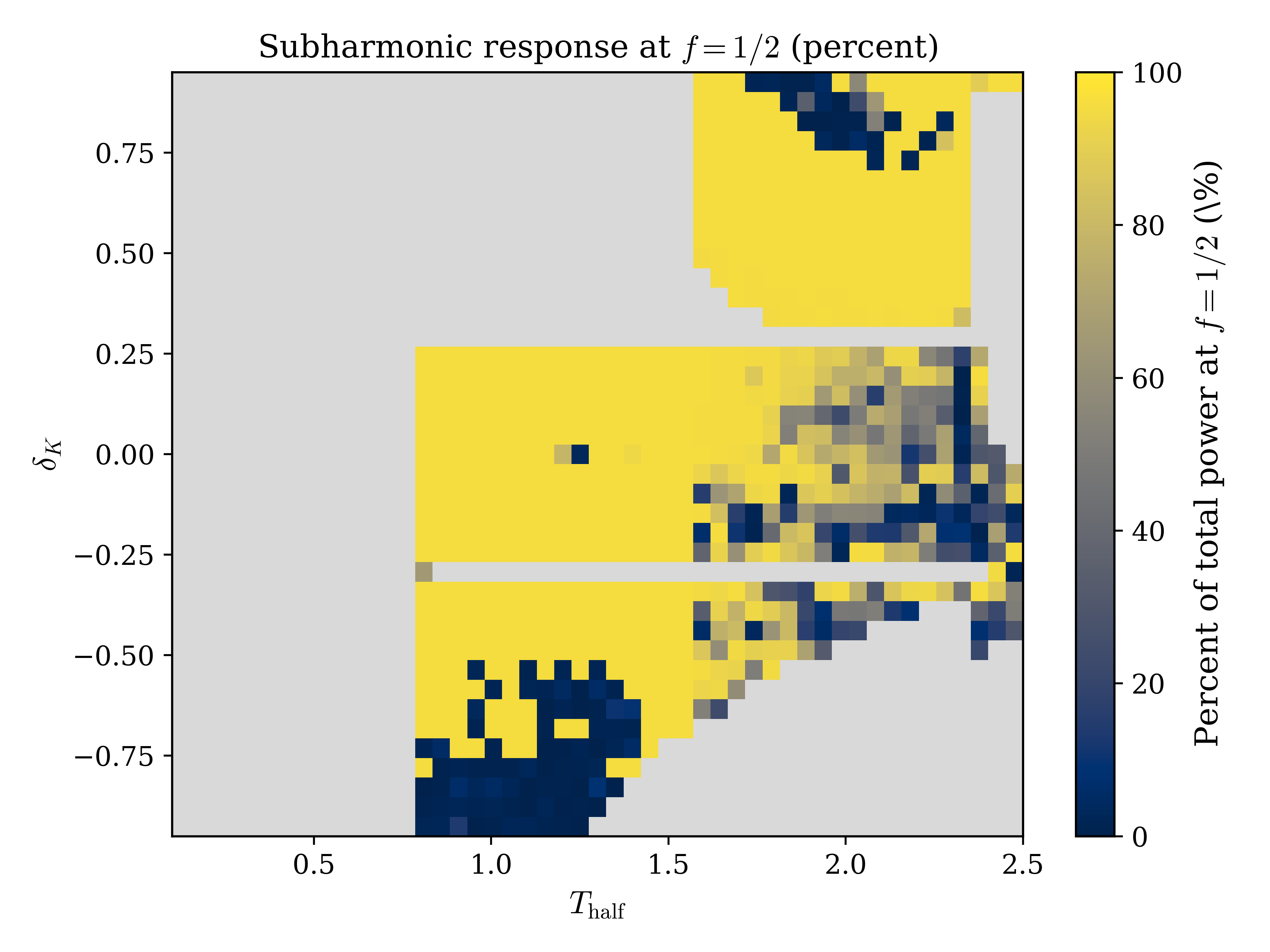}
        \caption{Coherent zero--$\pi$ superposition state.}
        \label{fig:wthr010_dtc}
    \end{subfigure}
    \hfill
    \begin{subfigure}{0.49\linewidth}
        \includegraphics[width=\linewidth]{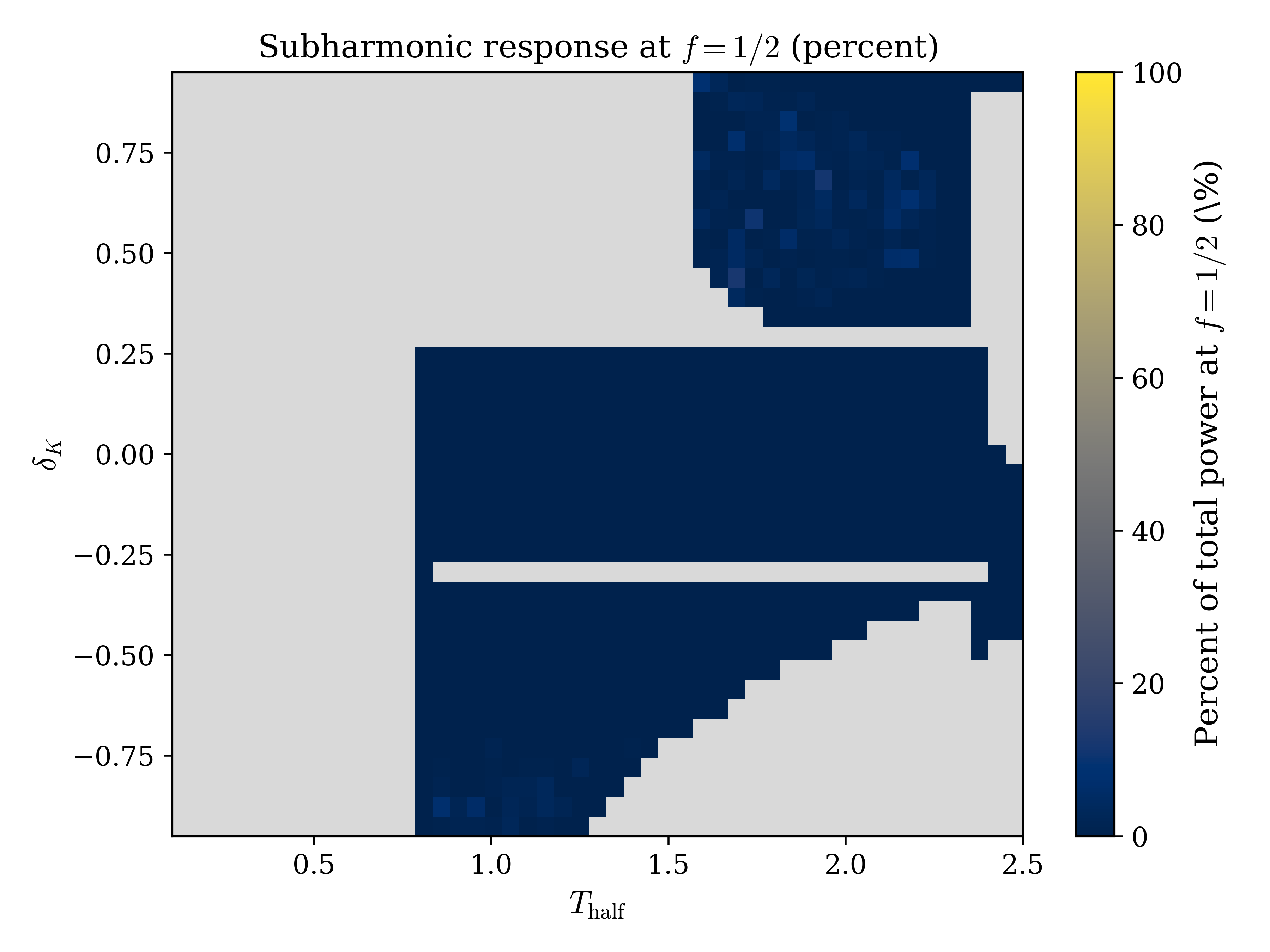}
        \caption{$\pi$-mode Floquet eigenstate (control).}
        \label{fig:wthr010_eig}
    \end{subfigure}
    \caption{Phase diagrams recomputed with $w_{\mathrm{thr}}=0.10$,
    in the same format as Fig.~2 of the main text (which used the default
    $w_{\mathrm{thr}}=0.30$).
    (a)~Coherent zero--$\pi$ superposition: the masked $\mathcal{F}_{1/2}$ map
    is indistinguishable from the main-text result, confirming that the
    topological window and the subharmonic response are unaffected by lowering
    the edge-weight threshold.
    (b)~$\pi$-mode Floquet eigenstate control: the map remains uniformly dark.
    The superposition--eigenstate contrast is preserved within grid resolution and numerical precision, demonstrating
    that $w_{\mathrm{thr}}=0.30$ is not fine-tuned.}
    \label{fig:wthr_robustness}
\end{figure}

\paragraph{Edge weight threshold $w_{\mathrm{thr}}$.}
A mode that has been classified as a zero or $\pi$ mode by the phase tolerance is further required to have edge weight $W_{\mathrm{edge}}(\alpha) \geq w_{\mathrm{thr}}$, where (for normalized eigenmodes)
\begin{equation}
W_{\mathrm{edge}}(\alpha)
=
\sum_{j=1}^{w}\bigl|\Phi_\alpha(j)\bigr|^2
+
\sum_{j=L-w+1}^{L}\bigl|\Phi_\alpha(j)\bigr|^2,
\qquad
w = \bigl\lfloor\sqrt{L}\bigr\rfloor,
\label{eq:Wedge_def}
\end{equation}
is the fraction of the mode's probability contained in two boundary windows of width $w$.
For a bulk-extended mode this weight is of order $2w/L$, whereas for a strongly edge-localized mode with localization length $\xi \ll w$ it remains close to unity.
We verify explicitly by recomputing the diagnostics with $w_{\mathrm{thr}} \in \{0.10,0.30\}$, which yields unchanged proxy and masked-response maps within numerical resolution; see Fig.~\ref{fig:wthr_robustness}.
The default choice $w_{\mathrm{thr}} = 0.30$ is therefore not fine-tuned.

\paragraph{System-size dependence and robustness to the subsystem choice.}
To check that the observed subharmonic response is not a small-system artifact, we repeated the main diagnostics for several chain lengths.
Inside the topological regime, the tracked time-domain signal and the Fourier metric $\mathcal{F}_{1/2}$ remain stable across the sizes studied, consistent with the boundary-localized character of the relevant Floquet modes once $L$ is sufficiently larger than their localization length.

\begin{figure}[htbp]
    \centering
    \begin{subfigure}{0.48\linewidth}
        \includegraphics[width=\linewidth]{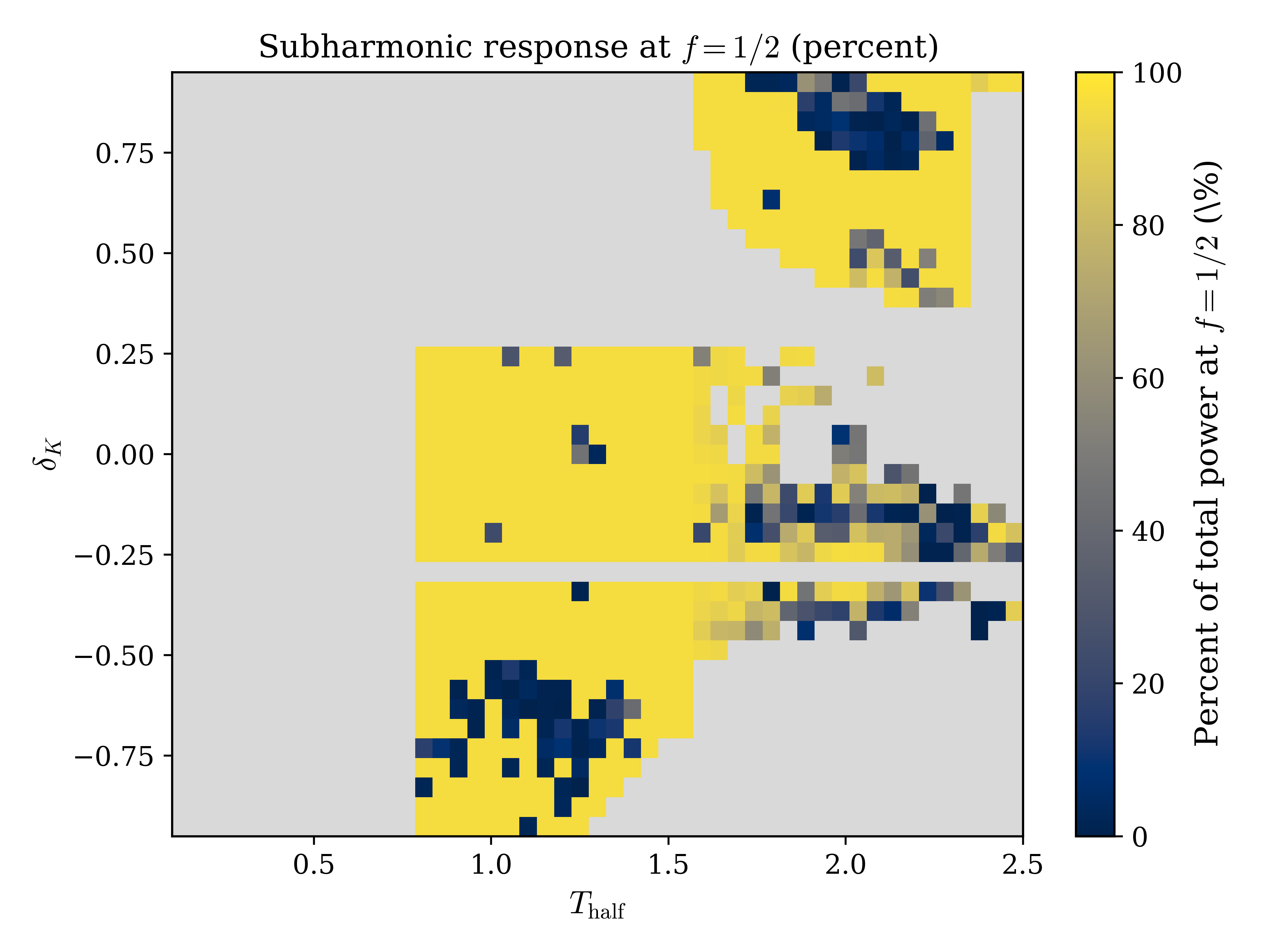}
        \caption{$L_A=12$, superposition.}
        \label{fig:LA12_dtc}
    \end{subfigure}
    \hfill
    \begin{subfigure}{0.48\linewidth}
        \includegraphics[width=\linewidth]{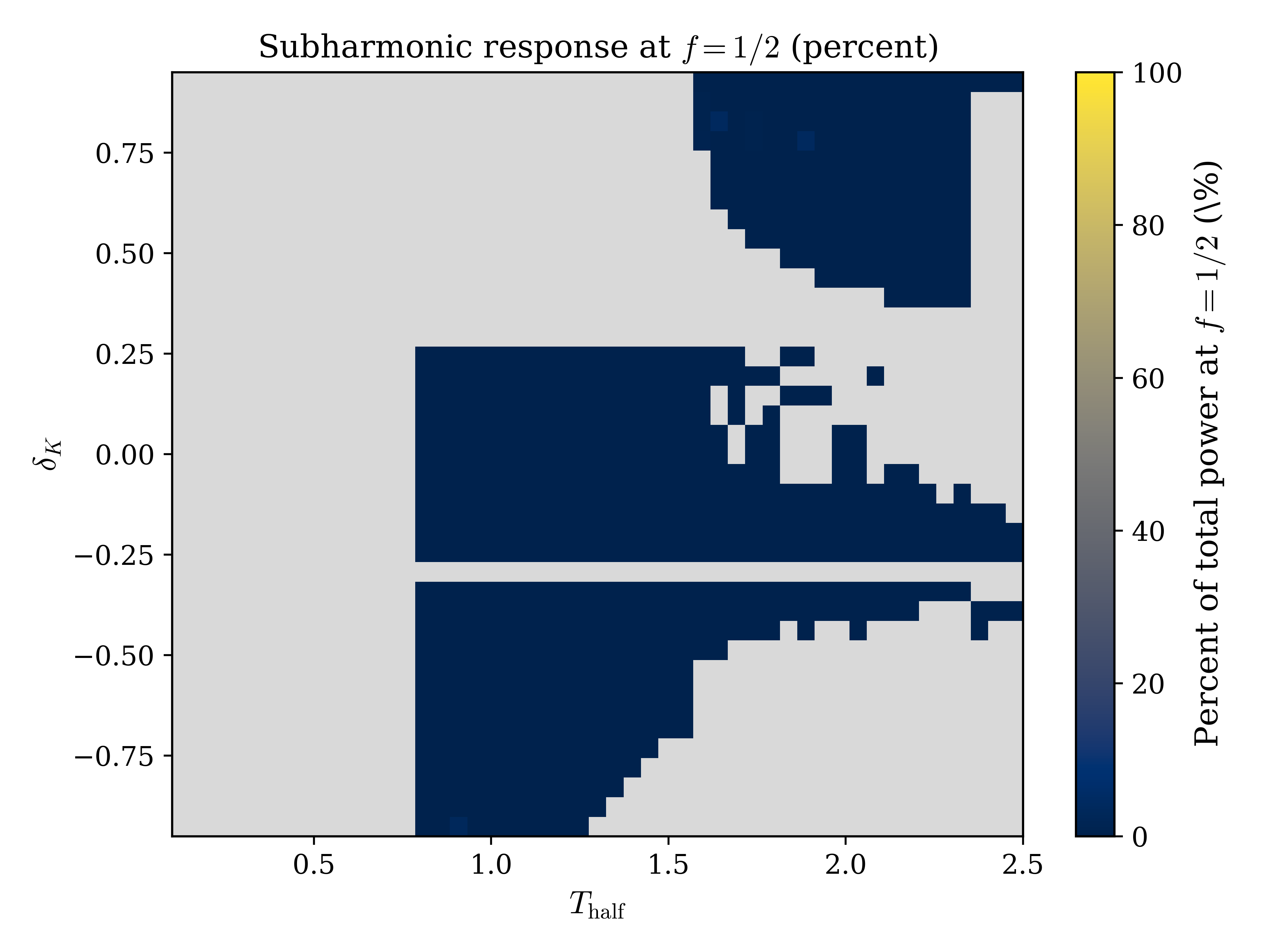}
        \caption{$L_A=12$, eigenstate control.}
        \label{fig:LA12_eig}
    \end{subfigure}

    \vspace{0.4em}

    \begin{subfigure}{0.48\linewidth}
        \includegraphics[width=\linewidth]{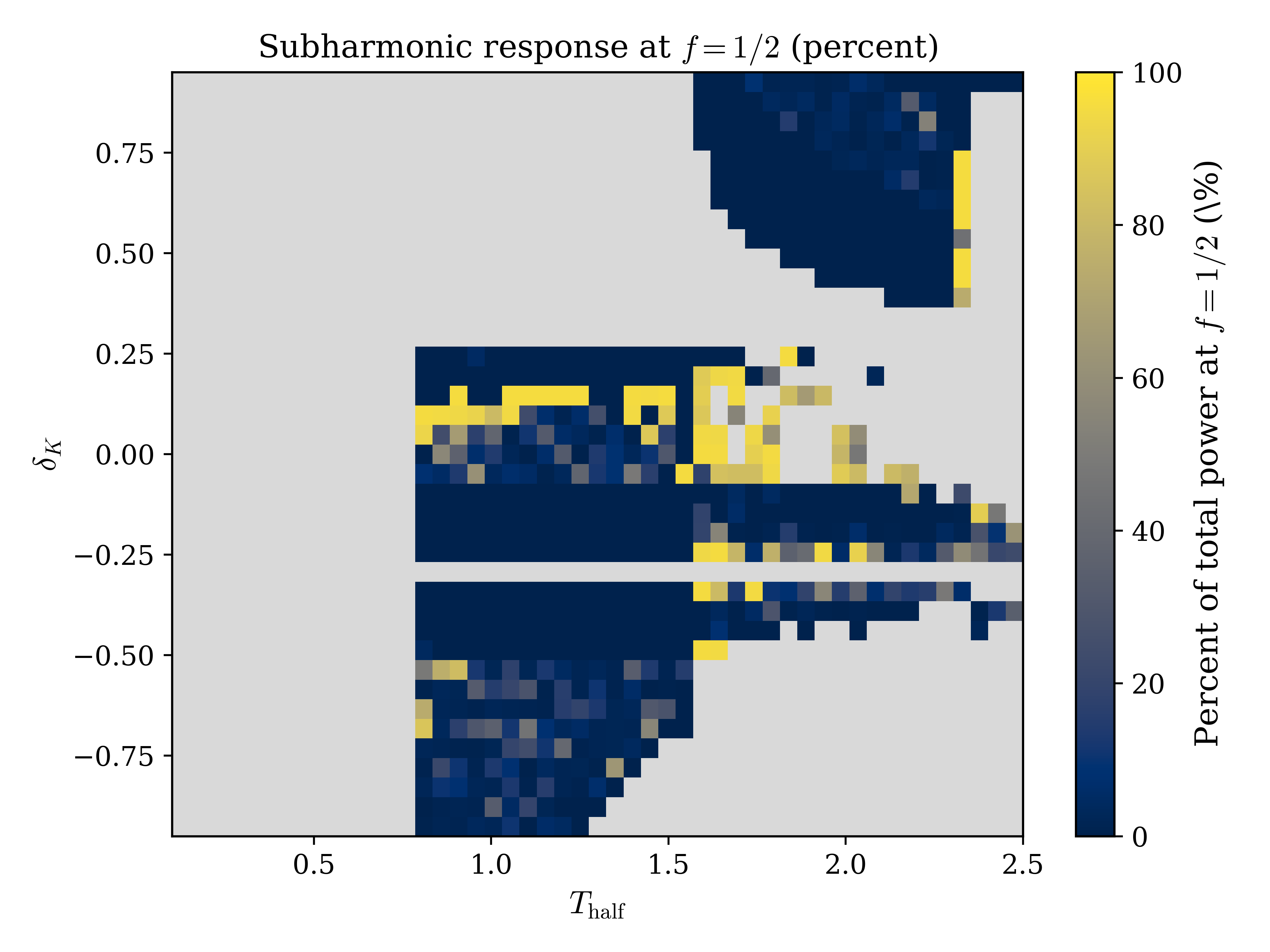}
        \caption{$L_A=100$, superposition.}
        \label{fig:LA100_dtc}
    \end{subfigure}
    \hfill
    \begin{subfigure}{0.48\linewidth}
        \includegraphics[width=\linewidth]{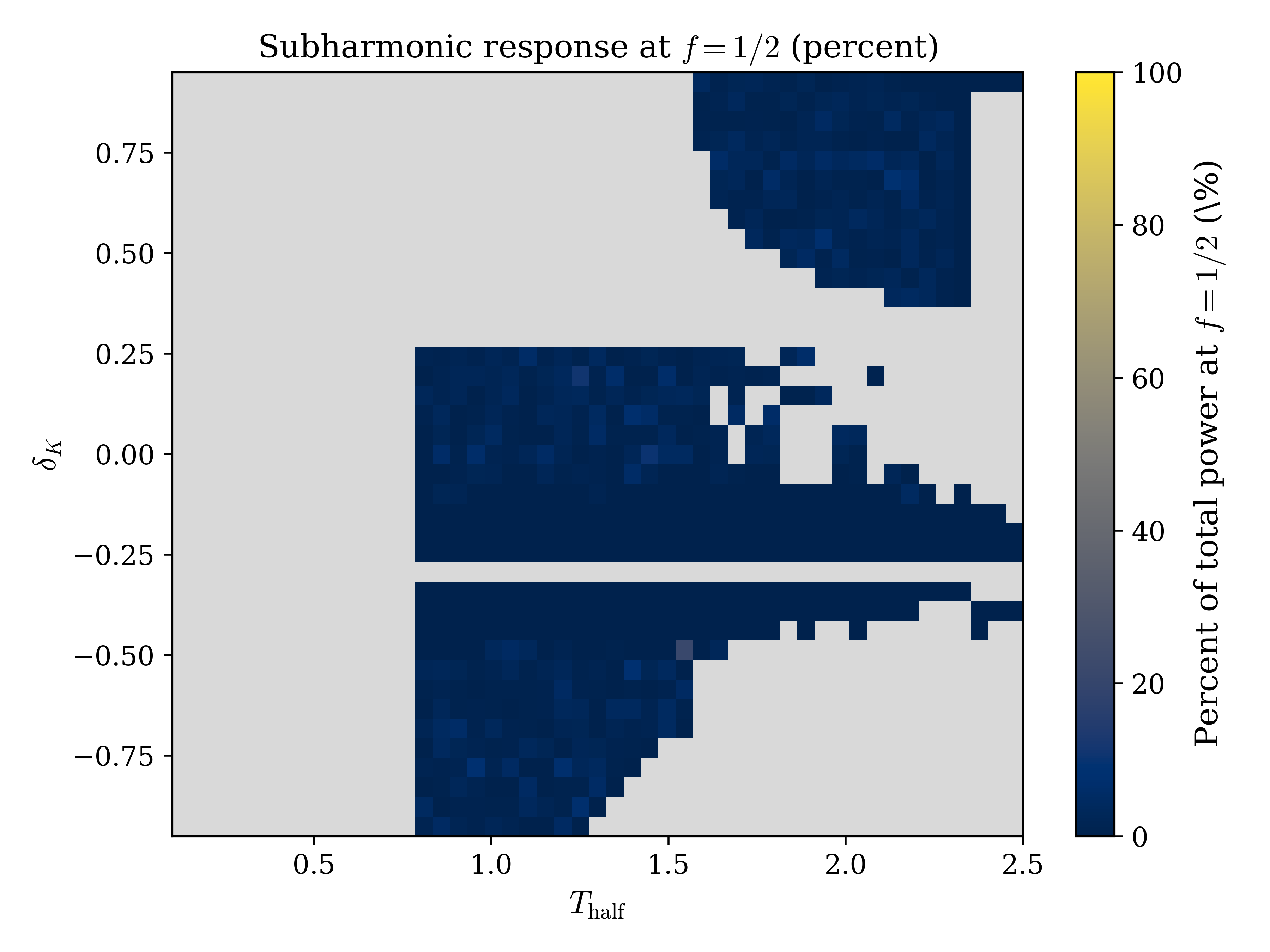}
        \caption{$L_A=100$, eigenstate control.}
        \label{fig:LA100_eig}
    \end{subfigure}

    \vspace{0.4em}

    \begin{subfigure}{0.48\linewidth}
        \includegraphics[width=\linewidth]{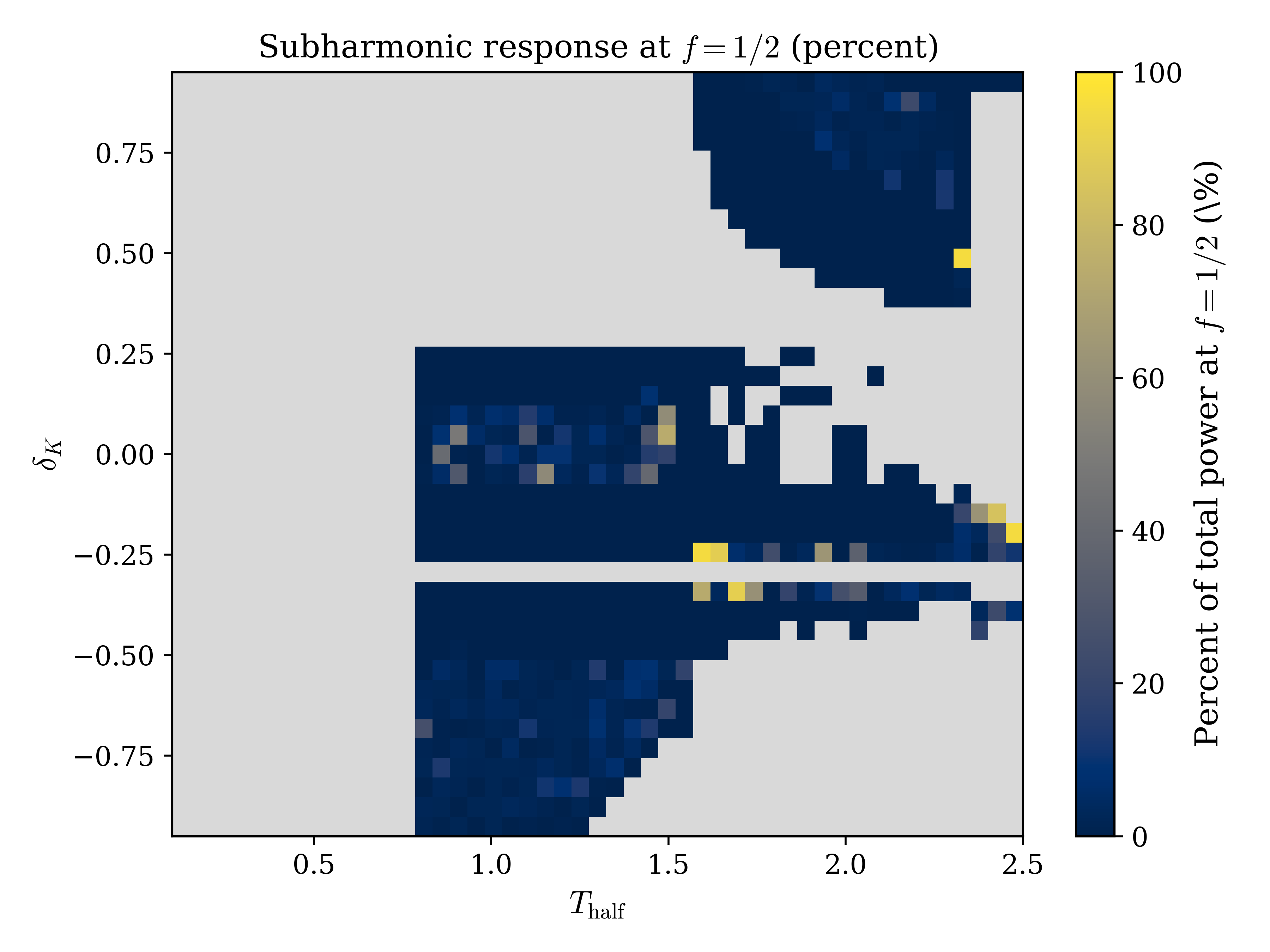}
        \caption{$L_A=200$, superposition.}
        \label{fig:LA200_dtc}
    \end{subfigure}
    \hfill
    \begin{subfigure}{0.48\linewidth}
        \includegraphics[width=\linewidth]{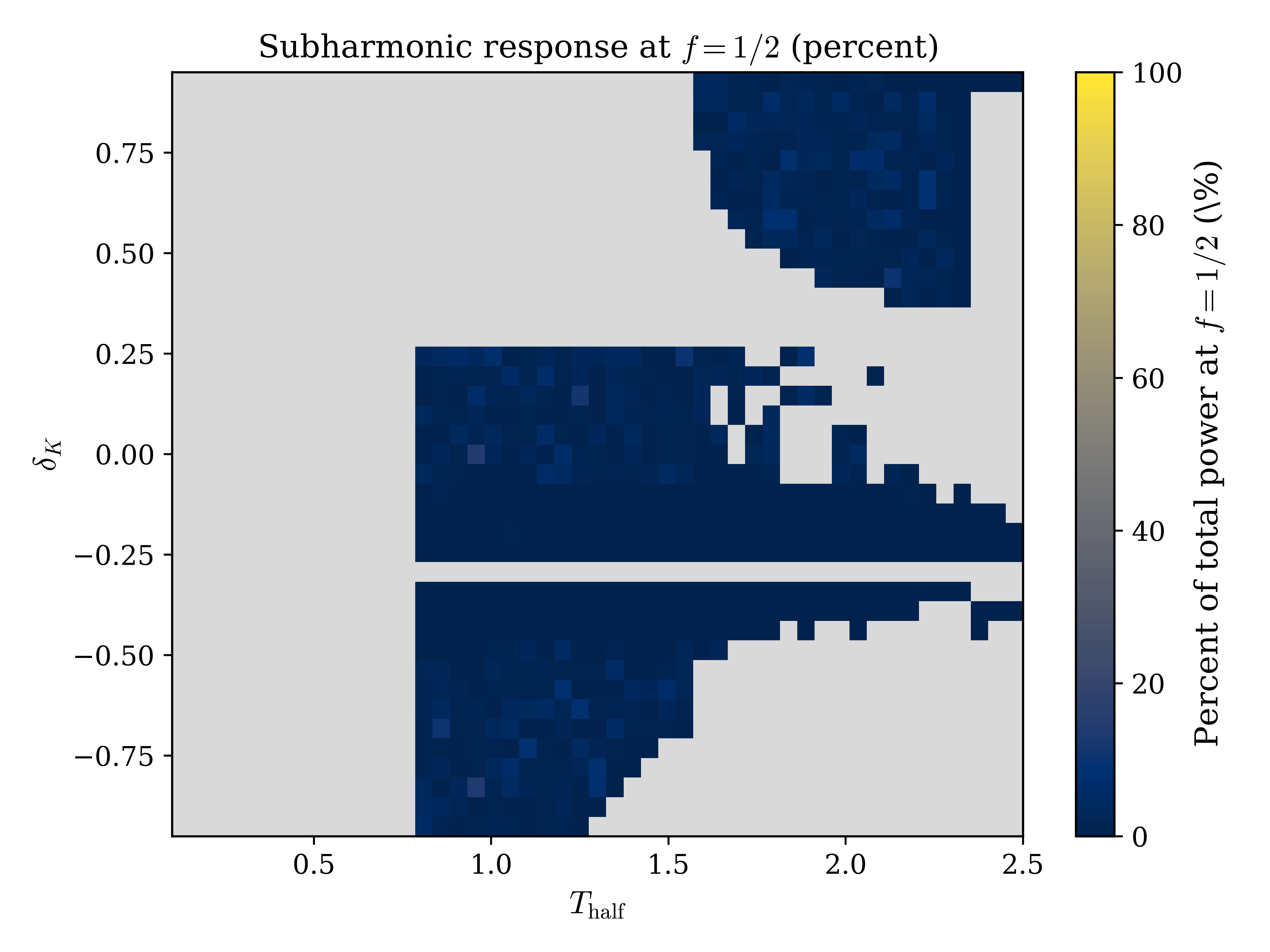}
        \caption{$L_A=200$, eigenstate control.}
        \label{fig:LA200_eig}
    \end{subfigure}
    \caption{Robustness of the masked $\mathcal{F}_{1/2}$ phase diagram to the
    subsystem size $L_A$ at fixed total size $L=500$.
    Each row corresponds to one value of $L_A$ (from top to bottom:
    $12$, $100$, $200$); left column is the coherent zero--$\pi$ superposition
    state and right column is the $\pi$-mode Floquet eigenstate control, in
    the same format as Fig.~2 of the main text (which already shows the
    default $L_A=22$).
    For $L_A=12$ the superposition map is indistinguishable from the
    main-text result, confirming that the default choice is well converged.
    Increasing $L_A$ to $100$ and $200$ progressively shrinks and eventually
    washes out the high-$\mathcal{F}_{1/2}$ region, consistent with bulk-dominated
    degrees of freedom diluting the edge-sector signal inside $C_A$.
    The eigenstate control columns remain uniformly dark at all $L_A$, as
    expected for a stroboscopically stationary state.}
    \label{fig:LA_robustness}
\end{figure}

We also tested the dependence on the subsystem size used in the entanglement construction.
For $L=500$, reducing $L_A$ from $22$ to $12$ leaves the response essentially unchanged, while increasing $L_A$ to $100$ and $200$ progressively weakens the isolated period-$2T$ signal by admixing more bulk-dominated degrees of freedom into $C_A$; see Fig.~\ref{fig:LA_robustness}.

\begin{figure}[htbp]
    \centering
    \begin{subfigure}{0.49\linewidth}
        \includegraphics[width=\linewidth]{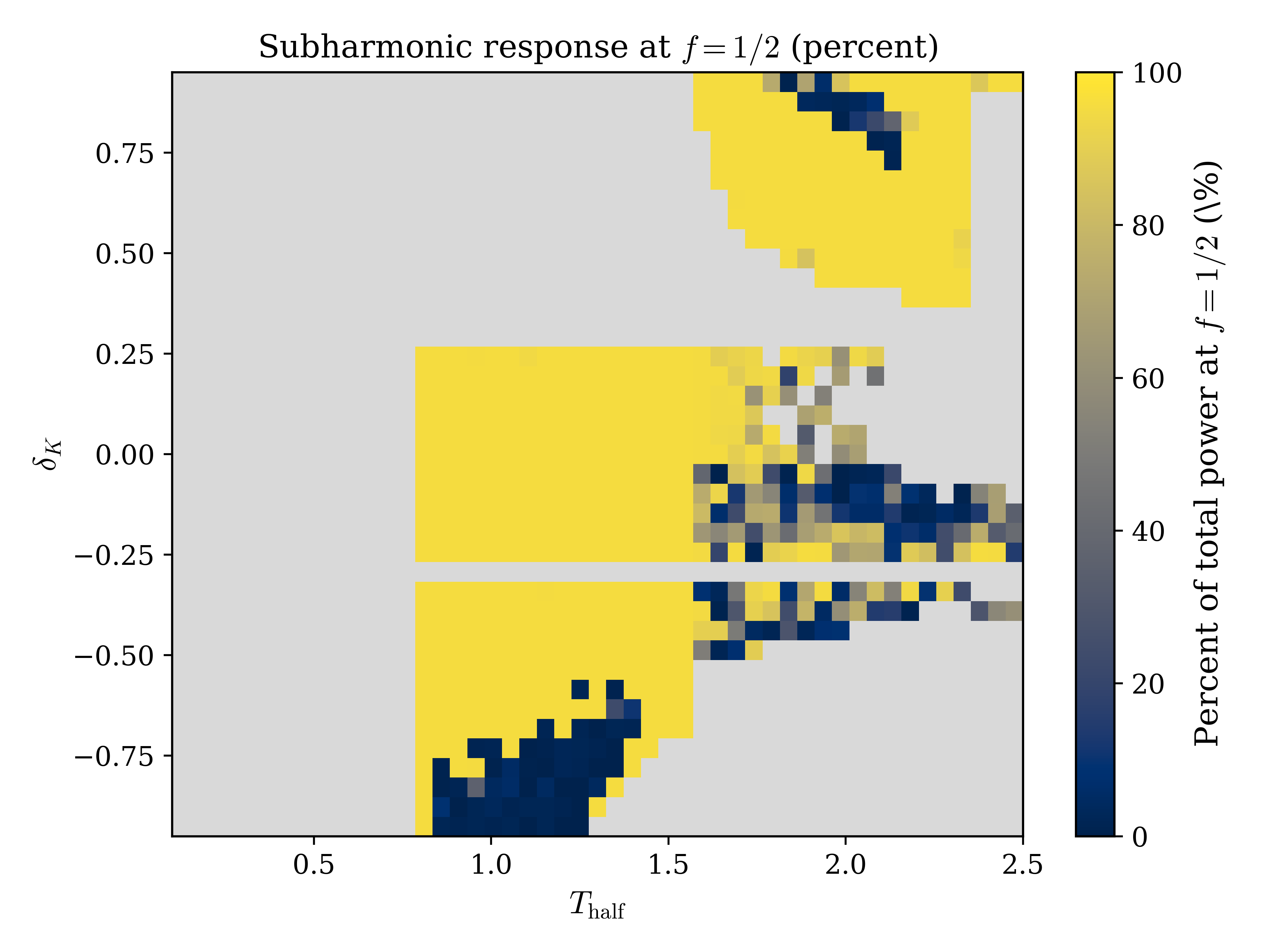}
        \caption{$L=400$, superposition.}
        \label{fig:L400_dtc}
    \end{subfigure}
    \hfill
    \begin{subfigure}{0.49\linewidth}
        \includegraphics[width=\linewidth]{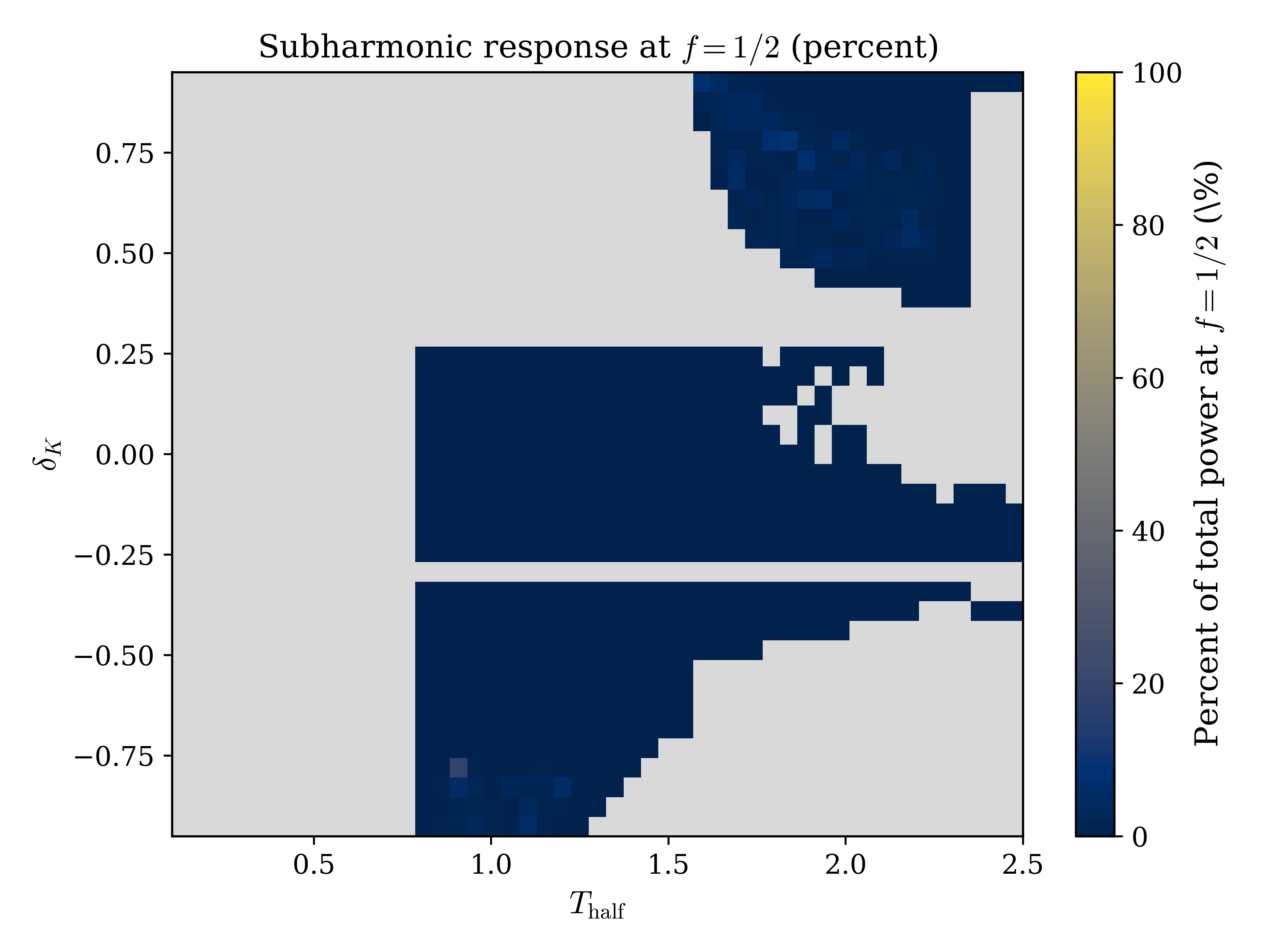}
        \caption{$L=400$, eigenstate control.}
        \label{fig:L400_eig}
    \end{subfigure}

    \vspace{0.4em}

    \begin{subfigure}{0.49\linewidth}
        \includegraphics[width=\linewidth]{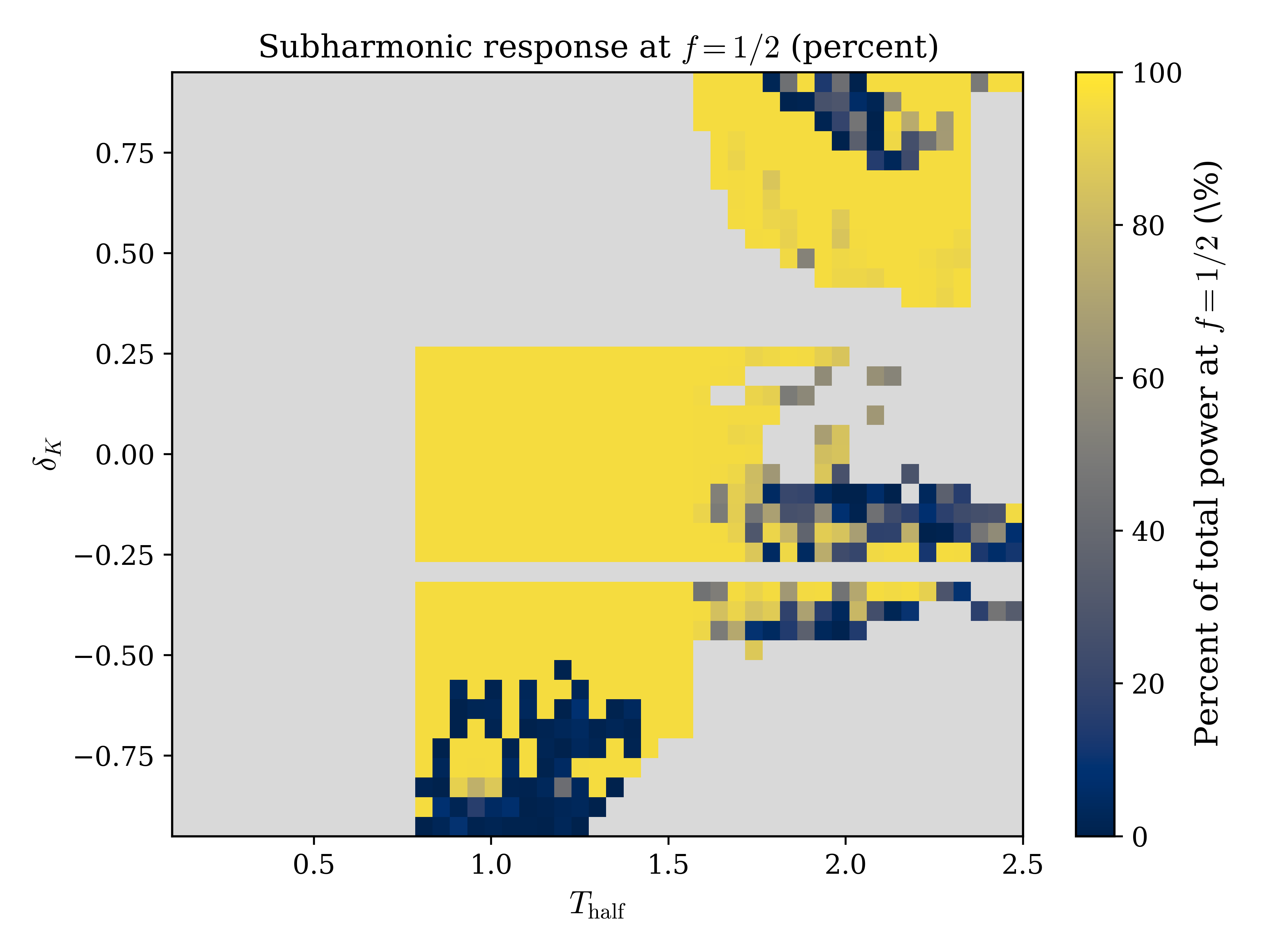}
        \caption{$L=450$, superposition.}
        \label{fig:L450_dtc}
    \end{subfigure}
    \hfill
    \begin{subfigure}{0.49\linewidth}
        \includegraphics[width=\linewidth]{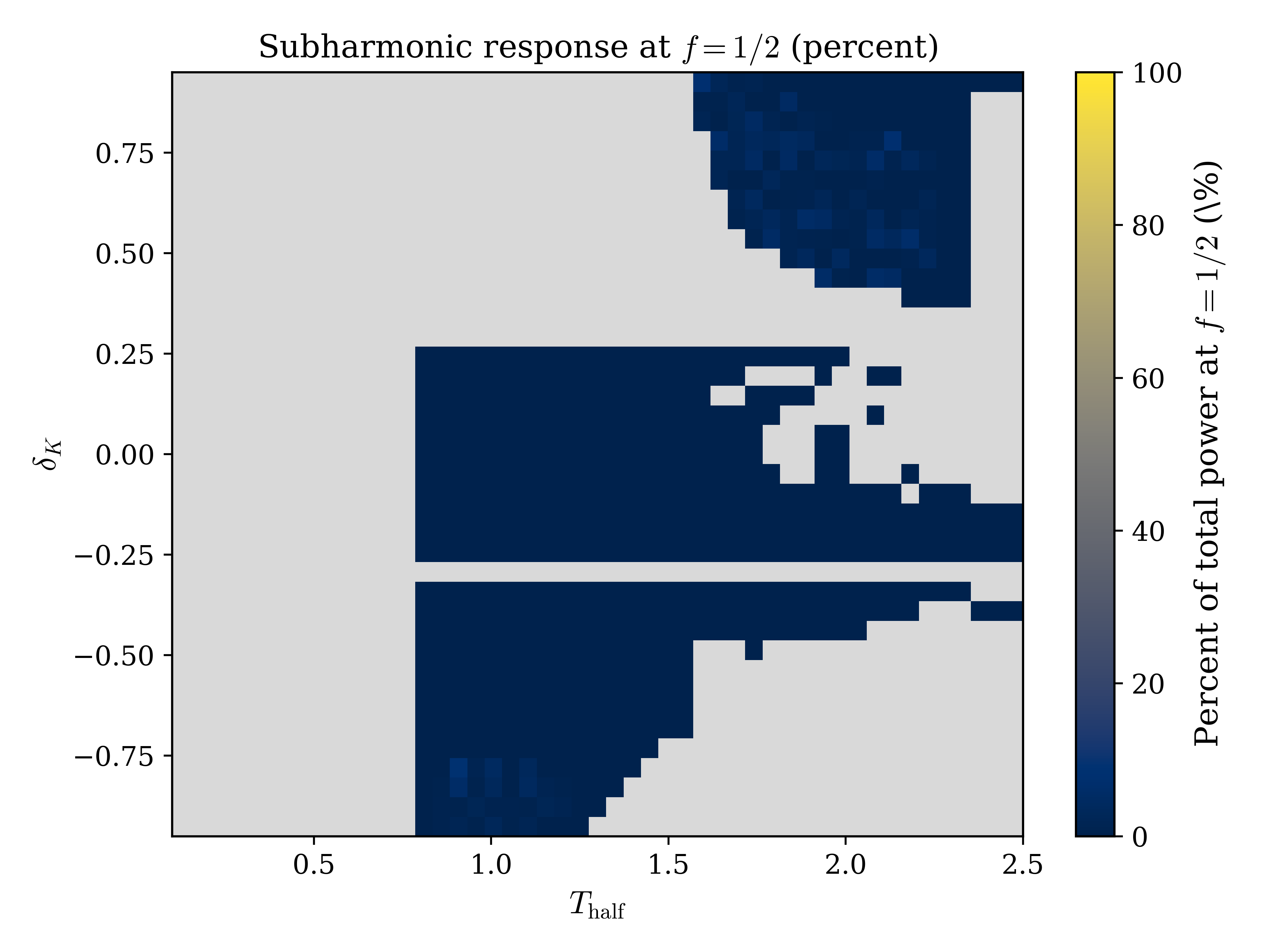}
        \caption{$L=450$, eigenstate control.}
        \label{fig:L450_eig}
    \end{subfigure}
    \caption{Robustness of the masked $\mathcal{F}_{1/2}$ phase diagram to the
    total system size at fixed subsystem size $L_A=22$.
    Each row corresponds to one value of $L$ (top: $L=400$; bottom: $L=450$);
    left column is the coherent zero--$\pi$ superposition state and right column
    is the $\pi$-mode Floquet eigenstate control, in the same format as Fig.~2
    of the main text (which already shows $L=500$, $L_A=22$).
    Across all sizes the superposition map is indistinguishable from the
    main-text result, consistent with the boundary-localized character of the
    relevant Floquet edge modes.
    The eigenstate control columns remain uniformly dark at all $L$, as expected
    for a stroboscopically stationary state.}
    \label{fig:L_robustness}
\end{figure}

Finally, fixing $L_A=22$ and varying the total chain length from $L=400$ to $450$ to $500$ leaves both the tracked time series and $\mathcal{F}_{1/2}$ unchanged within numerical resolution; see Fig.~\ref{fig:L_robustness}.

\section{Overlap-ranked levels and the even-odd diagnostic}
\label{sec:eo_diagnostic}

The Fourier metric $\mathcal{F}_{1/2}$ (Sec.~\ref{sec:observable}) quantifies how much of the total oscillatory power in a time series falls at frequency $f=1/2$. It does not, however, distinguish between a level that alternates cleanly between two stable values and one that merely has unequal mean values on even and odd stroboscopic periods while fluctuating substantially within each sub-sequence. We therefore introduce a complementary diagnostic that enforces genuine stationarity of both sub-sequences.

\subsection{Even-odd sub-sequence criterion}
\label{subsec:eo_criterion}

Let $\eta(nT)$, $n=0,1,\dots,N$, be a tracked entanglement-energy time series. Define the even and odd stroboscopic sub-sequences
\begin{equation}
\eta_{\rm e} = \bigl\{\eta(0),\,\eta(2T),\,\eta(4T),\dots\bigr\},
\qquad
\eta_{\rm o} = \bigl\{\eta(T),\,\eta(3T),\,\eta(5T),\dots\bigr\}.
\end{equation}
A genuine period-$2T$ oscillator alternates between two stationary values: $\eta_{\rm e}$ is approximately constant at one value and $\eta_{\rm o}$ is approximately constant at a second, different value.

To quantify this, define the signed amplitude and the signal-to-noise ratio (SNR):
\begin{equation}
\Delta_{eo} = \overline{\eta_{\rm e}} - \overline{\eta_{\rm o}},
\qquad
\mathrm{SNR} = \frac{|\Delta_{eo}|}{\max\!\bigl(\sigma(\eta_{\rm e}),\,\sigma(\eta_{\rm o})\bigr)},
\label{eq:snr_def}
\end{equation}
where $\overline{(\cdot)}$ and $\sigma(\cdot)$ denote the mean and standard deviation of a sub-sequence, respectively. The denominator measures within-sub-sequence fluctuations; a large SNR means each sub-sequence is individually flat, not merely that the two means differ.

We report the amplitude $|\Delta_{eo}|$ only when $\mathrm{SNR} \geq \mathrm{SNR}_{\min}$; otherwise it is set to \texttt{NaN}:
\begin{equation}
|\Delta_{eo}|_{\rm reported} =
\begin{cases}
|\Delta_{eo}|, & \mathrm{SNR} \geq \mathrm{SNR}_{\min}, \\
\texttt{NaN},  & \mathrm{SNR} < \mathrm{SNR}_{\min}.
\end{cases}
\label{eq:eo_reported}
\end{equation}
Throughout this work we use $\mathrm{SNR}_{\min} = 3.0$, a conservative threshold (given that we achieve values of order $10^8$ for the locked subharmonic response; see Tables~\ref{tab:overlap_L500} and~\ref{tab:overlap_L1000} below) that requires the gap between the two sub-sequence means to be at least three times the within-sub-sequence fluctuation. A \texttt{NaN} entry indicates that the level is not a clean stable period-$2T$ oscillator; it does not indicate missing data.

\subsection{Overlap ranking and table structure}
\label{subsec:eo_table}

At each stroboscopic time, the $L_A$ eigenvectors of $C_A$ are ranked in descending order of their squared overlap with the reference vector $\mathbf{v}_{\rm ref}$ (Eq.~\eqref{eq:ref_vec}). Rank 1 is the level with the largest overlap, rank 2 the next, and so on. If $\phi_r(nT)$ denotes the eigenvector at rank $r$ after this sorting, then the time-averaged overlap percentage is $\bar{o}_r = 100 \times \langle |\mathbf{v}_{\rm ref}^\dagger \phi_r(nT)|^2 \rangle_n$, which quantifies how consistently that rank tracks the reference.

For reporting, we retain the top 5 and bottom 5 overlap-ranked levels. This range is sufficient to show both the high-overlap levels, where the period--$2T$ signal is cleanest, and low-overlap levels, which serve as a baseline.

\subsection{Numerical results}
\label{subsec:eo_results}

Tables~\ref{tab:overlap_L500} and \ref{tab:overlap_L1000} summarize the
results for $L=500$, $L_A=22=\lfloor\sqrt{500}\rfloor$ and
$L=1000$, $L_A=31=\lfloor\sqrt{1000}\rfloor$, respectively, at
$\delta_0=-0.30$, $\delta_K=0.80$ (parameters as in Table~\ref{tab:params}).
Two initial conditions are compared in each table: the superposition state
and the $\pi$-mode Floquet eigenstate.

For the superposition state at $L=500$, ranks 1 and 2 are the cleanest
period-$2T$ oscillators: $\mathcal{F}_{1/2} \approx 0.960$,
$|\Delta_{eo}| \approx 3.94$, and $\mathrm{SNR} \sim 10^8$.
Both carry $\approx 24\%$ mean overlap with the reference vector.
Ranks 5 and 18 also pass the criterion ($\mathrm{SNR} \approx 6.5$) with
larger amplitudes ($|\Delta_{eo}| \approx 35.9$ and $26.7$, respectively)
but much lower overlap ($5.8\%$ and $0.08\%$).
The remaining reported ranks fail: their $|\Delta_{eo}|$ entries are
\texttt{NaN} because the within-sub-sequence fluctuations are too large
relative to the even-odd gap.

At $L=1000$ the subharmonic fraction $\mathcal{F}_{1/2}\approx0.960$ for
all top-ranked levels, identical to the $L=500$ result, confirming that the signal is stable under doubling the system size
with $L_A$ scaled as $\lfloor\sqrt{L}\rfloor$.
All five top-ranked levels now pass the even-odd criterion, including ranks 3
and 4 which failed at $L=500$; their SNR reaches $\sim10^6$, showing that the
larger system reinforces rather than weakens the period-$2T$ structure in the
edge-dominated sector.
The $|\Delta_{eo}|$ amplitudes are not directly comparable between the two
system sizes because the entanglement energies are not extensive in $L$ and
their scale depends on $L_A$; the appropriate finite-size diagnostic is
$\mathcal{F}_{1/2}$, which is unchanged.

For the $\pi$-mode eigenstate control at both system sizes, all reported
$|\Delta_{eo}|$ entries are \texttt{NaN}, the $\mathrm{SNR}$ remains small,
and $\mathcal{F}_{1/2}$ is essentially zero apart from tiny numerical residue.
This confirms that the existence of a $\pi$ mode in the Floquet spectrum is
not sufficient for period doubling; the nonequilibrium coherent superposition
is additionally required.
This selective pattern confirms both conceptual points of
Sec.~\ref{sec:tracking}: the largest--overlap rule isolates the cleanest
signal, but other levels can independently satisfy the period--$2T$ criterion,
and the effect is robust to finite-size scaling.

\begin{table}[h]
\centering
\caption{Overlap-ranked diagnostic for $L=500$, $L_A=22$, $\delta_0=-0.30$, $\delta_K=0.80$, $N=800$ periods. $\bar{o}$: mean overlap with reference vector (\%). $|\Delta_{eo}|$: even-odd amplitude (reported only when $\mathrm{SNR}\geq 3.0$; otherwise \texttt{NaN}). $\mathcal{F}_{1/2}$: subharmonic fraction. Top 5 and bottom 5 overlap-ranked levels are shown.}
\label{tab:overlap_L500}
\begin{tabular}{ccccccccc}
\toprule
& \multicolumn{4}{c}{Superposition state} & \multicolumn{4}{c}{$\pi$-mode eigenstate} \\
\cmidrule(lr){2-5}\cmidrule(lr){6-9}
Rank & $\bar{o}$ & $|\Delta_{eo}|$ & SNR & $\mathcal{F}_{1/2}$ & $\bar{o}$ & $|\Delta_{eo}|$ & SNR & $\mathcal{F}_{1/2}$ \\
\midrule
1  & 24.05 & 3.944  & $9.6\times10^8$ & 0.960 & 59.69 & NaN & 0.15 & 0.000 \\
2  & 24.05 & 3.944  & $3.8\times10^8$ & 0.960 & 21.34 & NaN & 0.06 & 0.000 \\
3  & 14.10 & NaN    & 0.89            & 0.288 & 11.48 & NaN & 0.03 & 0.000 \\
4  & 14.10 & NaN    & 0.89            & 0.288 &  4.05 & NaN & 0.03 & 0.000 \\
5  &  5.77 & 35.878 & 6.53            & 0.915 &  2.04 & NaN & 0.08 & 0.003 \\
\midrule
18 &  0.08 & 26.689 & 6.59            & 0.910 &  0.00 & NaN & 0.14 & 0.000 \\
19 &  0.05 & NaN    & 1.64            & 0.524 &  0.00 & NaN & 0.04 & 0.000 \\
20 &  0.05 & NaN    & 1.64            & 0.524 &  0.00 & NaN & 0.05 & 0.000 \\
21 &  0.00 & NaN    & 0.90            & 0.257 &  0.00 & NaN & 0.11 & 0.001 \\
22 &  0.00 & NaN    & 0.90            & 0.257 &  0.00 & NaN & 0.11 & 0.009 \\
\bottomrule
\end{tabular}
\end{table}

\begin{table}[h]
\centering
\caption{Overlap-ranked diagnostic for $L=1000$, $L_A=31=\lfloor\sqrt{1000}\rfloor$,
$\delta_0=-0.30$, $\delta_K=0.80$, $N=800$ periods. Columns as in Table~\ref{tab:overlap_L500}.
The subharmonic fraction $\mathcal{F}_{1/2}\approx0.960$ is identical to the
$L=500$ result to four significant figures, confirming finite-size robustness
of the subharmonic response when $L_A$ is scaled as $\lfloor\sqrt{L}\rfloor$.}
\label{tab:overlap_L1000}
\begin{tabular}{ccccccccc}
\toprule
& \multicolumn{4}{c}{Superposition state} & \multicolumn{4}{c}{$\pi$-mode eigenstate} \\
\cmidrule(lr){2-5}\cmidrule(lr){6-9}
Rank & $\bar{o}$ & $|\Delta_{eo}|$ & SNR & $\mathcal{F}_{1/2}$
     & $\bar{o}$ & $|\Delta_{eo}|$ & SNR & $\mathcal{F}_{1/2}$ \\
\midrule
1  & 23.87 & 45.635 & $3.7\times10^{6}$ & 0.960 & 42.83 & NaN & 0.04 & 0.002 \\
2  & 23.87 & 45.634 & $2.5\times10^{6}$ & 0.960 & 35.65 & NaN & 0.15 & 0.005 \\
3  & 16.46 & 2.221  & $1.5\times10^{6}$ & 0.960 & 11.16 & NaN & 0.03 & 0.000 \\
4  & 16.46 & 2.221  & $2.1\times10^{6}$ & 0.960 &  3.92 & NaN & 0.07 & 0.000 \\
5  &  4.83 & 42.385 & $2.2\times10^{5}$ & 0.960 &  2.53 & NaN & 0.00 & 0.000 \\
\midrule
27 &  0.00 & NaN    & 0.05 & 0.005 & 0.00 & NaN & 0.04 & 0.005 \\
28 &  0.00 & NaN    & 0.10 & 0.000 & 0.00 & NaN & 0.01 & 0.002 \\
29 &  0.00 & NaN    & 0.39 & 0.061 & 0.00 & NaN & 0.04 & 0.000 \\
30 &  0.00 & NaN    & 0.06 & 0.011 & 0.00 & NaN & 0.00 & 0.000 \\
31 &  0.00 & NaN    & 0.04 & 0.000 & 0.00 & NaN & 0.00 & 0.000 \\
\bottomrule
\end{tabular}
\end{table}

    \bibliography{refs.bib}